\documentclass[a4paper,11pt]{article}
\pdfoutput=1
\usepackage[colorlinks,bookmarksnumbered]{hyperref}
\usepackage{graphicx,xcolor}
\usepackage{amsmath,amsfonts}
\usepackage{authblk}   
\usepackage[right= 1in, left= 1in, top=1.15in, bottom=1.25in]{geometry}
\usepackage{fancyhdr}
\usepackage{tocbibind}
\usepackage{enumitem}
\usepackage{graphpap}
 
\DeclareMathOperator{\sym}{sym}
 
\DeclareMathOperator{\trace}{tr}
\DeclareMathOperator{\dev}{dev}          
\DeclareMathOperator{\divg}{div}         
\DeclareMathOperator{\grad}{\nabla\!}    
\DeclareMathOperator{\refgrad}{\nabla_{\!\oref}} 
\DeclareMathOperator{\refdivg}{\divg_{\oref}}    
\DeclareMathOperator{\reflapl}{\Delta_{\oref}}   
\DeclareMathOperator{\arctanh}{arctanh}
\newcommand{\oref}{{\mathrm{o}}}
\newcommand{\curshape}{\mathcal{R}}          
\newcommand{\refshape}{\mathcal{R}_\oref}    
\newcommand{\curshapeP}{\mathcal{P}}         
\newcommand{\refshapeP}{\mathcal{P}_\oref}   

\newcommand{\ve}{\mathbf{e}}             

\newcommand{\defm}{{\chi}}    
\newcommand{\defg}{{\grad\!\defm}}       
\newcommand{\refF}{\mathbf{F}_\oref}     
\newcommand{\eF}{\mathbf{F}}             
\newcommand{\cF}{\mathbf{G}}             
\newcommand{\refFdot}{\mathbf{\dot{F}}_\oref}
\newcommand{\eFdot}{\mathbf{\dot{F}}}    
\newcommand{\chp}{\mu}         
\newcommand{\chpf}{{\hat\chp}}       
\newcommand{\curhflux}{\mathbf{h}}
\newcommand{\refhflux}{{\mathbf{h}_\oref}}
\newcommand{\curhsrc}{h}
\newcommand{\refhsrc}{{h_\oref}}

\newcommand{\curn}{\mathbf{n}}          
\newcommand{\refn}{\mathbf{n}_\oref}    
\newcommand{\vel}{\mathbf{v}}

\newcommand{\reft}{{\mathbf{t}_\oref}}  
\newcommand{\curt}{\mathbf{t}}          
\newcommand{\refb}{{\mathbf{b}_\oref}}  
\newcommand{\refS}{{\mathbf{S}_\oref}}  
\newcommand{\eS}{\mathbf{S}}            
\newcommand{\eSf}{\mathbf{\hat{S}}}     
\newcommand{\iS}{\eS_\Is}   
\newcommand{\iSf}{\eSf_\Is}             
\newcommand{\freeE}{\psi}           
\newcommand{\strE}{\varphi}         
\newcommand{\Esh}{\mathbf{E}}           
\newcommand{\curMob}{\mathbf{M}}        
\newcommand{\refMob}{{\mathbf{M}_\oref}}
\newcommand{\refDif}{{\mathbf{D}_\oref}}  
\newcommand{\Id}{\mathbf{I}}
\newcommand{\tT}{\mathbf{T}}         
\newcommand{\tTf}{\mathbf{\hat{T}}}  
\newcommand{\iT}{\tT_\Is}         
\newcommand{\psf}{{\hat{p}}}         
\newcommand{\ps}{{p}}                
\newcommand{\etavis}{\eta_{\textsf{vis}}} 
\newcommand{\epsf}{{\psf_\elas}}     
\newcommand{\spm}{s}                 
\newcommand{\spmf}{\hat{s}}      
\newcommand{\isp}{\check{s}}         
\newcommand{\icg}{\check{\mathbf{g}}}          
\newcommand{\icgo}{{\check{\mathbf{g}}_\oref}} 
\newcommand{\icgof}{{\hat{\mathbf{g}}_\oref}} 
\newcommand{\cgo}{{\mathbf{g}_\oref}} 
\newcommand{\cgodiss}{{\mathbf{g}_\oref^{+}}} 
\newcommand{\ctro}{{\tau_\oref}}     
\newcommand{\cdo}{\mathbf{d}_\oref}  
\newcommand{\kdel}{{\delta}}         
\newcommand{\kcvx}{{k_\cvx}}         
\newcommand{\kspn}{{k_\spn}}         
\newcommand{\kg}{{k_{g}}}            
\newcommand{\kdg}{{k_{d}}}           
\newcommand{\kgf}{{k_\mathfrak{g}}}
\newcommand{\ficgo}{{\check{\mathbf{\mathfrak{g}}}_\oref}} 
\newcommand{\fcgo}{{\mathbf{\mathfrak{g}}_\oref}} 
\newcommand{\fctro}{{\mathfrak{t}_\oref}}     
\newcommand{\fdsb}{\mathfrak{s}}                 
\newcommand{\fdsi}{\check{\mathfrak{s}}}         

\newcommand{\icgos}{\check{\mathrm{g}}_\oref^\star}

\newcommand{\cgos}{\mathrm{g}_\oref^\star}
\newcommand{\ctros}{{\tau_\oref^\star}}     
\newcommand{\trac}{{\sigma}}           
\newcommand{\rhoo}{\rho_\oref}        
\newcommand{\rhob}{\rho_{\mathfrak{b}}} 
\newcommand{\cvx}{{\textsf{x}}}      
\newcommand{\spn}{{\textsf{s}}}      
\newcommand{\ches}{{ch}}             
\newcommand{\elas}{{e}}              
\newcommand{\gras}{{g}}              
\newcommand{\Is}{{\iota}}            
\newcommand{\Ve}{{\textsf{vol}}}              

\newcommand{\effs}{{\mathrm{tot}}}  
\newcommand{\augs}{{\mathrm{aug}}}  
\newcommand{\bn}[1]{{\scriptscriptstyle[#1]}}   
\newcommand{\bbn}[1]{{[#1]}}                
\newcommand{\jump}[1]{\ensuremath{[\![\,#1\,]\!]}}  
\newcommand{\refhfluxn}{\mathrm{h}_\oref}
\newcommand{\refhfluxs}{\mathrm{h}_\oref^\star}
\newcommand{\refV}{{V_\oref}}  
\newcommand{\refA}{{A_\oref}}  
\newcommand{\ith}{{\varepsilon}} 
\newcommand{\Li}{\mathfrak{b}}
\newcommand{\iF}{\mathbf{F}_\Is}

\newcommand{\iFdot}{\dot{\mathbf{F}}_\Is}

\newcommand{\test}[1]{\underline{#1}} 
\newcommand{\testnote}[1]{#1}         
\newcommand{\trp}{\mathsf{T}}            
\newcommand{\secref}[1]{Sec.~\ref{#1}}
\newcommand{\figref}[1]{Fig.~\ref{#1}}

\newcommand{\simulId}{}          
\newcommand{\simulLabel}{}       
\newcommand{\simulLabelS}{}      
\newcommand{\simulRef}[1]{\texttt{[#1]}} 
\newcommand{\simulDisplayRef}{\hyperref[\simulLabel]{\simulRef{\simulLabel}}}
\newcommand{\simulDisplayName}{\simulDisplayRef}

\newcommand{\dBrack}[1]{{\texttt{[\!\![{#1}]\!\!]}}}
\newcommand{\simColl}{}   
\newcommand{\NextColl}{}  
\newcommand{\BackColl}{}  
\newcommand{\goNextColl}{\hyperref[sec:\NextColl]{[next]}}
\newcommand{\goBackColl}{\hyperref[sec:\BackColl]{[previous]}}
\newcommand{\gotoc}{\hyperref[Tab:simCollns]{[collections]}}
\newcommand{\scale}{1}
\def\papertitle{Driving forces in cell migration and pattern formation\\ in a soft tissue}

\renewcommand\Affilfont{\small\itshape}
\renewcommand\Authsep{, }
\renewcommand\Authand{ and }
\renewcommand\Authands{ and }
\begin{document}

\title{\papertitle}

\author[1]{Amabile Tatone}
\author[2]{Filippo Recrosi}
\author[3]{Giuseppe Tomassetti}
\affil[1]{DISIM, University of L'Aquila, L'Aquila, Italy}
\affil[2]{Department of Engineering and Geology (INGEO), University of Chieti-Pescara, Pescara, Italy}
\affil[3]{Department of Industrial, Electronic, and Mechanical Engineering, Roma Tre University, Rome, Italy}
\date{}
\maketitle

\begin{abstract}
We give a description of cell diffusion in a soft tissue, paying special attention to the 
coupling of force, matter, and microforce balance laws through a suitable dissipation principle.
To this end, we cast our framework into a multi-level schematics,
comprising both kinematics and kinetics, based on a characterization of the free energy.
We lay down first a force balance law, where force and stress fields are defined as power conjugate
quantities to velocity fields and their gradients, then we give a species molar balance law,
with chemical potential test fields, as power conjugate quantities to the rate of change of the
species concentration, and finally a microforce balance law.
The main feature of this framework is the constitutive expression for the chemical potential
which is split into a term derived from the homogeneous convex part of the free energy
and an active external chemical potential giving rise to the spinodal decomposition.
The active part of the chemical potential is given an expression depending on the cell concentration and 
resembling the one defined in \cite{Oster-Murray-Harris-1983}
where it is meant to characterize an upward cell diffusion induced by cell motility.
Further we show how an external vector field, entering the microforce balance law as a power conjugate
quantity to the rate of change of the concentration gradient, can guide the diffusion process to a different 
limit stationary pattern.
This vector field could possibly model any directional cue or bias characterizing the interaction 
of the migrating cells and the surrounding tissue. 

\end{abstract}
\tableofcontents
\label{sec:tableofcontents}
\newpage
\clearpage
\section{Introduction}

Our goal was to construct a mechanical model to account 
for the fibrosis pattern formation in the liver \cite{Recrosi-et-al-Aimeta-2019}.
As a first step we chose 
to build on the basic framework in \cite{Tatone-et-al-2019} and \cite{Tatone-Recrosi-2024},
and describe an \emph{active cell diffusion}, 
\mbox{i.e.} an \emph{uphill diffusion} driven by \emph{cell motility},
while framing such a description as a Cahn-Hilliard equation.


Pattern formations in \emph{biological systems} and \emph{materials science} is a wide research area, 
comprising both the general framework setting and the evolution analysis, where both
analytical methods and numerical simulations are used for characterizing 
time evolution and pattern instabilities. 

Starting from the work of Cahn 
(\cite{Cahn-1961}, \cite{Cahn-Hilliard-1958}, \cite{Cahn-Hilliard-1959}, \cite{Allen-Cahn-1979}, \cite{Larche-Cahn-1973}), 
\emph{nucleation}, \emph{solidification}, and \emph{phase separation}, 
from topics in \emph{theoretical metallurgy}, have been extended over the following years to 
the general framework of \emph{phase-field} methods 
(\cite{Bates-Fife-1990}, \cite{Fife-Penrose-1995}),
comprising a wide range of microstructural evolutions.

A \emph{reaction-diffusion} theory for pattern formation was proposed by A. Turing 
in his work in 1952 on morphogenesis \cite{Turing-1952}, consisting in the description 
of the concurrent diffusion of two competing species. 

In the 1970s H. Meinhardt and A. Gierer introduced the notions of \emph{activating} and \emph{inhibiting}
chemicals to model pattern formation during the developmental processes 
\cite{Gierer-Meinhardt-1972, Meinhardt-2008}. 

A mechano-chemical theory of pattern formation was developed by 
J.D. Murray and G. Oster \cite{Murray-Oster-1984a} based on mechanical interactions 
as a mechanism for pattern formation in biological systems, described in \cite{Oster-Murray-Harris-1983}.
\section{Species diffusion in a crystal lattice\label{sect:diff-1}}

We outline here the basic setting for describing cell diffusion in a tissue
referring to the atomic diffusion in a crystal lattice \cite{Gurtin-2010}, as a prototype,
and drawing mainly from \cite{Tatone-et-al-2019}.

\subsection{Kinematics and kinetics}

Let us denote by 
\begin{equation}\label{spd-def:010}
   \defm : \refshape\to\curshape\,,
\end{equation}
a time dependent deformation of a crystal lattice from the reference shape to the current shape.
Describing the \emph{intercalation distortion} 
of a crystal lattice \cite{Larche-Cahn-1985} as a spherical tensor field
\begin{equation}\label{spd-def:020}
   \cF=\beta^{\frac{1}{3}}\;\Id\,,
\end{equation}
with $\;\det\cF=\beta\,$, and ruling out any plastic distortion, 
the accompanying \emph{elastic distortion} $\eF$ is defined by the deformation gradient decomposition
\begin{equation}\label{spd-def:030}
   \refF\equiv\defg = \eF\,\cF\,.
\end{equation}
It is convenient to describe the amount of intercalated $\Li$-atoms by the \emph{concentration}
\begin{equation}\label{spd-def:040}
c=\dfrac{\rhob}{\rhoo} =
\dfrac{\text{\small $<$molar density of species $\Li$ per unit reference volume$>$}}%
     {\text{\small $<$molar density of lattice sites per unit reference volume$>$}}\,,
\end{equation}
and make the assumption that it is related to the lattice volume change through
\begin{equation}\label{spd-def:050}
   \beta= 1 + \alpha\,(c-c_\oref)\,,
\end{equation}
where $\alpha$ is a stoichiometric positive constant coefficient and $\,c_\oref\,$ is a reference concentration.


\subsection{Species balance law and force balance law}

If we denote by $\,\rho\,$ 
the molar density of lattice sites \emph{per unit current volume}, then 
by \eqref{spd-def:040} the product \mbox{$\;c\,\rho\,$} is 
the molar density of species $\Li$ per unit current volume.
For any regular subset \mbox{$\,\curshapeP\subset\curshape\,$}
which is \emph{convected} from a reference subset \mbox{$\refshapeP\subset\refshape$} 
by the same deformation \eqref{spd-def:010},
the rate of change of the amount of species $\Li$ will be
\begin{equation}\label{spd-bal:0130}
   \frac{d}{dt}\int_{\curshapeP}c\,\rho\,dV = 
   \frac{d}{dt}\int_{\refshapeP}c\,\rho\,\det{\refF}\,dV =
   \frac{d}{dt}\int_{\refshapeP}c\,\rhoo\,dV =
   \int_{\refshapeP}\dot{c}\,\rhoo\,dV = \int_{\curshapeP}\dot{c}\,\rho\,dV\,. 
\end{equation}

Denoting by $\,\curhflux\,$ the \emph{molar flux} per unit current area,
and by $\,\curhsrc\,$ a \emph{supply} density per unit current volume, 
the species $\Li$ \emph{molar balance law} reads
\begin{equation}\label{spd-bal:0140}
  \int_{\curshapeP}\dot{c}\,\rho\,dV=
   -\int_{\partial\curshapeP}\curhflux\cdot\curn\,dA +\int_{\curshapeP}\curhsrc\,dV\,,
\end{equation}
and localizes to
\begin{equation}\label{spd-bal:0150}
  \dot{c}\,\rho = -\divg\curhflux +\curhsrc\,.
\end{equation}

Let us set now a scalar field
$\,\test{\chp}\,$,
power conjugate to the kinetic descriptor $\,\dot{c}\,\rho\,$,
transforming the \emph{molar balance law} \eqref{spd-bal:0150} into a \emph{power balance law}
\begin{equation}\label{spd-bal:010}
  \int_{\curshapeP}\test{\chp}\,\dot{c}\,\rho\,dV = 
  -\int_{\curshapeP}\test{\chp}\,\divg\curhflux\,dV
  +\int_{\curshapeP}\test{\chp}\,\curhsrc\,dV
  \qquad\forall\test{\chp}\,.
\end{equation}
Since
\begin{equation}\label{spd-bal:020}
   \divg(\chp\,\curhflux)
   =\chp\,\divg\curhflux+\curhflux\cdot\grad\chp\,,
\end{equation}
we get finally the \emph{molar balance law} \eqref{spd-bal:0140}
replaced by the \emph{species power balance law}
\begin{equation}\label{spd-bal:030}
  \int_{\curshapeP}\test{\chp}\,\dot{c}\,\rho\,dV = 
  -\int_{\partial\curshapeP}\test{\chp}\,\curhflux\cdot\curn\,dA
  +\int_{\curshapeP}\curhflux\cdot\grad\test{\chp}\,dV
  +\int_{\curshapeP}\test{\chp}\,\curhsrc\,dV
  \qquad\forall\test{\chp}\,.
\end{equation}
Notice that $\test{\chp}\,$ (\emph{energy per mole}) is a \emph{chemical potential}, 
acting here just as a test field.%
\testnote{\footnote{Throughout the paper we will consistently denote \emph{test} fields by underlying the corresponding symbol.}}
It is worth noting that equation \eqref{spd-bal:030} can be interpreted 
as the balance of an \emph{energy transport} \cite{Gurtin-2010}.


We can move the species power balance \eqref{spd-bal:030} back to the reference shape, 
and get \mbox{$\;\forall\refshapeP\subset\refshape\,$}
\begin{equation}\label{spd-bal:040}
  \int_{\refshapeP}\test{\chp}\,\dot{c}\,\rhoo\,dV = 
  -\int_{\partial\refshapeP}\test{\chp}\,%
  \refhflux\cdot\refn\,dA 
+ \int_{\refshapeP}\refhflux\cdot\refgrad\test{\chp}\,dV
+ \int_{\refshapeP}\test{\chp}\,\refhsrc\,dV
  \qquad\forall\test{\chp}\,,
\end{equation}
by using first the identity relating the reference and the current gradient of the scalar field $\,\chp\,$,
which we get from
\begin{equation}\label{spd-bal:051}
   \forall\ve\quad
   (\refgrad\chp)\cdot\ve=(\grad\chp)\cdot\refF\,\ve
   \quad\Rightarrow\quad
   \refgrad\chp = \refF^\trp\,\grad\chp\,,
\end{equation}
then replacing the current flux with the reference flux, according to the relation
\begin{align}
   \refhflux  &= (\det\refF)\,\refF^{-1}\,\curhflux\,,\label{spd-bal:052}\\
   \refhsrc   &= (\det\refF)\,\curhsrc\,.\label{spd-bal:052a}
\end{align}

The \emph{force power balance law}, \mbox{$\,\forall\refshapeP\subset\refshape\,$} reads 
\begin{equation}\label{spd-bal:060}
  \int_{\refshapeP}\refb\cdot\test{\vel}\,dV + \int_{\partial\refshapeP}\reft\cdot\test{\vel}\,dA =
  \int_{\refshapeP}\refS\cdot\refgrad\test{\vel}\,dV
  \qquad\forall\test{\vel}\,,
\end{equation}
where $\,\refb\,$ and $\,\reft\,$  stand for the reference bulk force density and
the reference boundary traction.
The \emph{reference Piola stress} $\,\refS\,$, the \emph{Cauchy stress} $\,\tT\,$, 
and the \emph{intermediate Piola stress} $\,\eS\,$, turn out to be related one another by
\begin{equation}\label{spd-bal:070}
\begin{aligned}
   \refS\cdot\refFdot 
   &= 
   (\det{\refF})\,\tT\cdot\refFdot\,\refF^{-1} 
   =
   \beta\,J\,\tT\,\refF^{-\trp}\cdot\refFdot 
   \\[\jot]
   &=
   \beta^\frac{2}{3}\,J\,\tT\,\eF^{-\trp}\cdot
   \big(\frac{1}{3}\,\beta^{-\frac{2}{3}}\,\dot{\beta}\,\eF+\beta^\frac{1}{3}\dot{\eF}\big) 
   \\
   &=
   \dot{\beta}\,J\,\frac{1}{3}\,\tT\cdot\Id 
   +
   \beta\,J\,\tT\,\eF^{-\trp}\cdot\dot{\eF}
   =
   \dot{\beta}\,J\,\frac{1}{3}\,\trace{\tT} 
   +
   \beta\,\eS\cdot\dot{\eF}
   \\[\jot]
   &=
   -\dot{\beta}\,J\,\ps 
   +
   \beta\,\eS\cdot\dot{\eF}
   \,,
\end{aligned}
\end{equation}
or, as an alternative, by
\begin{equation}\label{spd-bal:145}
\begin{aligned}
   \refS\cdot\refFdot 
   &=
   \dot{\beta}\,J\,\frac{1}{3}\,\tT\cdot\eF\,\eF^{-1}
   +
   \beta\,\eS\cdot\dot{\eF}
   \\[\jot]
   &=
   \dot{\beta}\,\frac{1}{3}\,\eF^\trp\,\eS\cdot\Id
   +
   \beta\,\eS\cdot\dot{\eF}
   =
   \dot{\beta}\,\frac{1}{3}\,\trace{(\eF^\trp\,\eS)}
   +
   \beta\,\eS\cdot\dot{\eF}
   \,,
\end{aligned}
\end{equation}
where
\begin{align}
   J   &:=\det{\eF}\,,\label{spd-bal:140}\\
   \ps &:= -\frac{1}{3}\,\trace{\tT}\,.\label{spd-bal:150}
\end{align}
The current gradient of the vector field $\,\vel\,$ is related to the reference one 
by the identity
\begin{equation}\label{spd-bal:075}
   \refgrad\vel=(\grad\vel)\,\refF\,.
\end{equation}
The standard \emph{frame-invariance} argument, stating that \mbox{$\,\tT\cdot\grad{\test{\vel}}=0\;$}
for any rigid test velocity field, leads to the symmetry property of tensor $\tT$.
\subsection{Free energy imbalance}

Let us consider now any evolution of the model we are defining, \hbox{i.e.} any \emph{constitutive process}, 
and the corresponding force power balance law
\begin{equation}\label{spd-bal:080}
  \underbrace{\int_{\refshapeP}\refb\cdot\vel\,dV + \int_{\partial\refshapeP}\reft\cdot\vel\,dA}_{ \text{\emph{(exchanged) external power}}} 
  = \int_{\refshapeP}\refS\cdot\refFdot\,dV\,,
\end{equation}
together with the species power balance law
\begin{equation}\label{spd-bal:090}
  \int_{\refshapeP}\chp\,\dot{c}\,\rhoo\,dV = 
  \underbrace{-\int_{\partial\refshapeP}\chp\,\refhflux\cdot\refn\,dA
  + \int_{\refshapeP}\chp\,\refhsrc\,dV
  }_{
  \text{\emph{(exchanged) external power}}} 
  + \int_{\refshapeP}\refhflux\cdot\refgrad\chp\,dV\,.
\end{equation}

Comparing\footnote{\cite{Gurtin-2010}, p.366}
the power exchanged between the matter inside any $\,{\refshapeP}\,$ and the outside 
with the rate of change of a free energy density per unit reference volume $\,\freeE\,$, we state the 
\emph{energy imbalance} or \emph{dissipation inequality} \cite{Gurtin-2010, Coleman-Noll-1963}
\begin{equation}  \label{spd-bal:130}
  \int_{\refshapeP}\refS\cdot\refFdot\,dV
  + \int_{\refshapeP}\chp\,\dot{c}\,\rhoo\,dV
  - \int_{\refshapeP}\refhflux\cdot\refgrad\chp\,dV
  - \frac{d}{dt}\int_{\refshapeP}\freeE\,dV \ge 0
  \qquad\forall\refshapeP
  \,,
\end{equation}
which can be localized to
\begin{equation}\label{spd-bal:160}
   \refS\cdot\refFdot
   + \chp\,\rhoo\,\dot{c}
   - \refhflux\cdot\refgrad\chp
   -\frac{d}{dt}\freeE \ge 0\,.
\end{equation}
By \eqref{spd-bal:070} and \eqref{spd-def:050}, the inequality above can be rewritten as
\begin{equation}\label{spd-bal:100}
   \beta\;\eS\cdot\eFdot
   +\big(\chp\,\rhoo-J\,\ps\,\alpha\big)\,\dot{c}
   - \refhflux\cdot\refgrad\chp
   -\frac{d}{dt}\freeE \ge 0\,,
\end{equation}
or, equivalently, by \eqref{spd-bal:145},
\begin{equation}\label{spd-bal:105}
   \beta\;\eS\cdot\eFdot
   +\big(\chp\,\rhoo+\frac{1}{3}\,\trace{(\eF^\trp\,\eS)}\,\alpha\big)\,\dot{c}
   - \refhflux\cdot\refgrad\chp
   -\frac{d}{dt}\freeE \ge 0\,.
\end{equation}

\subsection{Free energy expression and constitutive characterization\label{sect:spd-freeEn}}

Looking at \eqref{spd-bal:100} let us choose a free energy expression like the one given in 
\cite{Bower-et-al-2011, Cui-et-al-2012, Wu-2001}, and by \emph{(31)} in \cite{Tatone-et-al-2019}
\begin{equation}\label{spd-freeEn:010}
   \freeE=\hat\freeE(\eF,c) = \strE_\ches(c) + \beta\,\strE_\elas(\eF)\,,
\end{equation}
which is the sum of a \emph{chemical energy} density per unit reference volume,
and a \emph{strain energy} density per unit intermediate volume.
By chemical energy $\strE_\ches(c)$ we mean the free energy related to any form of
\emph{volumetric rearrangement}, be it atomic intercalation, 
particle inclusion, or cell migration. 
By contrast we make the simplifying assumption that the strain energy $\strE_\elas$ 
does not depend on $\,c\,$.

Defining the response functions $\chpf$ and $\eSf$ such that
\begin{align}
   \rhoo\,\chpf(c)\,\dot{c} &= \frac{d}{dt}\strE_\ches(c) \,,\label{spd-freeEn:025}
   \\[\jot]
   \eSf(\eF)\cdot\eFdot  &= \frac{d}{dt}\strE_\elas(\eF) \,,\label{spd-freeEn:020}
\end{align}
the rate of change of the free energy \eqref{spd-freeEn:010}, 
because of the decomposition \eqref{spd-def:030} and the assumption \eqref{spd-def:050}, turns out to be
\begin{equation}\label{spd-freeEn:030}
   \frac{d}{dt}\,\hat\freeE(\eF,c) = \beta\;\eSf(\eF)\cdot\eFdot + 
   \big(\rhoo\,\chpf(c) + \alpha\;\strE_\elas(\eF)\big)\,\dot{c}\,.
\end{equation}
If we finally substitute \eqref{spd-freeEn:030} into \eqref{spd-bal:100} we get%
\footnote{see \eqref{spd-freeEn:300} in \secref{sect:App-mat-response} for a more detailed strain energy density.}
\begin{equation}\label{spd-freeEn:050}
   \beta\,\big(
   \underbrace{\eS-\eSf(\eF)}_{\eS^{+}}
   \big)\cdot\eFdot + \big(
   \underbrace{\rhoo\,\big(\chp-\chpf(c)\big) -\alpha\,\big(J\,\ps +\strE_\elas(\eF)\big)}_{\rhoo\,\chp^{+}} 
   \big)\,\dot{c}
   -\refhflux\cdot\refgrad\chp
   \ge0\,.
\end{equation}
In order for the inequality \eqref{spd-freeEn:050} to hold for any constitutive process 
the following conditions must be fulfilled%
\footnote{The ratio $\,{\alpha}/{\rhoo}\,$ is nothing but the \emph{reference molar volume}, by \eqref{spd-def:050}.}
\begin{gather}
   \chp =  \chpf(c) + \frac{\alpha}{\rhoo}\big(J\,\ps+\strE_\elas(\eF)\big) 
   + \chp^{+}\,,\quad\chp^{+}\,\dot{c}\ge0\,,
   \label{spd-freeEn:060}\\[2\jot]
   \eS = \eSf(\eF) + \eS^{+}\,,\quad
   \eS^{+}\cdot\eFdot\ge0\,,
   \label{spd-freeEn:070}\\[2\jot]
   -\refhflux\cdot\refgrad\chp \ge 0\,,
   \label{spd-freeEn:080}
\end{gather}
with $\eS^{+}$ and $\chp^{+}$ possibly describing dissipative mechanisms.

It is worth noting that if we substitute 
\eqref{spd-freeEn:030} into the inequality \eqref{spd-bal:105} instead, 
we end up with \eqref{spd-freeEn:060} replaced by the equivalent expression
\begin{equation}
   \chp =  \chpf(c) + \frac{\alpha}{\rhoo}\big(\mbox{$-\dfrac{1}{3}\,\trace{(\eF^\trp\,\eS)}$}+\strE_\elas(\eF)\big) 
   + \chp^{+}\,,\quad\chp^{+}\,\dot{c}\ge0\,,
   \label{spd-freeEn:065}
\end{equation}
describing the coupling between diffusion and stress by 
the spherical part of the Eshelby stress \cite{Eshelby-1975, Wu-2001}
\begin{equation}\label{spd-freeEn:120}
   \Esh := -\eF^\trp\,\eS+\strE_\elas(\eF)\,\Id		\,.
\end{equation}

\subsection{Fick's law}

The condition \eqref{spd-freeEn:080} holds true if 
\begin{equation}\label{spd-Fick:090}
   \refhflux = -\refMob\,\refgrad\chp\,,
\end{equation}
with $\,\refMob\,$ a positive semi-definite tensor.
Expression \eqref{spd-Fick:090} is the reference form of \emph{Fick's law}. 
By \eqref{spd-bal:051} and \eqref{spd-bal:052} the reference flux 
and the reference chemical potential gradient can be transformed 
into the corresponding current quantities, leading to the new expression of \emph{Fick's law}
\begin{equation}\label{spd-Fick:100}
   \curhflux = -\curMob\,\grad\chp\,,
\end{equation}
where the reference and the current \emph{mobility} tensors are related by 
\begin{equation}\label{spd-Fick:110}
   \refMob  = (\det{\refF})\,\refF^{-1}\,\curMob\,\refF^{-\trp}\,.
\end{equation}
%


\section{Cahn-Hilliard equation\label{sect:C-H}}
The following derivation of the Cahn-Hilliard equation is based on \cite{Gurtin-1996}. 
Similar derivations can be found in \cite{Podio-2006}, \cite{Anand-2012}, 
\cite{Chen-Fan-Hong-et-al-2014} and \cite{DiLeo-et-al-2014}).

\subsection{Interfacial free energy\label{sect:C-H-energy}}

According to \cite{Cahn-Hilliard-1958}, let us consider the diffusion of a single species with 
the free energy density \eqref{spd-freeEn:010} modified by an additional term
\begin{equation}\label{spd-gfreeEn:010}
   \freeE=\hat\freeE_\gras(\eF,c,\refgrad{c}) = \strE_\ches(c) + \beta\,\strE_\elas(\eF) 
   +\strE_\gras(\refgrad{c})\,,
\end{equation}
where
\begin{equation}\label{spd-gfreeEn:020}
   \strE_\gras(\refgrad{c}) = \frac{1}{2}\,\kg\,\|\refgrad{c}\|^2
\end{equation}
is called the \emph{gradient energy} or the \emph{interfacial free energy} \cite{Cahn-Hilliard-1958}.
%
\subsection{Microforce balance law\label{sect:C-H-interfac}}

The rate of change of the free energy \eqref{spd-gfreeEn:010} turns out to be, 
by \eqref{spd-freeEn:025}, \eqref{spd-freeEn:020}, and \eqref{spd-gfreeEn:020},
\begin{equation}\label{spd-gfreeEn:030}
   \frac{d}{dt}\freeE 
   = \frac{d}{dt}\hat\freeE_\gras(\eF,c,\refgrad{c}) 
   = \beta\;\eSf(\eF)\cdot\eFdot 
   + \big(\rhoo\,\chpf(c) 
   + \alpha\;\strE_\elas(\eF)\big)\,\dot{c}
   + \kg\,\refgrad{c}\cdot\refgrad{\dot{c}} \,,
\end{equation}
where again the decomposition \eqref{spd-def:030}, together with the 
assumption \eqref{spd-def:050}, have been used.

Since the expression above differs from \eqref{spd-freeEn:030} by the last term,
we wonder what additional term could possibly complement the power expenditure in
\eqref{spd-bal:160}. 
Looking at the last term of \eqref{spd-gfreeEn:030}, we guess that additional term taking the expression
\begin{equation}\label{spd-gbal:040}
  \rhoo\,\isp\,\dot{c} + \icgo\cdot\refgrad{\dot{c}} \,.
\end{equation}
Recalling that \eqref{spd-bal:160} is based on the balance laws \eqref{spd-bal:080} and \eqref{spd-bal:090}, in the same spirit we should consider balancing \eqref{spd-gbal:040} by some external power and introduce
the \emph{microforce balance law} 
\begin{equation}\label{spd-gbal:050}
   \forall\refshapeP\subseteq\refshape\qquad
    \int_{\refshapeP}\rhoo\,\isp\,\test{\dot{c}}\,dV 
   +\int_{\refshapeP}\icgo\cdot\refgrad{\test{\dot{c}}}\,dV  
   =
   \underbrace{\int_{\partial\refshapeP}\ctro\,\test{\dot{c}}\,dA
   +\int_{\refshapeP}\rhoo\,\spm\,\test{\dot{c}}\,dV}_{
  \text{\emph{microforce external power}}}
   \qquad\forall\test{\dot{c}}\,. 
\end{equation}
By using the identity
\begin{equation}\label{spd-gbal:060}
  \refdivg(\test{\dot{c}}\,\icgo) = \test{\dot{c}}\,\refdivg{\icgo} + \icgo\cdot\refgrad{\test{\dot{c}}}
\end{equation}
we get \eqref{spd-gbal:050} transformed into
\begin{equation}\label{spd-gbal:070}
    \int_{\refshapeP}\rhoo\,\isp\,\test{\dot{c}}\,dV 
   +\int_{\partial\refshapeP}(\icgo\cdot\refn)\,\test{\dot{c}}\,dA  
   =\int_{\partial\refshapeP}\ctro\,\test{\dot{c}}\,dA
   +\int_{\refshapeP}(\refdivg{\icgo} + \rhoo\,\spm)\,\test{\dot{c}}\,dV\,,
\end{equation}
from which we derive the local form%
\footnote{see \cite{Gurtin-1996}, $(1.12)$ and $(2.1)$, where different symbols are used.}
\begin{gather}
   \refdivg{\icgo} + \rhoo\,(\spm - \isp) = 0 \qquad\text{on}\;\forall\refshapeP\,,\label{spd-gbal:080}\\[2\jot]
   \icgo\cdot\refn = \ctro \qquad\text{on}\;\partial\refshapeP \,.\label{spd-gbal:090}
\end{gather}
\subsection{Dissipation inequality\label{sect:C-H-diss}}

The dissipation inequality \eqref{spd-bal:130} with the additional terms \eqref{spd-gbal:040} will change to
\begin{equation}\label{spd-gbal:100}
\begin{split}
    \int_{\refshapeP}\refS\cdot\refFdot\,dV
  + \int_{\refshapeP}\rhoo\,(\chp+\isp)\,\dot{c}\,dV 
  + \int_{\refshapeP}\icgo\cdot\refgrad{\dot{c}}\,dV \\[\jot]
  - \int_{\refshapeP}\refhflux\cdot\refgrad\chp\,dV 
  - \frac{d}{dt}\int_{\refshapeP}\freeE\,dV \ge 0
  \qquad\forall\refshapeP \,.
\end{split}
\end{equation}
After localizing, by \eqref{spd-bal:070} we get the inequality \eqref{spd-bal:100} replaced by\footnote{see \cite{Gurtin-1996}, (3.6)}
\begin{equation}\label{spd-gbal:110}
     \beta\;\eS\cdot\eFdot
   + \big(\rhoo\,(\chp+\isp)-J\,\ps\,\alpha\big)\,\dot{c}
   + \icgo\cdot\refgrad{\dot{c}}
   - \refhflux\cdot\refgrad\chp
   - \frac{d}{dt}\freeE \ge 0\,.
\end{equation}
If we finally substitute \eqref{spd-gfreeEn:030} into \eqref{spd-gbal:110} we get\begin{equation}\label{spd-gfreeEn:050}
\begin{split}
   \beta\,\big(
   \underbrace{\eS-\eSf(\eF)}_{\eS^{+}}
   \big)\cdot\eFdot
   + \big(
   \underbrace{\rhoo\,\big(\chp+\isp-\chpf(c)\big) -\alpha\,\big(J\,\ps +\strE_\elas(\eF)\big)}_{\rhoo\,\spm^{+}} 
   \big)\,\dot{c}\\[\jot]
   + \big(
   \underbrace{\icgo-\icgof(\refgrad{c})}_{\cgodiss}
   \big)\cdot\refgrad{\dot{c}} 
   -\refhflux\cdot\refgrad\chp
   \ge0\,.
\end{split}
\end{equation}
In order for this inequality to be fulfilled for any constitutive process,
we get condition \eqref{spd-freeEn:060} updated to
\begin{equation} \label{spd-gfreeEn:060}
   \chp
   = \chpf(c) - \isp + \frac{\alpha}{\rhoo}\big(J\,\ps+\strE_\elas(\eF)\big) 
   + \spm^{+}\,,\qquad\spm^{+}\,\dot{c}\ge0\,,
\end{equation}
keep conditions \eqref{spd-freeEn:070} and \eqref{spd-freeEn:080} unchanged, 
and make the additional constitutive assumption
\begin{equation} \label{spd-gfreeEn:090}
   \icgo = \icgof(\refgrad{c}) + \cgodiss \,,\qquad\cgodiss\cdot\refgrad{\dot{c}}\ge0\,,
\end{equation}
such that  
\begin{equation} \label{spd-gfreeEn:095}
   \icgof(\refgrad{c})\cdot\refgrad{\dot{c}} = \frac{d}{dt}\strE_\gras(\refgrad{c})\,.
\end{equation}
Hence, by \eqref{spd-gfreeEn:020}, it turns out
\begin{equation} \label{spd-gfreeEn:100}
   \icgof(\refgrad{c}) = \kg\,\refgrad{c}\,.
\end{equation}

We could also define $\,\icg\,$, as a vector field on the current shape, by
\begin{equation} \label{spd-gbal:120}
   \int_{\curshapeP}\icg\cdot\grad\test{\dot{c}}\,dV 
  =\int_{\refshapeP}\icgo\cdot\refgrad\test{\dot{c}}\,dV \,,
\end{equation}
and get, as in \eqref{spd-bal:052},
\begin{equation} \label{spd-gbal:130}
   \icgo  = (\det\refF)\,\refF^{-1}\,\icg\,.
\end{equation}

\subsection{Summary of balance laws\label{sect:C-H-nbal}}

Summarizing, the current framework is characterized by the \emph{force balance law} \eqref{spd-bal:060}
\begin{equation*}
  \int_{\refshapeP}\refb\cdot\test{\vel}\,dV + \int_{\partial\refshapeP}\reft\cdot\test{\vel}\,dA =
  \int_{\refshapeP}\refS\cdot\refgrad\test{\vel}\,dV
  \qquad\forall\test{\vel}\,,
\end{equation*}
the \emph{species balance law} \eqref{spd-bal:040}
\begin{equation*}
  \int_{\refshapeP}\test{\chp}\,\dot{c}\,\rhoo\,dV = 
  -\int_{\partial\refshapeP}\test{\chp}\,%
  \refhflux\cdot\refn\,dA 
+ \int_{\refshapeP}\refhflux\cdot\refgrad\test{\chp}\,dV
+ \int_{\refshapeP}\test{\chp}\,\refhsrc\,dV
  \qquad\forall\test{\chp}\,,
\end{equation*}
and the \emph{microforce balance law} \eqref{spd-gbal:050} 
\begin{equation*}
    \int_{\refshapeP}\rhoo\,\isp\,\test{\dot{c}}\,dV 
   +\int_{\refshapeP}\icgo\cdot\refgrad{\test{\dot{c}}}\,dV  
   =\int_{\partial\refshapeP}\ctro\,\test{\dot{c}}\,dA
   +\int_{\refshapeP}\rhoo\,\spm\,\test{\dot{c}}\,dV\qquad\forall\test{\dot{c}}\,, 
\end{equation*}
supplemented by constitutive prescriptions about 
$\,\eS\,$, $\,\refhflux\,$, $\,\chp\,$, $\,\isp\,$, $\,\icgo\,$,
consistent with the assumptions 
\eqref{spd-freeEn:070} and 
\eqref{spd-freeEn:080}, possibly through Fick's law \eqref{spd-Fick:090}, 
as well as \eqref{spd-gfreeEn:060}
and \eqref{spd-gfreeEn:090}.

It is interesting to look at the local form of the balance laws above.
Let us derive first the local form of the \emph{species power balance law} \eqref{spd-bal:040}
\begin{equation}\label{spd-gbal:210}
  \dot{c}\,\rhoo = -\refdivg\refhflux +\refhsrc\,,
\end{equation}
corresponding to \eqref{spd-bal:0150},
and get, by replacing Fick's law \eqref{spd-Fick:090},
\begin{equation}\label{spd-gbal:220}
  \dot{c}\,\rhoo = \refdivg(\refMob\,\refgrad\chp) +\refhsrc\,.
\end{equation}

Further, let us note that the balance laws \eqref{spd-gbal:080} 
and \eqref{spd-gbal:210} turn out to be coupled
through $\,\isp\,$ by the constitutive expression \eqref{spd-gfreeEn:060}.
In order to make this coupling explicit let us replace $\,\isp\,$ in 
\eqref{spd-gfreeEn:060} with its expression from \eqref{spd-gbal:080} and get
\begin{equation}\label{spd-gfreeEn:170}
   \chp =  \chpf(c) - \spm + \frac{\alpha}{\rhoo}\big(J\,\ps+\strE_\elas(\eF)\big) 
   - \frac{1}{\rhoo}\refdivg{\icgo}
   \,,    
\end{equation}
where $\icgo$ is characterized by \eqref{spd-gfreeEn:090}. Eventually, the expression above
should be substituted into the species balance law \eqref{spd-gbal:220}.

As noted in \cite{Gurtin-1996}, we get the standard form of the Cahn-Hilliard equation by assuming 
\begin{equation}\label{spd-gfreeEn:180} 
   \spm=0\,,
\end{equation}
and removing the dissipative terms $\spm^{+}$ in \eqref{spd-gfreeEn:060}, and $\cgodiss$ in \eqref{spd-gfreeEn:090}, altogether.

We further assume \mbox{$\,\ctro=0\,$} on the outermost boundary $\,\partial\refshape\,$,
which is equivalent, by  \eqref{spd-gfreeEn:100} and \eqref{spd-gbal:090}, to the condition
\mbox{$\,\refgrad{c}\cdot\refn=0\,$} on $\,\partial\refshape\,$
(as in \cite{Cahn-Hilliard-1958}, where because of a small strain assumption there is no difference between 
$\,\refgrad{c}\cdot\refn\,$ and $\,\grad{c}\cdot\curn\,$).

\section{Active species diffusion\label{sect:Act-diff}}
\subsection{Uphill diffusion and aggregation}

If the chemical potential $\,\chp\,$ depends just on concentration,
Fick's law \eqref{spd-Fick:090} transforms into
\begin{equation}\label{spd-Fick-c:090}
   \refhflux = -\refMob\,\refgrad\chp 
   = -\frac{\partial\chp}{\partial c}\,\refMob\,\refgrad{c}
   = -\refDif\,\refgrad{c} \,.
\end{equation}
While the \emph{mobility} $\,\refMob\,$, by \eqref{spd-freeEn:080}, is a positive semi-definite tensor, 
the \emph{diffusivity} $\,\refDif\,$ is a positive or negative semi-definite tensor, depending on 
whether $\,{\partial\chp}/{\partial c}\,$ is positive or negative.

The expression \emph{uphill diffusion} \cite{Cahn-1968} refers to the case where
\begin{equation}\label{spd-Fick-c:095}
   \frac{\partial\chp}{\partial c}<0 \quad\Rightarrow\quad
   \refhflux\cdot\refgrad{c} = -\refDif\,\refgrad{c}\cdot\refgrad{c} \ge 0\,.
\end{equation}

In \cite{Oster-Murray-Harris-1983}, an uphill diffusion model 
relies on a motility mechanism based on \emph{haptotaxis}, 
whereby cells tend to move up a gradient of ``increasing substrate adhesion'' 
\cite{CarterSB-1965, CarterSB-1967}.

A different point of view relates an uphill diffusion to a motility mechanism based 
on the \emph{differential adhesion hypothesis} \cite{Steinberg-2007, Foty-Steinberg-2005}, 
whereby cell-cell adhesion is stronger for same type cells 
and weaker for dissimilar cells.

We rely on the differential adhesion mechanism by making the assumption that the
uphill diffusion is driven by the gradient of an \emph{active chemical potential}
depending on cell concentration. 
In doing so, the expression we found appropriate, 
\eqref{act-diff:120} through \eqref{act-diff:122},
turns out to be consistent with expression \emph{(7)} in \cite{Murray-Oster-1984b} 
describing a \emph{cell traction} due to the interaction with the extracellular matrix.

\subsection{Active chemical potential constitutive characterization\label{sect:act-chp}}

Let us consider the derivation of the Cahn-Hilliard equation in \secref{sect:C-H}
and look at the constitutive expression for the chemical potential \eqref{spd-gfreeEn:170}. 
Besides the assumption \eqref{spd-gfreeEn:180}, while removing the dissipative terms, 
we should use, as in \cite{Cahn-Hilliard-1958},
the \emph{regular solution model} for the free energy term \mbox{$\,\strE_\ches\,$} 
and get an expression for \mbox{$\,\chpf\,$} from \eqref{spd-freeEn:025}. 

Quoting from \cite{Cahn-Hilliard-1958}, p.258:
\emph{``Several different meanings are associated with the term \emph{regular solution}. 
We will use it to denote a solution having an ideal configurational entropy 
and an enthalpy of mixing which varies parabolically with composition.''}

We will assume instead that the free energy \mbox{$\,\strE_\ches\,$} is given by
\begin{equation}\label{act-diff:100}
\strE_\ches(c) =
\strE_\cvx(c)  =\frac{1}{2}\,\rhoo\,\kcvx\,c_{max}\big(\bar{c}\,\log(\bar{c}) 
               + (1 - \bar{c})\,\log(1 - \bar{c})\big) \,,
\end{equation}
with \mbox{$\bar{c}=c/c_{max}$},
which is just the (convex) \emph{entropic energy} in the regular solution model,
leading through \eqref{spd-freeEn:025} to the chemical potential term 
\begin{equation}\label{act-diff:110}
   \chpf(c) =
   \chpf_\cvx(c) = \frac{1}{\rhoo}\, \frac{d}{dc}\,\strE_\cvx(c) 
   = -\kcvx\,\arctanh(1-{2\,\bar{c}})\,, 
\end{equation}
and replace the assumption \eqref{spd-gfreeEn:180} for the microforce $\spm$ by
\begin{equation}\label{act-diff:120}
   \spm =\spmf(c,\gamma) = \kspn\,\frac{\gamma\,(c - c_\spn)}{\exp\big(\kdel\,(c - c_\spn)^2\big)}\,,
\end{equation}
meant to characterize cell interactions,%
\footnote{Inspired by the saturating function in the cell traction expression (7) in \cite{Murray-Oster-1984b}, 
formula \eqref{act-diff:120} implements sort of \emph{quorum sensing}, whereby cell interactions 
are based on their concentration, rather than on their mutual distances.}
modulated by an \emph{activity control parameter} $\,\gamma>0\,$.

It is also convenient to introduce the \emph{active chemical potential} 
\begin{equation}\label{act-diff:122}
   \chpf_\spn(c,\gamma)=-\spmf(c,\gamma)\,,
\end{equation}
as well as the corresponding energy
\begin{equation} \label{act-diff:150}
   \strE_\spn(c,\gamma)= \int{\rhoo\,\chpf_\spn(c,\gamma)}\,dc=
   \frac{1}{2}\,\rhoo\,\kspn\,\frac{\gamma}{\kdel\,\exp\big(\kdel\,(c - c_\spn)^2\big)}\,.
\end{equation}

It is worth noting that, by \eqref{act-diff:110}, \eqref{act-diff:120}, and \eqref{act-diff:122}, 
the expression \eqref{spd-gfreeEn:060} for the chemical potential becomes
\begin{equation} \label{act-diff:121}
   \chp =  \chpf_\cvx(c)+\chpf_\spn(c,\gamma) 
   + \frac{\alpha}{\rhoo}\big(J\,\ps+\strE_\elas(\eF)\big) 
   + \spm^{+}\,.
\end{equation}
\renewcommand{\scale}{0.47}
\begin{figure}[t!]
\setlength{\unitlength}{\scale pt} 
\centering
\boxed{
\begin{picture}(535,580)
  \put(  5,410){\includegraphics[viewport= 0 0 260 161, scale=\scale]{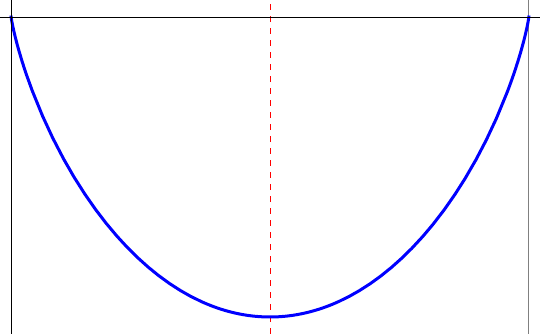}}
  \put(275,410){\includegraphics[viewport= 0 0 260 161, scale=\scale]{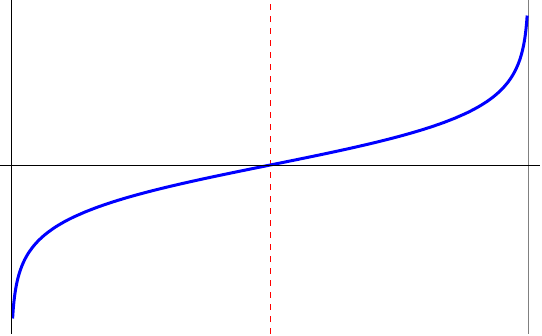}}
  \put(  5,210){\includegraphics[viewport= 0 0 260 161, scale=\scale]{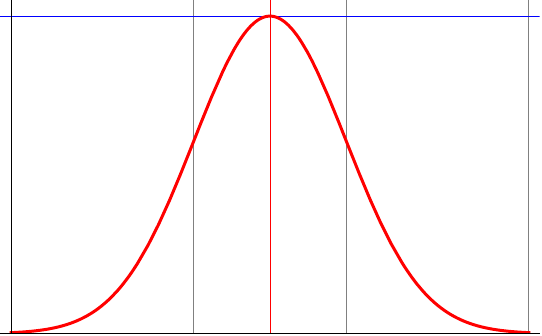}}
  \put(275,210){\includegraphics[viewport= 0 0 260 161, scale=\scale]{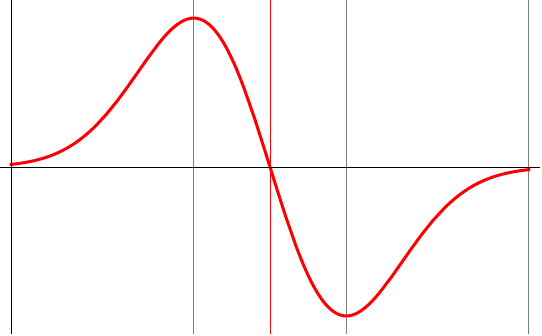}}
  \put(  5,005){\includegraphics[viewport= 0 0 260 161, scale=\scale]{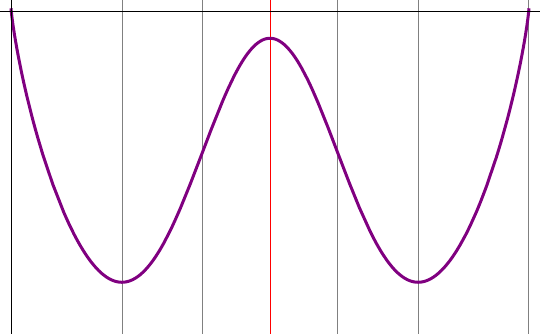}}
  \put(275,005){\includegraphics[viewport= 0 0 260 161, scale=\scale]{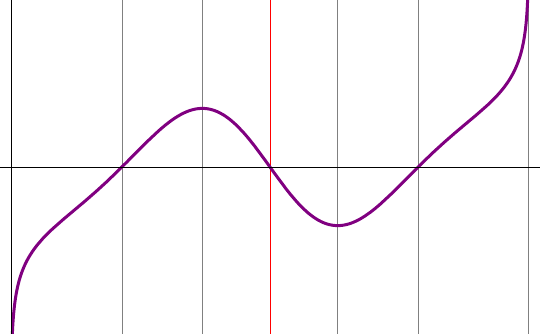}}
  \put(050,530){\small $\strE_\cvx(c)$}
  \put(025,330){\small $\strE_\spn(c,\gamma)$}
  \put(070,170){\small $\strE_\cvx(c)+\strE_\spn(c,\gamma)$}
  \put(320,530){\small $\chpf_\cvx(c)$}
  \put(430,330){\small $\chpf_\spn(c,\gamma)$}
  \put(330,150){\small $\chpf_\cvx(c)+\chpf_\spn(c,\gamma)$}
\end{picture}
}
\caption{Active chemical potential and double well free energy 
(\mbox{$c_{max} = 12$}, 
 \mbox{$c_\spn = 0.5\,c_{max}$}
 ).}
\label{fig-chp-sym}
\end{figure}

\begin{figure}[t]
\setlength{\unitlength}{\scale pt} 
\centering
\boxed{
\begin{picture}(535,580)
  \put(  5,410){\includegraphics[viewport= 0 0 260 161, scale=\scale]{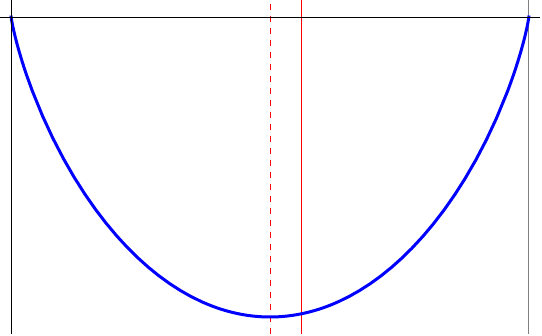}}
  \put(275,410){\includegraphics[viewport= 0 0 260 161, scale=\scale]{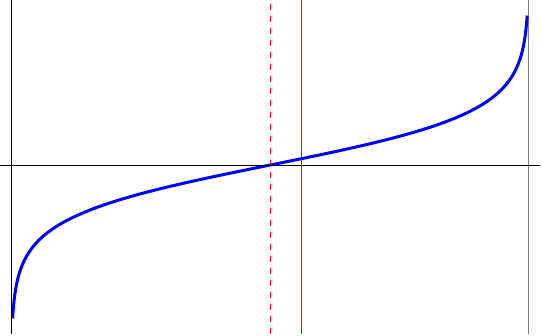}}
  \put(  5,210){\includegraphics[viewport= 0 0 260 161, scale=\scale]{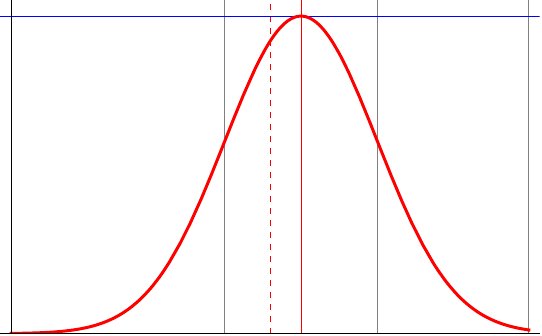}}
  \put(275,210){\includegraphics[viewport= 0 0 260 161, scale=\scale]{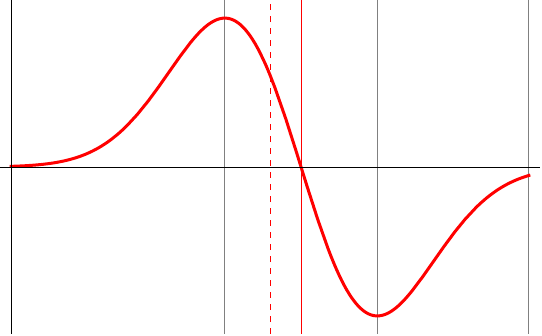}}
  \put(  5,005){\includegraphics[viewport= 0 0 260 161, scale=\scale]{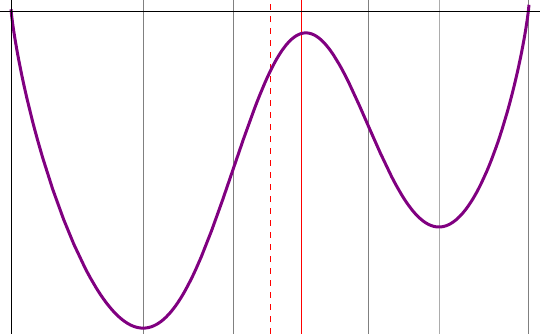}}
  \put(275,005){\includegraphics[viewport= 0 0 260 161, scale=\scale]{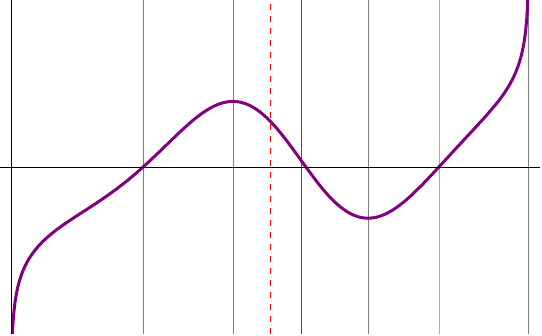}}
  \put(050,530){\small $\strE_\cvx(c)$}
  \put(025,330){\small $\strE_\spn(c,\gamma)$}
  \put(070,170){\small $\strE_\cvx(c)+\strE_\spn(c,\gamma)$}
  \put(320,530){\small $\chpf_\cvx(c)$}
  \put(430,330){\small $\chpf_\spn(c,\gamma)$}
  \put(330,150){\small $\chpf_\cvx(c)+\chpf_\spn(c,\gamma)$}
\end{picture}
}
\caption{Active chemical potential with shifted spinodal and double well free energy
(\mbox{$c_{max} = 12$}, 
 \mbox{$c_\spn = 0.56\,c_{max}$}
 ).}
\label{fig-chp-skw}
\end{figure}
Some properties of \eqref{act-diff:122} are worth noting:
the slope of its graph at $\,c_\spn\,$ 
\begin{equation} \label{act-diff:125}
    \left.\frac{d}{dc}\,\chpf_\spn(c,\gamma) =
    -\frac{d}{dc}\,\spmf(c,\gamma)\right|_{c=c_\spn} = -\kspn\,\gamma\,,
\end{equation}
is negative and independent of $\,\kdel\,$, while the $\,c\,$-interval where
\begin{equation} \label{act-diff:130}
    \frac{d}{dc}\,\chpf_\spn(c,\gamma) = -\frac{d}{dc}\spmf(c,\gamma) < 0\,,
\end{equation}
turns out to be independent of $\gamma$ 
\begin{equation} \label{act-diff:140}
(\,c_\spn-1/\sqrt{2\,\kdel}\,) < c < (\,c_\spn+1/\sqrt{2\,\kdel}\,) \,.
\end{equation}
As shown in Fig.\ref{fig-chp-sym} and Fig.\ref{fig-chp-skw} the composition 
\mbox{$\,\big(\chpf_\cvx(c)+\chpf_\spn(c,\gamma)\big)\,$} in \eqref{act-diff:121} 
can exhibit a \emph{spinodal interval} for the species concentration.

A straightforward, though approximate,%
\footnote{We are considering just the first two terms of the whole expression \eqref{act-diff:121}.}
rule for the spinodal decomposition to occur, according to \eqref{spd-Fick-c:095}, can be set as
\begin{equation} \label{act-bal:200}
    \left.\frac{d}{dc}\big(\chpf_\cvx(c)+\chpf_\spn(c,\gamma)\big)\right|_{c=c_\spn}
    =\frac{1}{2}\,\kcvx\,\frac{c_{max}}{c_\spn\,(c_{max}-c_\spn)} - \kspn\,\gamma < 0\,,
\end{equation}
or, equivalently,
\begin{equation} \label{act-bal:210}
    \gamma > \gamma_{cr} = \frac{1}{2}\,\frac{\kcvx}{\kspn}\,\frac{c_{max}}{c_\spn\,(c_{max}-c_\spn)} \,,
\end{equation}
showing the expression for the \emph{critical} value $\gamma_{cr}\,$ of the activity control parameter.

Within the same approximation, the concentration values of the separated phases are
selected by the two minima of
\mbox{$\,\big(\strE_\cvx(c)+\strE_\spn(c,\gamma)\big)\,$} or, equivalently, by the zeroes of
\mbox{$\,\big(\chpf_\cvx(c)+\chpf_\spn(c,\gamma)\big)\,$}.
\subsection{Microforce balance law extended\label{sect:go}}

In order to better characterize active diffusion, 
let us complement the \emph{external power} in the \emph{microforce balance law} \eqref{spd-gbal:050} 
with an additional term 
\begin{equation}\label{act-bal:150}
    \int_{\refshapeP}\rhoo\,\isp\,\test{\dot{c}}\,dV 
   +\int_{\refshapeP}\icgo\cdot\refgrad{\test{\dot{c}}}\,dV  
   =
   \underbrace{
   \int_{\partial\refshapeP}\ctro\,\test{\dot{c}}\,dA
   +\int_{\refshapeP}\rhoo\,\spm\,\test{\dot{c}}\,dV
   +\int_{\refshapeP}\cgo\cdot\refgrad{\test{\dot{c}}}\,dV}_{
  \text{\emph{microforce external power}}}  
   \qquad\forall\test{\dot{c}}\,,
\end{equation}
where the new vector field $\,\cgo\,$ is power conjugate to the concentration gradient rate of change.
This form of a balance law can be traced back to \cite{Germain-1973}, where a general theory
of a \emph{micromorphic} continuum is outlined. 

Just as we let $\,\spm\,$ characterize the spinodal decomposition,
we wonder if $\,\cgo\,$ can drive the diffusion in the coarsening process.

Because the additional term modifies solely the external power, the dissipation inequality
\eqref{spd-gbal:100} does not change, still leading to the constitutive expressions 
\eqref{spd-gfreeEn:060} and \eqref{spd-gfreeEn:090}. 
By using again the identity \eqref{spd-gbal:060} we can localize 
the \emph{microforce balance law} \eqref{act-bal:150} to
\begin{gather}
   \refdivg{(\icgo - \cgo)} 
   + \rhoo\,(\spm - \isp) = 0 \qquad\text{on}\;\forall\refshapeP\,,\label{act-bal:080}\\[2\jot]
   (\icgo-\cgo)\cdot\refn = \ctro \qquad\text{on}\;\partial\refshapeP \,.\label{act-bal:090}
\end{gather}
To show explicitly the expression for $\,\chp\,$, we replace $\isp $ in \eqref{spd-gfreeEn:060} 
with its expression from \eqref{act-bal:080} and get
\begin{equation} \label{act-gfreeEn:170}
   \chp =  \chpf(c) - \spm
   + \frac{\alpha}{\rhoo}\big(J\,\ps+\strE_\elas(\eF)\big) 
   - \frac{1}{\rhoo}\,\refdivg{(\icgo-\cgo)} \,,
\end{equation}
to be compared with \eqref{spd-gfreeEn:170}.
Again, by \eqref{act-diff:110}, \eqref{act-diff:120}, and \eqref{act-diff:122}, 
the expression above  for the chemical potential becomes
\begin{equation} \label{act-diff:123}
   \chp =  \chpf_\cvx(c)+\chpf_\spn(c,\gamma) 
   + \frac{\alpha}{\rhoo}\big(J\,\ps+\strE_\elas(\eF)\big) 
   - \frac{1}{\rhoo}\,\refdivg{(\icgo-\cgo)}
   \,.
\end{equation}
\subsection{Microforce balance law reinterpreted}

The \emph{microforce balance law} \eqref{act-bal:150} looks rather a 
\emph{chemical potential balance law}, because the test field has been given the meaning 
of species concentration rate of change.
Nevertheless, by \eqref{spd-def:050}, we can replace $\,\dot{c}\,$ by $\,\dot{\beta}/\alpha\,$ 
and get \eqref{act-bal:150} transformed into
\begin{equation}\label{act-bal:250}
    \int_{\refshapeP}\fdsi\,\test{\dot{\beta}}\,dV 
   +\int_{\refshapeP}\ficgo\cdot\refgrad{\test{\dot{\beta}}}\,dV  
   =\int_{\partial\refshapeP}\fctro\,\test{\dot{\beta}}\,dA
   +\int_{\refshapeP}\fdsb\,\test{\dot{\beta}}\,dV
   +\int_{\refshapeP}\fcgo\cdot\refgrad{\test{\dot{\beta}}}\,dV  
   \qquad\forall\test{\dot{\beta}}\,,
\end{equation}
where the new quantities defined by 
\begin{gather}
   \isp   = \frac{\alpha}{\rhoo}\,\fdsi \,,\quad
   \spm   = \frac{\alpha}{\rhoo}\,\fdsb \,,\label{act-gfreeEn:261} \\[\jot]
   \icgo = \alpha\,\ficgo             \,,\quad
   \cgo  = \alpha\,\fcgo              \,,\quad
   \ctro  = \alpha\,\fctro              \,,\label{act-gfreeEn:265}
\end{gather}
are power conjugate to $\,\dot{\beta}\,$ and its gradient.
As such  the scalar fields $\fdsi$ and $\fdsb$ can be given the meaning of internal and external \emph{pressure}, 
while the vector fields $\ficgo$ and $\fcgo$ can be given the meaning of internal and external \emph{double pressure} 
or \emph{pressure couple}. 
Substitution of \eqref{act-gfreeEn:261} and \eqref{act-gfreeEn:265} into the localized balance laws \eqref{act-bal:080} and \eqref{act-bal:090} leads to
\begin{gather}
   \refdivg{(\ficgo - \fcgo)} 
   + (\fdsb - \fdsi) = 0 \qquad\text{on}\;\forall\refshapeP\,,\label{act-bal:270}\\
   (\ficgo-\fcgo)\cdot\refn = \fctro \qquad\text{on}\;\partial\refshapeP \,,\label{act-bal:275}
\end{gather}
while the constitutive expression \eqref{act-gfreeEn:170} takes the form%
\footnote{As already noted in \secref{sect:spd-freeEn}, the ratio $\,{\alpha}/{\rhoo}\,$ is the \emph{species molar volume}. 
}
\begin{equation} \label{act-gfreeEn:280}
   \chp =  \chpf(c) - \frac{\alpha}{\rhoo}\,\fdsb
   + \frac{\alpha}{\rhoo}\big(J\,\ps+\strE_\elas(\eF)\big) 
   - \frac{\alpha}{\rhoo}\,\refdivg{(\ficgo - \fcgo)} \,,
\end{equation}
with the new coefficient $\kgf$ arising from \eqref{spd-gfreeEn:020}
when replacing $\,\refgrad{c}\,$ with $\,(\refgrad{\beta})/\alpha\,$ by \eqref{spd-def:050}%
\footnote{We rely on the constitutive assumption that the reference concentration $c_\oref$ is uniform:
$\,\refgrad{c_\oref}=0\,$.}
\begin{equation}\label{act-gfreeEn:290}
   \strE_\gras(\refgrad{c}) = \frac{1}{2}\,\kg\,\|\refgrad{c}\|^2 
   = \frac{1}{2}\,\frac{\kg}{\alpha^2}\,\|\refgrad{\beta}\|^2
   = \frac{1}{2}\,\kgf\,\|\refgrad{\beta}\|^2
   \,.
\end{equation}
\subsection{Active micro couple $\,\cgo\,$ characterization\label{sect:go-char}}

Choosing a constitutive expression for $\,\cgo\,$ is a matter of modeling.
Let us consider the case where $\,\cgo\,$ is characterized by a vector field $\cdo$ as in
\begin{equation}\label{act-bal:110}
   \cgo = \kdg\,\cdo\,.
\end{equation}
It is worth noting that even if $\cgo$ is a uniform vector field,
despite the chemical potential expression \eqref{act-diff:123} 
is not affected because $\refdivg{\cgo}=0$, 
the diffusion will be affected nevertheless, 
since the boundary condition \eqref{act-bal:090} does depend on $\,\cgo\,$.

Another simple expression, describing $\,\cgo\,$ as a vector field depending on $\,c\,$, could be
\begin{equation}\label{act-bal:140}
   \cgo = \hat{k}(c)\,\cdo
\end{equation}
from which we get
\begin{equation}\label{act-bal:160}
   \begin{aligned}
   \refdivg{\cgo} &= \trace{\refgrad\big(\hat{k}(c)\,\cdo\big)}
   = \trace{(\cdo\otimes\refgrad{\hat{k}(c)} })\\
   &= \trace{(\cdo\otimes(\hat{k}'(c)\refgrad{c}) })\\
   &= \hat{k}'(c)\,\cdo\cdot\refgrad{c} 
   \,.
   \end{aligned}
\end{equation}

A very different choice consists of the expression
\begin{equation}\label{act-bal:120}
   \cgo = \hat{k}(c)\,\refgrad{c}\,,
\end{equation}
from which we get
\begin{equation}\label{act-bal:130}
   \begin{aligned}
   \refdivg{\cgo} &= \trace{\refgrad\big(\hat{k}(c)\,\refgrad{c}\big)}
   = \trace{(\refgrad{c}\otimes\refgrad{\hat{k}(c)} + \hat{k}(c)\refgrad{(\refgrad{c})}})\\
   &= \trace{(\refgrad{c}\otimes(\hat{k}'(c)\refgrad{c}) + \hat{k}(c)\refgrad{(\refgrad{c})}})\\
   &= \hat{k}'(c)\,\refgrad{c}\cdot\refgrad{c} + \hat{k}(c)\,\refdivg{\refgrad{c}}\\
   &= \hat{k}'(c)\,\refgrad{c}\cdot\refgrad{c} + \hat{k}(c)\,\reflapl{c}
   \,.
   \end{aligned}
\end{equation}
Hence the vector field \eqref{act-bal:120} results in a chemical potential
expression with a correction of the coefficient $\kg\,$ in \eqref{spd-gfreeEn:100}
and the addition of a new term.
Such an expression looks very close to the \emph{(nearest integrable) Active Model B} 
in \cite{Wittkowski-et-al-2014}.
\section{Conclusion\label{sect:conclusion}}
The gradient energy leads to the emergence 
of microforces, subject to a microforce balance law,
which allow the characterization of the interactions as depending on the
concentration of the diffusing particles.

The phase separation process results from a double well energy made up of the sum of
a convex entropic energy $\strE_\cvx(c)$ and a non convex energy  $\strE_\spn(c,\gamma)$ 
arising from the integration of a microforce expression $\spmf(c,\gamma)$ 
inspired by the \emph{differential adhesion hypothesis}.

Some numerical simulations illustrate how the evolution of an initial uniform concentration
up to a non uniform stationary pattern is characterized by moving interfaces 
guided by a prescribed microforce $\cgo$ in the form \eqref{act-bal:110},
possibly with a non uniform vector field $\cdo$.
\section*{Acknowledgments}
The authors acknowledge the free access to bibliographic services and computing facilities
granted by the Department of Information Engineering, Computer Science and Mathematics (DISIM)
of the University of L'Aquila.
\newpage
\newcommand{\SI}[1]{$\mathrm{#1}$}
\begin{table}[ht]
   \centering
   \begin{tabular}{l@{\quad}l}
   \hline\\[-8pt]
   $[\,c\,]$                & \SI{1}                                               \\[\jot]
   $[\,\dot{c}\,]$          & \SI{1/s}                                             \\[\jot]
   $[\,\refgrad{\dot{c}}\,]$& \SI{1/(m\,s)}                                        \\[\jot]
   $[\,\dot{\beta}\,]$      & \SI{1/s}                                             \\[\jot]
   $[\,\rho\,           ]$  & \SI{(mol/m^3)}                                       \\[\jot]    
   $[\,\dot\rho\,       ]$  & \SI{(mol/m^3)/s}                                     \\[\jot]   
   \hline\\[-8pt]
   $[\,\curhsrc\,       ]$  & \SI{mol/(m^3\,s)}                                    \\[\jot]   
   $[\,\curhsrc/\rho\,  ]$  & \SI{mol/(m^3\,s)/(mol/m^3)=1/s}                      \\[\jot]
   $[\,\curhflux\,,\,\refhflux\,]$  & \SI{mol/(m^2\,s)}                            \\[\jot]   
   $[\,\curhflux/\rho\, ]$  & \SI{mol/(m^2\,s)/(mol/m^3)=m/s}                      \\[\jot]   
   $[\,\chp\,           ]$  & \SI{(J/mol)=(N\,m)/mol=(N/m^2)(m^3/mol)=Pa/(mol/m^3)}\\[\jot] 
   $[\,\chp\,\curhsrc\, ]$  & \SI{(J/mol)\,(mol/(m^3\,s))=(J/s)/m^3}               \\[\jot]
   $[\,\chp\,\rho\,     ]$  & \SI{(J/mol)\,(mol/m^3)=(J/m^3)=(N\,m)/m^3=N/m^2=Pa}  \\[\jot]
   $[\,\refhflux\cdot\refgrad\chp\,]$ & \SI{(mol/(m^2\,s)(J/mol)/m)=(J/s)/m^3}     \\[\jot]
   \hline\\[-8pt]
   $[\,\icgo\,,\,\cgo\,]$   & \SI{(J/m^2)=(N/m)=(Pa\,m)}                           \\[\jot]
   $[\,\ctro\,]$            & \SI{(J/m^2)=(N/m)=(Pa\,m)}                           \\[\jot]
   $[\,\ctro\,{\dot{c}}\,]$ & \SI{(J/m^2)/s=(Pa\,m)/s}                             \\[\jot]
   $[\,\icgo\,\refgrad{\dot{c}}\,]$ & \SI{(J/m^2)/(m\,s)=(J/s)/m^3=(Pa\,m)/(m\,s)=Pa/s} \\[\jot]
   $[\,\refdivg\icgo\,  ]$  & \SI{(Pa\,m)/m=Pa}                                    \\[\jot]
   $[\,\spm\,           ]$  & \SI{(J/mol)}                                         \\[\jot]
   $[\,\rhoo\,\spm\,]$      & \SI{(J/mol)(mol/m^3)=(N\,m/m^3)=Pa}                  \\[\jot]
   $[\,\rhoo\,\spm\,{\dot{c}}\,]$  & \SI{Pa/s}                                     \\[\jot]
   \hline\\[-8pt]
   $[\,\dot{\freeE}\,   ]$  & \SI{(J/m^3)/s=Pa/s}                                  \\[\jot]
   \hline\\[-8pt]
   $[\,\tT\,,\,\refS\,]$    & \SI{Pa}                                              \\[\jot]
   $[\,\refS\cdot\refFdot\,]$  & \SI{Pa/s=(J/m^3)/s}                               \\[\jot]
   $[\,(\refS\,\refn)\cdot\vel\,]$  & \SI{Pa\,(m/s)=(J/m^2)/s}                     \\[\jot]
   \hline\\[-8pt]
   $[\,\kg\,            ]$  & \SI{J/m}                                         \\[\jot]   
   $[\,\kdg\,           ]$  & \SI{J/m^2}                                       \\[\jot]   
   \hline
   \end{tabular}
\caption{Unit of measure (SI) for main quantities and basic expressions [between brackets].}
\label{Tab:var-def}
\end{table}
\newpage
\bibliographystyle{elsarticle-num-url-fix} 
\bibliography{patterns}

\begin{thebibliography}{10}
\expandafter\ifx\csname url\endcsname\relax
  \def\url#1{\texttt{#1}}\fi
\expandafter\ifx\csname href\endcsname\relax
  \def\href#1#2{#2} \def\path#1{#1}\fi

\bibitem{Oster-Murray-Harris-1983}
G.~F. Oster, J.~D. Murray, A.~K. Harris, Mechanical aspects of mesenchymal
  morphogenesis, J. Embryol. Exp. Morph. 78 (1983) 83--125.
\newblock \href {http://dx.doi.org/10.1242/dev.78.1.83}
  {\path{doi:10.1242/dev.78.1.83}}.

\bibitem{Recrosi-et-al-Aimeta-2019}
F.~Recrosi, R.~Repetto, A.~Tatone, G.~Tomassetti, Mechanical model of fiber
  morphogenesis in the liver, in: A.~Carcaterra, A.~Paolone, G.~Graziani
  (Eds.), Proceedings of XXIV AIMETA Conference 2019, Springer International
  Publishing, Cham CH, 2020, pp. 671--688.
\newblock \href {http://dx.doi.org/10.1007/978-3-030-41057-5_55}
  {\path{doi:10.1007/978-3-030-41057-5_55}}.

\bibitem{Tatone-et-al-2019}
A.~Tatone, F.~Recrosi, R.~Repetto, G.~Guidoboni, From species diffusion to
  poroelasticity and the modeling of lamina cribrosa, J. Mech. Phys. Solids 124
  (2019) 849--870.
\newblock \href {http://dx.doi.org/10.1016/j.jmps.2018.11.017}
  {\path{doi:10.1016/j.jmps.2018.11.017}}.

\bibitem{Tatone-Recrosi-2024}
A.~Tatone, F.~Recrosi, Volumetric growth, microstructure, and kinetic energy,
  European Journal of Mechanics - A/Solids 103 (2024) 105154.
\newblock \href {http://dx.doi.org/10.1016/j.euromechsol.2023.105154}
  {\path{doi:10.1016/j.euromechsol.2023.105154}}.

\bibitem{Cahn-1961}
J.~W. Cahn, On spinodal decomposition, Acta Metallurgica 9 (1961) 795--801.
\newblock Available from:
  \url{https://www.materialstoday.com/materials-chemistry/news/in-memory-of-dr-john-w-cahn/},
  \href {http://dx.doi.org/10.1016/0001-6160(61)90182-1}
  {\path{doi:10.1016/0001-6160(61)90182-1}}.

\bibitem{Cahn-Hilliard-1958}
J.~W. Cahn, J.~E. Hilliard, Free energy of a nonuniform system. {I}.
  {I}nterfacial free energy, J. of Chemical Physics 28 (1958) 258--267.
\newblock \href {http://dx.doi.org/10.1063/1.1744102}
  {\path{doi:10.1063/1.1744102}}.

\bibitem{Cahn-Hilliard-1959}
J.~W. Cahn, J.~E. Hilliard, Free energy of a nonuniform system. {III}.
  {N}ucleation in a two-component incompressible fluid, J. of Chemical Physics
  31 (1959) 688--699.
\newblock \href {http://dx.doi.org/10.1063/1.1730447}
  {\path{doi:10.1063/1.1730447}}.

\bibitem{Allen-Cahn-1979}
S.~M. Allen, J.~W. Cahn, A microscopic theory for antiphase boundary motion and
  its application to antiphase domain coarsening, Acta Metallurgica 27 (1979)
  1085--1095.
\newblock \href {http://dx.doi.org/10.1016/0001-6160(79)90196-2}
  {\path{doi:10.1016/0001-6160(79)90196-2}}.

\bibitem{Larche-Cahn-1973}
F.~Larch\'e, J.~W. Cahn, Linear theory of thermomechanical equilibrium of
  solids under stress, Acta Metall. 21 (1973) 1051--1063.
\newblock \href {http://dx.doi.org/10.1016/0001-6160(73)90021-7}
  {\path{doi:10.1016/0001-6160(73)90021-7}}.

\bibitem{Bates-Fife-1990}
P.~W. Bates, P.~C. Fife, Spectral comparison principles for the
  {C}ahn-{H}illiard and phase-field equations, and time scales for coarsening,
  Physica {D} 43 (1990) 335--348.
\newblock \href {http://dx.doi.org/10.1016/0167-2789(90)90141-B}
  {\path{doi:10.1016/0167-2789(90)90141-B}}.

\bibitem{Fife-Penrose-1995}
P.~C. Fife, O.~Penrose, Interfacial dynamics for thermodynamically consistent
  phase-field models with nonconserved order parameter, Electronic Journal of
  Differential Equations 1995 (1995) 1--49.
\newblock Available from:
  \url{https://ejde.math.txstate.edu/Volumes/1995/16/Fife.pdf}.

\bibitem{Turing-1952}
A.~M. Turing, The chemical basis of morphogenesis, Philosophical Transactions
  of the Royal Society of London. Series B, Biological Sciences 237 (1952)
  37--72.
\newblock \href {http://dx.doi.org/10.1098/rstb.1952.0012}
  {\path{doi:10.1098/rstb.1952.0012}}.

\bibitem{Gierer-Meinhardt-1972}
A.~Gierer, H.~Meinhardt, A theory of biological pattern formation, Kybernetik
  12 (1972) 30--39.
\newblock \href {http://dx.doi.org/10.1007/BF00289234}
  {\path{doi:10.1007/BF00289234}}.

\bibitem{Meinhardt-2008}
H.~Meinhardt, Models of biological pattern formation: from elementary steps to
  the organization of embryonic axes, Curr. Top. Dev. Biol. 81 (2008) 1--63.
\newblock \href {http://dx.doi.org/10.1016/S0070-2153(07)81001-5}
  {\path{doi:10.1016/S0070-2153(07)81001-5}}.

\bibitem{Murray-Oster-1984a}
J.~D. Murray, G.~F. Oster, Generation of biological pattern and form, IMA J.
  Math. Appl. Med. Biol. 1 (1984) 51--75.
\newblock \href {http://dx.doi.org/10.1093/imammb/1.1.51}
  {\path{doi:10.1093/imammb/1.1.51}}.

\bibitem{Gurtin-2010}
M.~E. Gurtin, E.~Fried, L.~Anand, The Mechanics and Thermodynamics of Continua,
  Cambridge University Press, 2010.
\newblock \href {http://dx.doi.org/10.1017/CBO9780511762956}
  {\path{doi:10.1017/CBO9780511762956}}.

\bibitem{Larche-Cahn-1985}
F.~Larch\'e, J.~W. Cahn, Overview no. 41 {T}he interactions of composition and
  stress in crystalline solids, Acta Metallurgica 33 (1985) 331--357.
\newblock Available from:
  \url{https://www.materialstoday.com/crystalline-materials/features/overview-no-41/},
  \href {http://dx.doi.org/10.1016/0001-6160(85)90077-X}
  {\path{doi:10.1016/0001-6160(85)90077-X}}.

\bibitem{Coleman-Noll-1963}
B.~D. Coleman, W.~Noll, The thermodynamics of elastic materials with heat
  conduction and viscosity, Arch. Rational. Mech. Anal. 13 (1963) 167--178.
\newblock \href {http://dx.doi.org/10.1007/BF01262690}
  {\path{doi:10.1007/BF01262690}}.

\bibitem{Bower-et-al-2011}
A.~F. Bower, P.~R. Guduru, V.~A. Sethuraman, A finite strain model of stress,
  diffusion, plastic flow, and electrochemical reactions in a lithium-ion
  half-cell, J. Mech. Phys. Solids 59 (2011) 804--828.
\newblock \href {http://dx.doi.org/10.1016/j.jmps.2011.01.003}
  {\path{doi:10.1016/j.jmps.2011.01.003}}.

\bibitem{Cui-et-al-2012}
Z.~W. Cui, F.~Gao, J.~M. Qu, A finite deformation stress-dependent chemical
  potential and its applications to lithium ion batteries, J. Mech. Phys.
  Solids 60 (2012) 1280--1295.
\newblock \href {http://dx.doi.org/10.1016/j.jmps.2012.03.008}
  {\path{doi:10.1016/j.jmps.2012.03.008}}.

\bibitem{Wu-2001}
C.~H. Wu, The role of {E}shelby stress in composition-generated and
  stress-assisted diffusion, J. Mech. Phys. Solids 49 (2001) 1771--1794.
\newblock \href {http://dx.doi.org/10.1016/S0022-5096(01)00011-4}
  {\path{doi:10.1016/S0022-5096(01)00011-4}}.

\bibitem{Eshelby-1975}
J.~D. Eshelby, Elastic energy-momentum tensor, J. Elasticity 5 (1975) 321--335.
\newblock \href {http://dx.doi.org/10.1007/BF00126994}
  {\path{doi:10.1007/BF00126994}}.

\bibitem{Gurtin-1996}
M.~E. Gurtin, Generalized {G}inzburg-{L}andau and {C}ahn-{H}illiard equations
  based on a microforce balance, Physica D 92 (1996) 178--192.
\newblock \href {http://dx.doi.org/10.1016/0167-2789(95)00173-5}
  {\path{doi:10.1016/0167-2789(95)00173-5}}.

\bibitem{Podio-2006}
P.~Podio-Guidugli, Models of phase segregation and diffusion of atomic species
  on a lattice, Ricerche di Matematica 55 (2006) 105--118.
\newblock \href {http://dx.doi.org/10.1007/s11587-006-0008-8}
  {\path{doi:10.1007/s11587-006-0008-8}}.

\bibitem{Anand-2012}
L.~Anand, A {C}ahn-{H}illiard-type phase-field theory for species diffusion
  coupled with large elastic deformations, J. Mech. Phys. Solids 60 (2012)
  1983--2002.
\newblock \href {http://dx.doi.org/10.1016/j.jmps.2012.08.001}
  {\path{doi:10.1016/j.jmps.2012.08.001}}.

\bibitem{Chen-Fan-Hong-et-al-2014}
L.~Chen, F.~Fan, L.~Hong, J.~Chen, Y.~Z. Ji, S.~L. Zhang, T.~Zhu, L.~Q. Chen, A
  phase-field model coupled with large elasto-plastic deformation:
  {A}pplication to lithiated silicon electrodes, J. Electrochem. Soc. 161
  (2014) F3164--F3172.
\newblock \href {http://dx.doi.org/10.1149/2.0171411jes}
  {\path{doi:10.1149/2.0171411jes}}.

\bibitem{DiLeo-et-al-2014}
C.~V. {Di Leo}, E.~Rejovitzky, L.~Anand, A {C}ahn-{H}illiard-type phase-field
  theory for species diffusion coupled with large elastic deformations:
  {A}pplication to phase-separating {Li}-ion electrode materials, J. Mech.
  Phys. Solids 70 (2014) 1--29.
\newblock \href {http://dx.doi.org/10.1016/j.jmps.2014.05.001}
  {\path{doi:10.1016/j.jmps.2014.05.001}}.

\bibitem{Cahn-1968}
J.~W. Cahn, Spinodal decomposition, Transactions of the Metallurgical Society
  of AIME 242 (1968) 89--103.
\newblock \href {http://dx.doi.org/10.1002/9781118788295.ch12}
  {\path{doi:10.1002/9781118788295.ch12}}.

\bibitem{CarterSB-1965}
S.~B. Carter, Principles of cell motility: the direction of cell movement and
  cancer invasion, Nature 208 (1965) 1183--1187.
\newblock \href {http://dx.doi.org/10.1038/2081183a0}
  {\path{doi:10.1038/2081183a0}}.

\bibitem{CarterSB-1967}
S.~B. Carter, Haptotaxis and the mechanism of cell motility, Nature 213 (1967)
  256--260.
\newblock \href {http://dx.doi.org/10.1038/213256a0}
  {\path{doi:10.1038/213256a0}}.

\bibitem{Steinberg-2007}
M.~S. Steinberg, Differential adhesion in morphogenesis: a modern view, Current
  Opinion in Genetics \& Development 17 (2007) 281--286.
\newblock \href {http://dx.doi.org/10.1016/j.gde.2007.05.002}
  {\path{doi:10.1016/j.gde.2007.05.002}}.

\bibitem{Foty-Steinberg-2005}
R.~A. Foty, M.~S. Steinberg, The differential adhesion hypothesis: a direct
  evaluation, Developmental Biology 278 (2005) 255--263.
\newblock \href {http://dx.doi.org/10.1016/j.ydbio.2004.11.012}
  {\path{doi:10.1016/j.ydbio.2004.11.012}}.

\bibitem{Murray-Oster-1984b}
J.~D. Murray, G.~F. Oster, Cell traction models for generating pattern and form
  in morphogenesis, Journal of Mathematical Biology 19 (1984) 265--279.
\newblock \href {http://dx.doi.org/10.1007/BF00277099}
  {\path{doi:10.1007/BF00277099}}.

\bibitem{Germain-1973}
P.~Germain, The method of virtual power in continuum mechanics. {P}art 2:
  {M}icrostructure, SIAM Journal on Applied Mathematics 25 (1973) 556--575.
\newblock \href {http://dx.doi.org/10.1137/0125053}
  {\path{doi:10.1137/0125053}}.

\bibitem{Wittkowski-et-al-2014}
R.~Wittkowski, A.~Tiribocchi, J.~Stenhammar, R.~J. Allen, D.~Marenduzzo, M.~E.
  Cates, Scalar {$\varphi^4$} field theory for active-particle phase
  separation, Nature Communications 5 (2014) 4351.
\newblock \href {http://dx.doi.org/10.1038/ncomms5351}
  {\path{doi:10.1038/ncomms5351}}.

\bibitem{COMSOL}
COMSOL, Inc, {COMSOL} {M}ultiphysics$^\text{\tiny\textregistered}$ Reference
  Manual, version 5.4.
\newblock Available from: \url{https://www.comsol.com}.

\end{thebibliography}
\clearpage
\appendix
\section{Appendix -- Material response\label{sect:App-mat-response}}
\subsection{Strain energy splitting}

Let us assume that the strain energy in \eqref{spd-freeEn:010} and \eqref{spd-gfreeEn:010}
can be described as the sum of an isochoric part and a volumetric part
\begin{equation}\label{spd-freeEn:200}
   \strE_\elas(\eF) = \strE_\Is(\iF) + \strE_\Ve(J)\,,
\end{equation}
where $\,\iF\,$ is the \emph{isochoric} part of the elastic distortion $\,\eF\,$
defined by
\begin{equation}\label{spd-freeEn:210}
    \eF = J^\frac{1}{3}\,\iF\,.
\end{equation}

The corresponding velocity gradient decomposition
\begin{equation}\label{spd-freeEn:220}
    \dot{\eF}\,\eF^{-1} = \iFdot\,\iF^{-1} + \frac{1}{3}\,\frac{\dot{J}}{J}\,\Id\,,
\end{equation}
together with the definition of $\,\eS\,$ in \eqref{spd-bal:070}, leads to 
\begin{equation}\label{spd-freeEn:230}
   \eS\cdot\eFdot 
   = J\,\tT\cdot\eFdot\,\eF^{-1} 
   = J\,\tT\cdot\iFdot\,\iF^{-1} + \frac{1}{3}\tT\cdot\Id\,\dot{J} 
   = J\,\dev{\tT}\cdot\iFdot\,\iF^{-1} + \frac{1}{3}\,\trace{\tT}\,\dot{J} \,,
\end{equation}
as well as
\begin{equation}\label{spd-freeEn:240}
   \eSf(\eF)\cdot\eFdot 
   = J\,\dev{\tTf(\eF)}\cdot\iFdot\,\iF^{-1} + \frac{1}{3}\trace{\tTf(\eF)}\,\dot{J}\,.
\end{equation}

Denoting, just for convenience, by
\begin{align}
   \iS &= (\dev{\tT})\,\iF^{-\trp} \,, \label{spd-freeEn:250} \\
   \iSf(\eF) &= (\dev{\tTf(\eF)})\,\iF^{-\trp}  \,, \label{spd-freeEn:260}
\end{align}
and, consistent with the definition of $\,\ps\,$ in \eqref{spd-bal:150},
\begin{equation}\label{spd-freeEn:265}
   \epsf(J) = -\frac{1}{3}\trace{\tTf(\eF)} \,,
\end{equation}
the expressions \eqref{spd-freeEn:230} and \eqref{spd-freeEn:240} can be written in a simpler form
\begin{align}
   \eS\cdot\eFdot 
   &= J\,\iS\cdot\iFdot - \ps\,\dot{J} \,,\label{spd-freeEn:270}\\[\jot]
   \eSf(\eF)\cdot\eFdot 
   &= J\,\iSf(\eF)\cdot\iFdot - \epsf(J)\,\dot{J}\,.\label{spd-freeEn:280}
\end{align}

Looking back at \eqref{spd-freeEn:020} we can now derive a more detailed 
constitutive characterization consistent with the strain energy decomposition \eqref{spd-freeEn:200}.
To this end, let us replace the left-hand side of \eqref{spd-freeEn:020} by \eqref{spd-freeEn:280} and 
the right-hand side by \eqref{spd-freeEn:200}
\begin{equation}\label{spd-freeEn:290}
   J\,\iSf(\eF)\cdot\iFdot - \epsf(J)\,\dot{J}=\frac{d}{dt}\strE_\Is(\iF) + \frac{d}{dt}\strE_\Ve(J)\,,
\end{equation}
and make the more detailed assumptions, consistent with \eqref{spd-freeEn:020},
\begin{align}
      J\,\iSf(\eF)\cdot\iFdot &= \frac{d}{dt}\,\strE_\Is(\iF)\,,\label{spd-strainEn:295}\\[\jot]
      - \epsf(J)\,\dot{J} &= \frac{d}{dt}\strE_\Ve(J)\,.\label{spd-strainEn:300}
\end{align}

\subsection{Constitutive characterization\label{sect:neo-Hook}}

Replacing \eqref{spd-freeEn:270} and \eqref{spd-freeEn:280} respectively
into the inequalities \eqref{spd-bal:100} and \eqref{spd-freeEn:030} 
leads to the new form of \eqref{spd-freeEn:050}
\begin{equation}\label{spd-freeEn:300}
\begin{aligned}
   \beta\,J\,\big(
   \underbrace{
   \iS-\iSf(\eF)}_{\iS^{+}}
   \big)\cdot\iFdot
   &-\big(
   \underbrace{
   \ps-\epsf(J)}_{\ps^{+}}
   \big)\,\beta\,\dot{J}\\
   &+\big(
   \underbrace{
     \rhoo\,\big(\chp-\chpf(c)\big)
   - \alpha\,\big(J\,\ps + \strE_\elas(\eF)\big)}_{\rhoo\,\chp^{+}} 
   \big)\,\dot{c}
   -\refhflux\cdot\refgrad\chp
   \ge0\,,
\end{aligned}
\end{equation}
from which we get again \eqref{spd-freeEn:060} and \eqref{spd-freeEn:080} unchanged, 
and a more detailed constitutive characterization for the stress \eqref{spd-freeEn:070},
\begin{align}
   \iS &= \iSf(\eF) + \iS^{+}\,,\quad
   \iS^{+}\cdot\iFdot\ge0\,, \label{spd-freeEn:310}\\[\jot]
   \ps &= \epsf(J) + \ps^{+}\,,\quad
   -\ps^{+}\cdot\dot{J}\ge0\,, \label{spd-freeEn:320}
\end{align}
with $\iS^{+}$ and $\ps^{+}$ possibly describing different dissipative mechanisms. 
Consistent with the characterization of the dissipative stress in \eqref{spd-freeEn:310}, we set here
\begin{equation}\label{spd-freeEn:400}
\begin{aligned}
    \iS^{+} &= \iT^{+}\,\iF^{-\trp}\,,
    \\
    \iT^{+} &= 2\,\etavis\,\dev\sym\grad\vel\,,
\end{aligned}
\end{equation}
where $\etavis>0$ is the \emph{shear viscosity}.


Let us consider a neo-Hookean strain energy
\begin{equation}\label{spd-strainEn:200}
   \strE_\Is(\iF) = k_\Is\,(\bar{I}_1-3)\,,
\end{equation}
with
\begin{equation}\label{spd-strainEn:210}
\bar{I}_1=\trace{(\iF^\trp\,\iF)}=\iF\cdot\iF\,,
\end{equation}
to be substituted into \eqref{spd-freeEn:200}. 
Therefore from \eqref{spd-strainEn:295} we get
\begin{equation}\label{spd-strainEn:220}
      J\,\iSf(\eF)\cdot\iFdot = 2\,k_\Is\,\iF\cdot\iFdot\,,
\end{equation}
which by \eqref{spd-freeEn:260} can be conveniently transformed into
\begin{equation}\label{spd-strainEn:245}
      J\,\dev{\tTf(\eF)}\cdot\iFdot\,\iF^{-1} = 2\,k_\Is\,\dev(\iF\,\iF^\trp)\cdot\iFdot\,\iF^{-1}\,.
\end{equation}
We get this way the response function
\begin{equation}\label{spd-strainEn:260}
   \dev\tTf(\eF)=
   2\,k_\Is\,J^{-1}\,\dev(\iF\,\iF^\trp) =
   2\,k_\Is\,J^{-5/3}\,\dev(\eF\,\eF^\trp)\,,
\end{equation}
with, by \eqref{spd-freeEn:265},
\begin{equation}
   \tTf(\eF) = \dev\tTf(\eF) -\epsf(J)\,\Id\,.\label{spd-strainEn:250}
\end{equation}
For the Piola stress $\,\eS\,$ response we get from its definition in \eqref{spd-bal:070},
\begin{equation}\label{spd-strainEn:270}
   \begin{aligned}
      \eSf(\eF)\cdot\dot{\eF} &= J\,\tTf(\eF)\,\eF^{-\trp}\cdot\dot{\eF} \\
      &= 2\,k_\Is\,J^{-2/3}\,\big(\dev(\eF\,\eF^\trp)\big)\,\eF^{-\trp}\cdot\dot{\eF} 
      -J\,\epsf(J)\,\eF^{-\trp}\cdot\dot{\eF} \\
      &= 2\,k_\Is\,J^{-2/3}\,\big(\eF-\frac{1}{3}\trace(\eF\,\eF^\trp)\,\eF^{-\trp}\big)\cdot\dot{\eF}
      -J\,\epsf(J)\,\eF^{-\trp}\cdot\dot{\eF}\,.
   \end{aligned}
\end{equation}

Finally if we set
\begin{equation}\label{spd-strainEn:320}
   \strE_\Ve(J) = k_\Ve\,(J-1)^2\,,
\end{equation}
from \eqref{spd-strainEn:300} we get
\begin{equation}\label{spd-strainEn:330}
   -\epsf(J) = 2\,k_\Ve\,(J-1)\,.
\end{equation}

It is worth noting that, should the elastic distortion $\,\eF\,$ be isochoric then, 
as can be seen from the dissipation inequality \eqref{spd-freeEn:300}, the incompressibility
condition \mbox{$\,J=1\,$} would leave the constitutive characterization of $\,\ps\,$ void,
thus qualifying $\,\ps\,$ as a \emph{purely reactive pressure}.

%
\clearpage
\section{Appendix -- Numerical simulations}
\newcommand{\simMG}{s-4-g-2}
\newcommand{\jumptoSimColl}[1]{{\hyperref[sec:\simMG-#1]{\color{blue}\smalleyeball}{\;\;}{\dBrack{{#1}}}}}
\newcommand{\gosimColl}{\hyperref[sec:\simMG-\simColl]{\dBrack{\simColl}}}
\newcommand{\smalleyeball}{{$\bigcirc\hspace{-8.2pt}\bullet$}}
\newcommand{\jumptoSimul}[1]{{\hyperref[\simMG-#1]{\color{red}\smalleyeball}{\;\;}{\simulRef{{#1}}}}}
\newcommand{\jumptoSimulFig}[1]{{\hyperref[simul-\simMG-#1_bnd-traction]{\color{red}\smalleyeball}{\;\;}{\simulRef{{#1}}}}}
\newcommand{\cmB}{c_{\tiny\textsc{b}}} 
\newcommand{\cmR}{c_{\tiny\textsc{r}}} 
\newcommand{\cxwave}{{w_a}} 
\newcommand{\cmwave}{{w_m}} 
\newcommand{\crwave}{{w_r}} 
\newcommand{\cwave}{{w_c}}  
\newcommand{\inputParamTable}[1]{}
\newcommand{\kilo}{{\color{red}\text{k}}}

We conducted some numerical simulations on a toy model in the shape of a solid cylinder
sliding over a straight rigid core, with the end faces and the cylindrical lateral boundary free to deform.

We used COMSOL Multiphysics$^\text{\tiny\textregistered}$ software \cite{COMSOL} to get a finite element
solution by implementing directly in their original form the expressions for:
\begin{itemize}[nosep]
\item{}\emph{the force power balance law} \eqref{spd-bal:060},
\item{}\emph{the species power balance law} \eqref{spd-bal:040},
\item{}\emph{the microforce balance law} \eqref{act-bal:150}.
\end{itemize}
We didn't select values for the material parameters as appropriate to 
any specific physical problem.
Nevertheless our aim was to get some insights about the behavior of the mechanical model
arising from the framework outlined above, and devised for describing cell
diffusion and phase separation in a soft tissue.

The strain energy $\,\strE_\elas\,$ in \eqref{spd-freeEn:010}
has been chosen to characterize an almost incompressible neo-Hookean material,
as in \secref{sect:neo-Hook}, through $k_\Is$ and $k_\Ve$.  
The mobility tensor in \eqref{spd-Fick:100} has been given the simple form
$\curMob=M\,\Id$, with the scalar mobility coefficient $M$.
The shear viscosity $\etavis$ has been defined by \eqref{spd-freeEn:400}
while introducing the dissipative stress.

\begin{table}[h!]
\centering
\renewcommand{\arraystretch}{1.3}
\begin{tabular}{|c|c|c|c|c|c|c|c|}
\hline\rule{0pt}{2.8ex}
  \makebox[10pt][c]{$\rhoo$}
& \makebox[10pt][c]{$k_\Is$}
& \makebox[10pt][c]{$k_\Ve$}
& \makebox[10pt][c]{$\etavis$}
& \makebox[20pt][c]{$M$} 
\\ 
\hline\hline\rule{0pt}{2.5ex}
  $6\times10^4\,\texttt{[mol/m$^3$]}$
& $1\,\texttt{[GPa]}$
& $20\,\texttt{[GPa]}$
& $10^{-3}\texttt{[Pa$\times$s]}$
& $3\times10^{-4}\texttt{[mol$^2$/({m$^4$}$\times$Pa$\times$s)]}$
\\ 
\hline
\end{tabular}
\caption{Material constitutive parameters}
\label{Tab:const-param}
\end{table}


\begin{table}[h!]
\centering
\renewcommand{\arraystretch}{1.3}
\begin{tabular}{|c|c|c|c|c|c|c|c|}
  \hline\rule{0pt}{2.8ex}
  \makebox[10pt][c]{$\kcvx$}
& \makebox[10pt][c]{$\kspn$}
& \makebox[10pt][c]{$\kdel$}
& \makebox[10pt][c]{$\gamma$}
& \makebox[20pt][c]{$c_{max}$}
& \makebox[10pt][c]{$c_\spn$}
& \makebox[10pt][c]{$c_1$}
& \makebox[10pt][c]{$c_2$}
\\ 
\hline\hline\rule{0pt}{2.5ex}
  $10^5\texttt{[J/mol]}$
& $5\times10^5$\texttt{[J/mol]}
& $0.1$
& $0.25$
& $12$
& $6$
& $1.76$
& $10.24$
\\ 
\hline\rule{0pt}{2.5ex}
  $10^5\texttt{[J/mol]}$
& $5\times10^5$\texttt{[J/mol]}
& $0.1$
& $0.75$
& $12$
& $6$
& $0.81$
& $11.19$
\\ 
\hline
\end{tabular}
\caption{Spinodal parameters}
\label{Tab:spin-param}
\end{table}

\begin{table}[h!]
\centering
\renewcommand{\arraystretch}{1.3}
\begin{tabular}{|c|c|}
\hline\rule{0pt}{2.8ex}
  \makebox[10pt][c]{$\kg$}
& \makebox[10pt][c]{$\kdg$}
\\ 
\hline\hline\rule{0pt}{2.5ex}
  $3\times10^3$\texttt{[J/m]}
& $6\times10^5$\texttt{[J/m$^2$]}
\\ 
\hline
\end{tabular}
\caption{Interfacial energy and directional cue coefficients}
\label{Tab:g-param}
\end{table}

Each simulation consists in starting from an initial uniform cell concentration 
\mbox{$\,c_\spn=6\,$}, possibly modified by a small perturbation, 
$\cxwave$ (axial) or $\cwave$ (circumferential), 
and following the cells migrating inside the tissue according to the Cahn-Hilliard equations,
as described here by \eqref{spd-bal:060}, \eqref{spd-bal:040}, and \eqref{act-bal:150}.

The concentration value \mbox{$\,c_\spn\,$} marks the onset of the spinodal decomposition,
characterized by the parameters $\,\gamma\,$ and $\,\delta\,$ in \eqref{act-diff:120},
besides $\,\kcvx$ and $\,\kspn$, as well as by $\,\rhoo\,$ in \eqref{spd-def:040} 
and $\,\alpha\,$ in \eqref{spd-def:050},
while the coarsening evolution will depend on $\,\kg\,$ in \eqref{spd-gfreeEn:020}
and possibly on $\,\kdg\,$ in \eqref{act-bal:110}.
The corresponding energy graph in \figref{fig-chp-sym} exhibits two minima 
at $c_1$ and $c_2$, as in Table~\ref{Tab:spin-param}.
We make the assumption that during the phase separation and aggregation process 
the concentration cannot exceed \mbox{$\,c_{max}=12\,$}.
The \emph{critical} value for $\,\gamma\,$, according to \eqref{act-bal:210}, turns out to be 
\begin{equation} \label{num-simul:100}
    \gamma_{cr} = \frac{\kcvx}{\kspn}\,\frac{c_{max}}{2\,c_\spn\,(c_{max}-c_\spn)}
    =\frac{\kcvx}{\kspn}\,\frac{1}{6}\,.
\end{equation}
It is worth recalling what we mean by a concentration value like $c=6$. 
By \eqref{spd-def:040}, the \emph{molar density} per unit reference volume
of the guest species $\rhob$ is 6 times as much as $\rhoo$, 
the \emph{molar density} of the host tissue.
So the \emph{particles} (or cells, or molecules) of the guest 
species, per unit reference volume, are 6 times as many as the \emph{particles} of the host tissue.
Since we should look at $\alpha$ in \eqref{spd-def:040} as the ratio between 
the molar reference volume of the guest species and 
the molar reference volume of the host tissue,
taking a value like that in Table~\ref{Tab:micro-param},
we get the much smaller value $(\alpha\,c)=6\times10^{-2}$ to be replaced 
in the volume change expression \eqref{spd-def:050}.

\begin{table}[h!]
\centering
\renewcommand{\arraystretch}{1.3}
\begin{tabular}{|c|c|c|c|c|c|c|c|}
\hline\rule{0pt}{2.8ex}
  \makebox[30pt][c]{shape} 
& \makebox[10pt][c]{$D_{ext}$} 
& \makebox[10pt][c]{$D_{int}$} 
& \makebox[10pt][c]{$L$}
\\ 
\hline\hline\rule{0pt}{2.5ex}
  {\small\texttt{cylinder}} 
& $5\,\texttt{[mm]}$ 
& $2.5\,\texttt{[mm]}$
& $10\,\texttt{[mm]}$
\\ 
\hline\rule{0pt}{2.5ex}
  {\small\texttt{disk}} 
& $10\,\texttt{[mm]}$ 
& $2.5\,\texttt{[mm]}$
& $5/3\,\texttt{[mm]}$
\\ 
\hline
\end{tabular}
\caption{Geometrical parameters}
\label{Tab:geom-param}
\end{table}

\begin{table}[h!]
\centering
\renewcommand{\arraystretch}{1.3}
\begin{tabular}{|c|c|c|c|c|c|}
\hline\rule{0pt}{2.8ex}
  \makebox[30pt][c]{shape} 
& \makebox[10pt][c]{$\alpha$} 
\\ 
\hline\hline\rule{0pt}{2.5ex}
  {\small\texttt{cylinder}} 
& $0.01$ 
\\ 
\hline\rule{0pt}{2.5ex}
  {\small\texttt{disk}} 
& $0.01$ 
\\ 
\hline
\end{tabular}
\caption{Molar reference volume ratio}
\label{Tab:micro-param}
\end{table}

\subsection{Numerical simulation collections \label{sec:simCollns}} 

\begin{table}[h]
\renewcommand{\arraystretch}{1.5}
\centering
\begin{tabular}{|l|}
\hline\rule{0pt}{2.5ex}\jumptoSimColl{cyl}\\
\hline\rule{0pt}{2.5ex}\jumptoSimColl{disk}\\
\hline
\end{tabular}
\caption{Numerical simulation collections}
\label{Tab:simCollns}
\end{table}

The outcome of a few numerical simulations is reported in the following pages. 
For the sake of brevity, those simulations have been chosen from a larger collection 
where several sets of parameters have been considered. 
That is why some assessments may be based on the larger set of results.
Nevertheless, the aim of the numerical simulations was only illustrative of the 
predictive capability of the general framework.

\clearpage
\renewcommand{\simColl}{cyl}
\subsection*{Numerical simulation collection \dBrack{\simColl}\label{sec:\simMG-\simColl}}

\renewcommand{\BackColl}{}
\renewcommand{\NextColl}{\simMG-disk}
\gotoc---\goNextColl

\renewcommand{\simulDisplayRef}{\hyperref[Tab:\simColl]{\simulRef{\simulLabelS}}}
\renewcommand{\simulLabel}{\simMG-\simulLabelS}

\renewcommand{\inputParamTable}[1]{\input{#1}}
\medskip\noindent

For the cylinder, it can be shown, from  a larger set of simulations, that a smaller $\kg$ makes
both the concentration jump \mbox{$\Delta{c}=(\cmR-\cmB)$} 
and the maximum slope $|c'_{max}|$ higher, 
while the interface thickness \mbox{$\ell=\Delta{c}/|c'_{max}|$} gets smaller.
Notice how the concentration jump depends on $\cgo$ through $\cdo$.

\begin{table}[h!]
\hspace{-20pt}
\renewcommand{\arraystretch}{1.3}
\begin{tabular}{|l|c|c|c|c||c||c|c||c|c||c|c|}
\hline\rule{0pt}{2.8ex}
  \makebox[80pt][c]{\dBrack{\simColl}} 
& \makebox[30pt][c]{$c(0)$}
& \makebox[20pt][c]{$\gamma$}
& \makebox[20pt][c]{$\kg$}
& \makebox[30pt][c]{$\cdo$}
& \makebox[20pt][c]{\texttt{Fig.}}
& \makebox[20pt][c]{$\cmB$}
& \makebox[20pt][c]{$\cmR$}
& \makebox[30pt][c]{$|c'_{max}|$}
& \makebox[20pt][c]{$\ell$}
& \makebox[20pt][c]{\texttt{Fig.}}
& \makebox[20pt][c]{\texttt{Fig.}}
\\ 
\hline
\hline\rule{0pt}{2.5ex}
\jumptoSimulFig{mixdd-13a1}
& $c_\spn\;\cxwave$ 
& $0.25$
& $3.6\,\kilo$
& $(0,0,0)$
& \ref{simul-\simMG-mixdd-13a1_bnd-traction}
& $1.79$
& $10.2$
& $3.5\,\kilo$
& $2.5/\kilo$
& \ref{simul-\simMG-mixdd-13a1_cyl-surface}
& \ref{simul-\simMG-mixdd-13a1_A-line-graphs}
\\ 
\hline\rule{0pt}{2.5ex}
\jumptoSimulFig{mixdd-13b1}
& $c_\spn\;\cxwave$ 
& $0.25$
& $3.6\,\kilo$
& $2\,(1,0,0)$
& \ref{simul-\simMG-mixdd-13b1_bnd-traction}
& $1.51$
& $10.5$
& $3.5\,\kilo$
& $2.6/\kilo$
& \ref{simul-\simMG-mixdd-13b1_cyl-surface}
& \ref{simul-\simMG-mixdd-13b1_A-line-graphs}
\\ 
\hline
\end{tabular}
\caption{Simulation collection \dBrack{\simColl} {\color{red} [$\kilo=10^3$]}}.
\label{Tab:\simColl}
\end{table}
%
The outcome of each numerical simulation, characterized by the parameter values
in a row of Table~\ref{Tab:\simColl}, is described by a collection of figures as
detailed below for the first simulation. 
%
Fig.~\ref{simul-\simMG-mixdd-13a1_bnd-traction}
shows the last time frame in Fig.~\ref{simul-\simMG-mixdd-13a1_cyl-surface}, 
to be seen as an approximate stationary pattern where the two phases 
turn out to be characterized by concentration values $\cmB$ and $\cmR$, 
which can be found in small print on top and below the sidebar legend.
%
Fig.~\ref{simul-\simMG-mixdd-13a1_g-arrows}
complements the figure above by showing some sections of the same $c$ pattern. 
(It may show also a depiction of the $\cgo$ vector field, when applied).
Fig.~\ref{simul-\simMG-mixdd-13a1_end-face-disp} shows in a log-time graph
the axial displacement history of the left and right end faces.
Fig.~\ref{simul-\simMG-mixdd-13a1_strain-energy} shows in a log-time graph 
the history of the total strain energy. 
The relaxed state has been assumed to be characterized by $c=0$, not by 
the initial concentration $c(0)$.
%
The time frames in 
Fig.~\ref{simul-\simMG-mixdd-13a1_cyl-surface},
arranged in a grid with left-to-right, top-to-bottom (lr-tb), row wise ordering,
describe by color maps the time evolution of the concentration $c$ on the boundary surface.
The time frames in 
Fig.~\ref{simul-\simMG-mixdd-13a1_xy-slices},
arranged the same way as  in
Fig.~\ref{simul-\simMG-mixdd-13a1_cyl-surface},
describe by color maps the time evolution of $c$ 
on some sections.
The time frames in 
Fig.~\ref{simul-\simMG-mixdd-13a1_tau-line-graph},
show the time evolution graphs of $\ctro$ in \eqref{act-bal:090}, 
along a longitudinal line.
The time frames in 
Fig.~\ref{simul-\simMG-mixdd-13a1_isp-spm-line-graph},
show the time evolution graphs of 
$\spm$ and $\isp$ (dashed lines), and $(\spm-\isp)$ (solid line) 
along a longitudinal line.
The grid in 
Fig.~\ref{simul-\simMG-mixdd-13a1_A-line-graphs}
shows on the left column the time evolution, 
along a longitudinal line,
of $c$ (top row), 
and $\chp$ (bottom row), 
with the last frame highlighted on the right column (the dashed line describes the slope $c'$ of the $c$ graph). 
The grid in 
Fig.~\ref{simul-\simMG-mixdd-13a1_B-line-graphs}
shows the corresponding graphs for
$\ctro$ (top row), 
$\isp$ and
$\spm$ (bottom row, dashed lines),
$(\spm-\isp)$ (bottom row, solid line).
\renewcommand{\simulLabelS}{mixdd-13a1}
\clearpage
\renewcommand{\simulId}{2024-05-08_cyl-Cahn-Hill-\simMG-\simulLabelS}
\renewcommand{\scale}{0.30}
\begin{figure}
   \setlength{\unitlength}{\scale pt} 
   \centering
   \boxed{
   \begin{picture}(482,362) 
      \put(0,0){\includegraphics[viewport= 0 0 481 361, scale=\scale, clip]
      {flat-\simulId__bnd-traction.pdf}}
   \end{picture}
   }
   \caption{Last frame with $c$ pattern (\simulDisplayName).}
   \label{simul-s-4-g-2-mixdd-13a1_bnd-traction}
\end{figure}
\renewcommand{\simulId}{2024-05-08_cyl-Cahn-Hill-\simMG-\simulLabelS}
\renewcommand{\scale}{0.30}
\begin{figure}
   \setlength{\unitlength}{\scale pt} 
   \centering
   \boxed{
   \begin{picture}(482,362) 
      \put(0,0){\includegraphics[viewport= 0 0 481 361, scale=\scale, clip]
      {flat-\simulId__g-arrows.pdf}}
   \end{picture}
   }
   \caption{Last frame of a section with $c$ pattern and $\cgo$ field (\simulDisplayName).}
   \label{simul-s-4-g-2-mixdd-13a1_g-arrows}
\end{figure}
\renewcommand{\simulId}{2024-05-08_cyl-Cahn-Hill-\simMG-\simulLabelS}
\renewcommand{\scale}{0.30}
\begin{figure}
   \setlength{\unitlength}{\scale pt} 
   \centering
   \begin{picture}(482,362)  
      \put(0,0){\includegraphics[viewport= 0 12 481 361, clip, scale=\scale]
      {flat-\simulId__u1_Log.pdf}}
   \end{picture}
   \caption{Log time history of the average axial displacement at the end faces (solid line: right face, dashed line: left face) (\simulDisplayName).}
   \label{simul-s-4-g-2-mixdd-13a1_end-face-disp}
\end{figure}
\renewcommand{\simulId}{2024-05-08_cyl-Cahn-Hill-\simMG-\simulLabelS}
\renewcommand{\scale}{0.30}
\begin{figure}
   \setlength{\unitlength}{\scale pt} 
   \centering
   \begin{picture}(482,362)  
      \put(0,0){\includegraphics[viewport= 0 12 481 361, clip, scale=\scale]
      {flat-\simulId__strain-energy_Log.pdf}}
   \end{picture}
   \caption{Log time history of the total strain energy (\simulDisplayName).}
   \label{simul-s-4-g-2-mixdd-13a1_strain-energy}
\end{figure}
\clearpage
\renewcommand{\simulId}{2024-05-08_cyl-Cahn-Hill-\simMG-\simulLabelS}
\renewcommand{\scale}{0.2}
\begin{figure}
   \setlength{\unitlength}{\scale pt} 
   \centering
   \boxed{
   \begin{picture}(1285,1443)
      \put(000,962){\includegraphics[viewport= 0 0 641 481, scale=\scale]
      {flat-frames-\simulId__cyl-surface-001.pdf}}
      \put(641,962){\includegraphics[viewport= 0 0 641 481, scale=\scale] 
      {flat-frames-\simulId__cyl-surface-002.pdf}}
      \put(000,481){\includegraphics[viewport= 0 0 641 481, scale=\scale]
      {flat-frames-\simulId__cyl-surface-003.pdf}}
      \put(641,481){\includegraphics[viewport= 0 0 641 481, scale=\scale] 
      {flat-frames-\simulId__cyl-surface-004.pdf}}
      \put(000,000){\includegraphics[viewport= 0 0 641 481, scale=\scale]
      {flat-frames-\simulId__cyl-surface-005.pdf}}
      \put(641,000){\includegraphics[viewport= 0 0 641 481, scale=\scale] 
      {flat-frames-\simulId__cyl-surface-006.pdf}}
   \end{picture}
   }
   \caption{$c$ pattern time evolution on the boundary surface (lr-tb) (\simulDisplayName).}
   \label{simul-s-4-g-2-mixdd-13a1_cyl-surface}
\end{figure}
\renewcommand{\simulId}{2024-05-08_cyl-Cahn-Hill-\simMG-\simulLabelS}
\renewcommand{\scale}{0.2}
\begin{figure}
   \setlength{\unitlength}{\scale pt} 
   \centering
   \boxed{
   \begin{picture}(1285,1443)
      \put(  0,962){\includegraphics[viewport= 0 0 641 481, scale=\scale]
      {flat-frames-\simulId__xy-slices-001.pdf}}
      \put(641,962){\includegraphics[viewport= 0 0 641 481, scale=\scale] 
      {flat-frames-\simulId__xy-slices-002.pdf}}
      \put(  0,481){\includegraphics[viewport= 0 0 641 481, scale=\scale]
      {flat-frames-\simulId__xy-slices-003.pdf}}
      \put(641,481){\includegraphics[viewport= 0 0 641 481, scale=\scale] 
      {flat-frames-\simulId__xy-slices-004.pdf}}
      \put(  0,000){\includegraphics[viewport= 0 0 641 481, scale=\scale]
      {flat-frames-\simulId__xy-slices-005.pdf}}
      \put(641,000){\includegraphics[viewport= 0 0 641 481, scale=\scale] 
      {flat-frames-\simulId__xy-slices-006.pdf}}
   \end{picture}
   }
   \caption{$c$ pattern time evolution on longitudinal sections (lr-tb) (\simulDisplayName).}
   \label{simul-s-4-g-2-mixdd-13a1_xy-slices}
\end{figure}
\clearpage
\renewcommand{\simulId}{2024-05-08_cyl-Cahn-Hill-\simMG-\simulLabelS}
\renewcommand{\scale}{0.19}
\begin{figure}
\setlength{\unitlength}{\scale pt} 
\centering
\boxed{
\begin{picture}(1300,1440)   
   \put(000,980){\includegraphics[viewport= 0 20 641 481, clip, scale=\scale]
   {flat-frames-\simulId__tau-line-graph-001.pdf}}
   \put(660,980){\includegraphics[viewport= 0 20 641 481, clip, scale=\scale] 
   {flat-frames-\simulId__tau-line-graph-002.pdf}}
   \put(000,490){\includegraphics[viewport= 0 20 641 481, clip, scale=\scale]
   {flat-frames-\simulId__tau-line-graph-003.pdf}}
   \put(660,490){\includegraphics[viewport= 0 20 641 481, clip, scale=\scale] 
   {flat-frames-\simulId__tau-line-graph-004.pdf}}
   \put(000,000){\includegraphics[viewport= 0 20 641 481, clip, scale=\scale]
   {flat-frames-\simulId__tau-line-graph-005.pdf}}
   \put(660,000){\includegraphics[viewport= 0 20 641 481, clip, scale=\scale] 
   {flat-frames-\simulId__tau-line-graph-006.pdf}}
\end{picture}}
\caption{Time evolution graphs of $\ctro$ in \eqref{act-bal:090}, 
along a longitudinal line (lr-tb) (\simulDisplayName).}
\label{simul-s-4-g-2-mixdd-13a1_tau-line-graph}
\end{figure}
\renewcommand{\simulId}{2024-05-08_cyl-Cahn-Hill-\simMG-\simulLabelS}
\renewcommand{\scale}{0.19}
\begin{figure}
\setlength{\unitlength}{\scale pt} 
\centering
\boxed{
\begin{picture}(1300,1440)   
   \put(000,980){\includegraphics[viewport= 0 20 641 481, clip, scale=\scale]
   {flat-frames-\simulId__isp-spm-line-graph-001.pdf}}
   \put(660,980){\includegraphics[viewport= 0 20 641 481, clip, scale=\scale] 
   {flat-frames-\simulId__isp-spm-line-graph-002.pdf}}
   \put(000,490){\includegraphics[viewport= 0 20 641 481, clip, scale=\scale]
   {flat-frames-\simulId__isp-spm-line-graph-003.pdf}}
   \put(660,490){\includegraphics[viewport= 0 20 641 481, clip, scale=\scale] 
   {flat-frames-\simulId__isp-spm-line-graph-004.pdf}}
   \put(000,000){\includegraphics[viewport= 0 20 641 481, clip, scale=\scale]
   {flat-frames-\simulId__isp-spm-line-graph-005.pdf}}
   \put(660,000){\includegraphics[viewport= 0 20 641 481, clip, scale=\scale] 
   {flat-frames-\simulId__isp-spm-line-graph-006.pdf}}
\end{picture}}
\caption{Time evolution of 
$\spm$ and $\isp$ (dashed lines), $(\spm-\isp)$ (solid line),
along the top longitudinal line (lr-tb) (\simulDisplayName).}
\label{simul-s-4-g-2-mixdd-13a1_isp-spm-line-graph}
\end{figure}
\clearpage
\renewcommand{\simulId}{2024-05-08_cyl-Cahn-Hill-\simMG-\simulLabelS}
\renewcommand{\scale}{0.25}
\begin{figure}
\setlength{\unitlength}{\scale pt} 
\centering
\boxed{
\begin{picture}(990,725)     
   \put(000,378){\includegraphics[viewport= 00 15 481 361, clip, scale=\scale]
   {flat-\simulId__c-line-graph_All.pdf}}
   \put(511,378){\includegraphics[viewport= 00 15 481 361, clip, scale=\scale] 
   {flat-\simulId__c-line-graph_Last.pdf}}
   \put(000,000){\includegraphics[viewport= 00 15 481 361, clip, scale=\scale]
   {flat-\simulId__chp-line-graph_All.pdf}}
   \put(511,000){\includegraphics[viewport= 00 15 481 361, clip, scale=\scale] 
   {flat-\simulId__chp-line-graph_Last.pdf}}
\end{picture}}
\caption{Time evolution graphs of $c$ (top), $\chp$ (bottom), 
along the top longitudinal line (\simulDisplayName).}
\label{simul-s-4-g-2-mixdd-13a1_A-line-graphs}
\end{figure}
\renewcommand{\simulId}{2024-05-08_cyl-Cahn-Hill-\simMG-\simulLabelS}
\renewcommand{\scale}{0.25}
\begin{figure}
\setlength{\unitlength}{\scale pt} 
\centering
\boxed{
\begin{picture}(990,725)     
   \put(000,378){\includegraphics[viewport= 00 15 481 361, clip, scale=\scale]
   {flat-\simulId__tau-line-graph_All.pdf}}
   \put(511,378){\includegraphics[viewport= 00 15 481 361, clip, scale=\scale] 
   {flat-\simulId__tau-line-graph_Last.pdf}}
   \put(000,000){\includegraphics[viewport= 00 15 481 361, clip, scale=\scale]
   {flat-\simulId__isp-spm-line-graph_All.pdf}}
   \put(511,000){\includegraphics[viewport= 00 15 481 361, clip, scale=\scale] 
   {flat-\simulId__isp-spm-line-graph_Last.pdf}}
\end{picture}}
\caption{Time evolution graphs of $\ctro$ (top), 
and $\spm$, $\isp$, $(\spm-\isp)$ (bottom), 
along the top longitudinal line (\simulDisplayName).}
\label{simul-s-4-g-2-mixdd-13a1_B-line-graphs}
\end{figure}
\clearpage
\renewcommand{\simulLabelS}{mixdd-13b1}
\clearpage
\renewcommand{\simulId}{2024-05-08_cyl-Cahn-Hill-\simMG-\simulLabelS}
\renewcommand{\scale}{0.30}
\begin{figure}
\setlength{\unitlength}{\scale pt} 
\centering
\boxed{
\begin{picture}(482,362)
   \put(0,0){\includegraphics[viewport= 0 0 481 361, scale=\scale, clip]
   {flat-\simulId__bnd-traction.pdf}}
\end{picture}
}
\caption{Last frame with $c$ pattern (\simulDisplayName).}
\label{simul-s-4-g-2-mixdd-13b1_bnd-traction}
\end{figure}
\renewcommand{\simulId}{2024-05-08_cyl-Cahn-Hill-\simMG-\simulLabelS}
\renewcommand{\scale}{0.30}
\begin{figure}
\setlength{\unitlength}{\scale pt} 
\centering
\boxed{
\begin{picture}(482,362)
   \put(0,0){\includegraphics[viewport= 0 0 481 361, scale=\scale, clip]
   {flat-\simulId__g-arrows.pdf}}
\end{picture}
}
\caption{Last frame of a section with $c$ pattern and $\cgo$ field (\simulDisplayName).}
\label{simul-s-4-g-2-mixdd-13b1_g-arrows}
\end{figure}
\renewcommand{\simulId}{2024-05-08_cyl-Cahn-Hill-\simMG-\simulLabelS}
\renewcommand{\scale}{0.30}
\begin{figure}
\setlength{\unitlength}{\scale pt} 
\centering
\begin{picture}(482,362)    
   \put(  0,0){\includegraphics[viewport= 0 12 481 361, clip, scale=\scale]
   {flat-\simulId__u1_Log.pdf}}
\end{picture}
\caption{Log time history of the average axial displacement at the end faces (solid line: right face, dashed line: left face) (\simulDisplayName).}
\label{simul-s-4-g-2-mixdd-13b1_end-face-disp}
\end{figure}
\renewcommand{\simulId}{2024-05-08_cyl-Cahn-Hill-\simMG-\simulLabelS}
\renewcommand{\scale}{0.30}
\begin{figure}
\setlength{\unitlength}{\scale pt} 
\centering
\begin{picture}(482,362)    
   \put(  0,0){\includegraphics[viewport= 0 12 481 361, clip, scale=\scale]
   {flat-\simulId__strain-energy_Log.pdf}}
\end{picture}
\caption{Log time history of the total strain energy (\simulDisplayName).}
\label{simul-s-4-g-2-mixdd-13b1_strain-energy}
\end{figure}
\clearpage
\renewcommand{\simulId}{2024-05-08_cyl-Cahn-Hill-\simMG-\simulLabelS}
\renewcommand{\scale}{0.2}
\begin{figure}
\setlength{\unitlength}{\scale pt} 
\centering
\boxed{
\begin{picture}(1285,1443)
   \put(000,962){\includegraphics[viewport= 0 0 641 481, scale=\scale]
   {flat-frames-\simulId__cyl-surface-001.pdf}}
   \put(641,962){\includegraphics[viewport= 0 0 641 481, scale=\scale] 
   {flat-frames-\simulId__cyl-surface-002.pdf}}
   \put(000,481){\includegraphics[viewport= 0 0 641 481, scale=\scale]
   {flat-frames-\simulId__cyl-surface-003.pdf}}
   \put(641,481){\includegraphics[viewport= 0 0 641 481, scale=\scale] 
   {flat-frames-\simulId__cyl-surface-004.pdf}}
   \put(000,000){\includegraphics[viewport= 0 0 641 481, scale=\scale]
   {flat-frames-\simulId__cyl-surface-005.pdf}}
   \put(641,000){\includegraphics[viewport= 0 0 641 481, scale=\scale] 
   {flat-frames-\simulId__cyl-surface-006.pdf}}
\end{picture}
}
\caption{$c$ pattern time evolution on the boundary surface (lr-tb) (\simulDisplayName).}
\label{simul-s-4-g-2-mixdd-13b1_cyl-surface}
\end{figure}
\renewcommand{\simulId}{2024-05-08_cyl-Cahn-Hill-\simMG-\simulLabelS}
\renewcommand{\scale}{0.2}
\begin{figure}
\setlength{\unitlength}{\scale pt} 
\centering
\boxed{
\begin{picture}(1285,1443)
   \put(  0,962){\includegraphics[viewport= 0 0 641 481, scale=\scale]
   {flat-frames-\simulId__xy-slices-001.pdf}}
   \put(641,962){\includegraphics[viewport= 0 0 641 481, scale=\scale] 
   {flat-frames-\simulId__xy-slices-002.pdf}}
   \put(  0,481){\includegraphics[viewport= 0 0 641 481, scale=\scale]
   {flat-frames-\simulId__xy-slices-003.pdf}}
   \put(641,481){\includegraphics[viewport= 0 0 641 481, scale=\scale] 
   {flat-frames-\simulId__xy-slices-004.pdf}}
   \put(  0,000){\includegraphics[viewport= 0 0 641 481, scale=\scale]
   {flat-frames-\simulId__xy-slices-005.pdf}}
   \put(641,000){\includegraphics[viewport= 0 0 641 481, scale=\scale] 
   {flat-frames-\simulId__xy-slices-006.pdf}}
\end{picture}
}
\caption{$c$ pattern time evolution on longitudinal sections (lr-tb) (\simulDisplayName).}
\label{simul-s-4-g-2-mixdd-13b1_xy-slices}
\end{figure}
\clearpage
\renewcommand{\simulId}{2024-05-08_cyl-Cahn-Hill-\simMG-\simulLabelS}
\renewcommand{\scale}{0.19}
\begin{figure}
\setlength{\unitlength}{\scale pt} 
\centering
\boxed{
\begin{picture}(1300,1440)   
   \put(000,980){\includegraphics[viewport= 0 20 641 481, clip, scale=\scale]
   {flat-frames-\simulId__tau-line-graph-001.pdf}}
   \put(660,980){\includegraphics[viewport= 0 20 641 481, clip, scale=\scale] 
   {flat-frames-\simulId__tau-line-graph-002.pdf}}
   \put(000,490){\includegraphics[viewport= 0 20 641 481, clip, scale=\scale]
   {flat-frames-\simulId__tau-line-graph-003.pdf}}
   \put(660,490){\includegraphics[viewport= 0 20 641 481, clip, scale=\scale] 
   {flat-frames-\simulId__tau-line-graph-004.pdf}}
   \put(000,000){\includegraphics[viewport= 0 20 641 481, clip, scale=\scale]
   {flat-frames-\simulId__tau-line-graph-005.pdf}}
   \put(660,000){\includegraphics[viewport= 0 20 641 481, clip, scale=\scale] 
   {flat-frames-\simulId__tau-line-graph-006.pdf}}
\end{picture}}
\caption{Time evolution graphs of $\ctro$ in \eqref{act-bal:090}, 
along a longitudinal line (lr-tb) (\simulDisplayName).}
\label{simul-s-4-g-2-mixdd-13b1_tau-line-graph}
\end{figure}
\renewcommand{\simulId}{2024-05-08_cyl-Cahn-Hill-\simMG-\simulLabelS}
\renewcommand{\scale}{0.19}
\begin{figure}
\setlength{\unitlength}{\scale pt} 
\centering
\boxed{
\begin{picture}(1300,1440)   
   \put(000,980){\includegraphics[viewport= 0 20 641 481, clip, scale=\scale]
   {flat-frames-\simulId__isp-spm-line-graph-001.pdf}}
   \put(660,980){\includegraphics[viewport= 0 20 641 481, clip, scale=\scale] 
   {flat-frames-\simulId__isp-spm-line-graph-002.pdf}}
   \put(000,490){\includegraphics[viewport= 0 20 641 481, clip, scale=\scale]
   {flat-frames-\simulId__isp-spm-line-graph-003.pdf}}
   \put(660,490){\includegraphics[viewport= 0 20 641 481, clip, scale=\scale] 
   {flat-frames-\simulId__isp-spm-line-graph-004.pdf}}
   \put(000,000){\includegraphics[viewport= 0 20 641 481, clip, scale=\scale]
   {flat-frames-\simulId__isp-spm-line-graph-005.pdf}}
   \put(660,000){\includegraphics[viewport= 0 20 641 481, clip, scale=\scale] 
   {flat-frames-\simulId__isp-spm-line-graph-006.pdf}}
\end{picture}}
\caption{Time evolution of 
$\spm$ and $\isp$ (dashed lines), $(\spm-\isp)$ (solid line),
along the top longitudinal line (lr-tb) (\simulDisplayName).}
\label{simul-s-4-g-2-mixdd-13b1_isp-spm-line-graph}
\end{figure}
\clearpage
\renewcommand{\simulId}{2024-05-08_cyl-Cahn-Hill-\simMG-\simulLabelS}
\renewcommand{\scale}{0.25}
\begin{figure}
\setlength{\unitlength}{\scale pt} 
\centering
\boxed{
\begin{picture}(990,725)     
   \put(000,378){\includegraphics[viewport= 00 15 481 361, clip, scale=\scale]
   {flat-\simulId__c-line-graph_All.pdf}}
   \put(511,378){\includegraphics[viewport= 00 15 481 361, clip, scale=\scale] 
   {flat-\simulId__c-line-graph_Last.pdf}}
   \put(000,000){\includegraphics[viewport= 00 15 481 361, clip, scale=\scale]
   {flat-\simulId__chp-line-graph_All.pdf}}
   \put(511,000){\includegraphics[viewport= 00 15 481 361, clip, scale=\scale] 
   {flat-\simulId__chp-line-graph_Last.pdf}}
\end{picture}}
\caption{Time evolution graphs of $c$ (top), $\chp$ (bottom), 
along the top longitudinal line (\simulDisplayName).}
\label{simul-s-4-g-2-mixdd-13b1_A-line-graphs}
\end{figure}
\renewcommand{\simulId}{2024-05-08_cyl-Cahn-Hill-\simMG-\simulLabelS}
\renewcommand{\scale}{0.25}
\begin{figure}
\setlength{\unitlength}{\scale pt} 
\centering
\boxed{
\begin{picture}(990,725)     
   \put(000,378){\includegraphics[viewport= 00 15 481 361, clip, scale=\scale]
   {flat-\simulId__tau-line-graph_All.pdf}}
   \put(511,378){\includegraphics[viewport= 00 15 481 361, clip, scale=\scale] 
   {flat-\simulId__tau-line-graph_Last.pdf}}
   \put(000,000){\includegraphics[viewport= 00 15 481 361, clip, scale=\scale]
   {flat-\simulId__isp-spm-line-graph_All.pdf}}
   \put(511,000){\includegraphics[viewport= 00 15 481 361, clip, scale=\scale] 
   {flat-\simulId__isp-spm-line-graph_Last.pdf}}
\end{picture}}
\caption{Time evolution graphs of $\ctro$ (top), 
and $\spm$, $\isp$, $(\spm-\isp)$ (bottom), 
along the top longitudinal line (\simulDisplayName).}
\label{simul-s-4-g-2-mixdd-13b1_B-line-graphs}
\end{figure}
\clearpage
\renewcommand{\inputParamTable}[1]{}

\clearpage
\renewcommand{\simColl}{disk}\subsection*{Numerical simulation collection \dBrack{\simColl}\label{sec:\simMG-\simColl}}

\renewcommand{\BackColl}{\simMG-cyl}
\renewcommand{\NextColl}{}
\gotoc---\goBackColl

\renewcommand{\simulDisplayRef}{\hyperref[Tab:\simColl]{\simulRef{\simulLabelS}}}
\renewcommand{\simulLabel}{\simMG-\simulLabelS}

\renewcommand{\inputParamTable}[1]{\input{#1}}
\medskip\noindent

For a slice of the cylinder in the previous section,
with a larger external diameter as in Table~\ref{Tab:geom-param},
we can see how a directional cue vector field \mbox{$\,\cgo=\kdg\,\cdo\,$} 
can guide the phase separation process to a stationary pattern, by
interacting with the spinodal decomposition. 

\begin{table}[h!]
\hspace{-20pt}
\renewcommand{\arraystretch}{1.3}
\begin{tabular}{|l|c|c|c|c||c||c|c||c|c|}
\hline\rule{0pt}{2.8ex}
  \makebox[100pt][c]{\dBrack{\simColl}} 
& \makebox[30pt][c]{$c(0)$}
& \makebox[20pt][c]{$\gamma$}
& \makebox[20pt][c]{$\kg$}
& \makebox[30pt][c]{$\cdo$}
& \makebox[20pt][c]{\texttt{Fig.}}
& \makebox[20pt][c]{$\cmB$}
& \makebox[20pt][c]{$\cmR$}
& \makebox[20pt][c]{\texttt{Fig.}}
& \makebox[20pt][c]{\texttt{Fig.}}
\\ 
\hline
\hline\rule{0pt}{2.5ex}
\jumptoSimulFig{mixic-disk-15a}
& $c_\spn$ 
& $0.25$
& $3\,\kilo$
& $-5\,(0,\sin\theta,\cos\theta)$
& \ref{simul-\simMG-mixic-disk-15a_bnd-traction}
& $1.04$
& $11.0$
& \ref{simul-\simMG-mixic-disk-15a_cyl-surface}
& \ref{simul-\simMG-mixic-disk-15a_A-circumf-graphs}
\\ 
\hline\rule{0pt}{2.5ex}
\jumptoSimulFig{mixdd-disk-15d}
& $c_\spn$ 
& $0.25$
& $3\,\kilo$
& $5\,(0,\cos2\theta,-\sin2\theta)$
& \ref{simul-\simMG-mixdd-disk-15d_bnd-traction}
& $1.39$
& $10.6$
& \ref{simul-\simMG-mixdd-disk-15d_cyl-surface}
& \ref{simul-\simMG-mixdd-disk-15d_A-circumf-graphs}
\\ 
\hline
\end{tabular}
\caption{Simulation collection \dBrack{\simColl} {\color{red} [$\kilo=10^3$]}}.
\label{Tab:\simColl}
\end{table}
%
The outcome of each numerical simulation, characterized by the parameter values
in a row of Table~\ref{Tab:\simColl}, is described by a collection of figures as
detailed below.
%
Fig.~\ref{simul-\simMG-mixic-disk-15a_bnd-traction}
is the 3D view of the last time frame in 
Fig.~\ref{simul-\simMG-mixic-disk-15a_cyl-surface}, 
to be seen as an approximate stationary pattern where the two phases 
turn out to be characterized by concentration values $\cmB$ and $\cmR$, 
which can be found in small print on top and below the sidebar legend.
%
Fig.~\ref{simul-\simMG-mixic-disk-15a_g-arrows}
shows the last frame of the right face with the $c$ pattern, together with 
a depiction of the $\cgo$ vector field.
%
Fig.~\ref{simul-\simMG-mixic-disk-15a_diff-vel-arrows}
shows an intermediate frame with the $c$ pattern and the diffusion velocity 
$\,\refgrad{\dot{c}}\,$ vector field.
%
Fig.~\ref{simul-\simMG-mixic-disk-15a_strain-energy} shows in a log-time graph 
the history of the total strain energy. 
The relaxed state has been assumed to be characterized by $c=0$, not by 
the initial concentration $c(0)$.
%
The time frames in 
Fig.~\ref{simul-\simMG-mixic-disk-15a_cyl-surface},
arranged in a grid with left-to-right, top-to-bottom (lr-tb), row wise ordering,
are meant to describe by color maps the time evolution of the concentration $c$ on the right face,
while 
Fig.~\ref{simul-\simMG-mixic-disk-15a_diff-vel-frames}
describes, in a shorter time interval, the evolution of the $c$ pattern 
together with the diffusion velocity $\,\refgrad{\dot{c}}\,$ vector field.
%
The time frames in 
Fig.~\ref{simul-\simMG-mixic-disk-15a_tau-circumf-graph}
describe the time evolution graphs of $\ctro$  \eqref{act-bal:090}, 
along the right face boundary,
%
while in
Fig.~\ref{simul-\simMG-mixic-disk-15a_isp-spm-circumf-graph}
they describe the time evolution of 
$\spm$ and $\isp$ (dashed lines), $(\spm-\isp)$ (solid line),
along the right face boundary.
%
The grid in 
Fig.~\ref{simul-\simMG-mixic-disk-15a_A-circumf-graphs}
shows on the left column the time evolution 
along the right face boundary of
$c$ (top row), 
and $\chp$ (bottom row),
with the last time graph highlighted on the right column.
The grid in 
Fig.~\ref{simul-\simMG-mixic-disk-15a_B-circumf-graphs}
shows the corresponding graphs for
$\ctro$ (top row), 
and $\isp$ (bottom row, dashed line), 
$\spm$ (bottom row, dashed line),
$(\spm-\isp)$ (bottom row, solid line).

\clearpage
\renewcommand{\simulLabelS}{mixic-disk-15a}
\clearpage
\renewcommand{\simulId}{2024-01-03_cyl-Cahn-Hill-\simMG-\simulLabelS}
\renewcommand{\scale}{0.30}
\begin{figure}
\setlength{\unitlength}{\scale pt} 
\centering
\boxed{
\begin{picture}(482,362)
   \put(0,0){\includegraphics[viewport= 0 0 481 361, scale=\scale, clip]
   {flat-\simulId__bnd-traction.pdf}}
\end{picture}
}
\caption{Last frame with $c$ pattern (\simulDisplayName).}
\label{simul-s-4-g-2-mixic-disk-15a_bnd-traction}
\end{figure}
\renewcommand{\simulId}{2024-01-03_cyl-Cahn-Hill-\simMG-\simulLabelS}
\renewcommand{\scale}{0.30}
\begin{figure}
\setlength{\unitlength}{\scale pt} 
\centering
\boxed{
\begin{picture}(482,362)
   \put(0,0){\includegraphics[viewport= 0 0 481 361, scale=\scale, clip]
   {flat-\simulId__g-arrows.pdf}}
\end{picture}
}
\caption{Last frame with $c$ pattern and $\cgo$ vector field
on the right face (\simulDisplayName).}
\label{simul-s-4-g-2-mixic-disk-15a_g-arrows}
\end{figure}
\renewcommand{\simulId}{2024-01-03_cyl-Cahn-Hill-\simMG-\simulLabelS}
\renewcommand{\scale}{0.30}
\begin{figure}
\setlength{\unitlength}{\scale pt} 
\centering
\boxed{
\begin{picture}(482,362)
   \put(0,0){\includegraphics[viewport= 0 0 481 361, scale=\scale, clip]
   {flat-\simulId__diff-vel-arrows.pdf}}
\end{picture}
}
\caption{Intermediate frame with $c$ pattern and $\,\refgrad{\dot{c}}\,$ vector field (\simulDisplayName).}
\label{simul-s-4-g-2-mixic-disk-15a_diff-vel-arrows}
\end{figure}
\renewcommand{\simulId}{2024-01-03_cyl-Cahn-Hill-\simMG-\simulLabelS}
\renewcommand{\scale}{0.30}
\begin{figure}
\setlength{\unitlength}{\scale pt} 
\centering
\begin{picture}(482,362)    
   \put(  0,0){\includegraphics[viewport= 0 12 481 361, clip, scale=\scale]
   {flat-\simulId__strain-energy_Log.pdf}}
\end{picture}
\caption{Log time history of the total strain energy (\simulDisplayName).}
\label{simul-s-4-g-2-mixic-disk-15a_strain-energy}
\end{figure}
\clearpage
\renewcommand{\simulId}{2024-01-03_cyl-Cahn-Hill-\simMG-\simulLabelS}
\renewcommand{\scale}{0.2}
\begin{figure}
\setlength{\unitlength}{\scale pt} 
\centering
\boxed{
\begin{picture}(1285,1443)
   \put(  0,962){\includegraphics[viewport= 0 0 641 481, scale=\scale]
   {flat-frames-\simulId__cyl-surface-001.pdf}}
   \put(641,962){\includegraphics[viewport= 0 0 641 481, scale=\scale] 
   {flat-frames-\simulId__cyl-surface-002.pdf}}
   \put(  0,481){\includegraphics[viewport= 0 0 641 481, scale=\scale]
   {flat-frames-\simulId__cyl-surface-003.pdf}}
   \put(641,481){\includegraphics[viewport= 0 0 641 481, scale=\scale] 
   {flat-frames-\simulId__cyl-surface-004.pdf}}
   \put(  0,000){\includegraphics[viewport= 0 0 641 481, scale=\scale]
   {flat-frames-\simulId__cyl-surface-005.pdf}}
   \put(641,000){\includegraphics[viewport= 0 0 641 481, scale=\scale] 
   {flat-frames-\simulId__cyl-surface-006.pdf}}
\end{picture}
}
\caption{$c$ pattern time evolution on the right face (lr-tb) (\simulDisplayName).}
\label{simul-s-4-g-2-mixic-disk-15a_cyl-surface}
\end{figure}
\renewcommand{\simulId}{2024-01-03_cyl-Cahn-Hill-\simMG-\simulLabelS}
\renewcommand{\scale}{0.2}
\begin{figure}
\setlength{\unitlength}{\scale pt} 
\centering
\boxed{
\begin{picture}(1285,1443)
   \put(  0,962){\includegraphics[viewport= 0 0 641 481, scale=\scale]
   {flat-frames-\simulId__diff-vel-arrows-001.pdf}}
   \put(641,962){\includegraphics[viewport= 0 0 641 481, scale=\scale] 
   {flat-frames-\simulId__diff-vel-arrows-002.pdf}}
   \put(  0,481){\includegraphics[viewport= 0 0 641 481, scale=\scale]
   {flat-frames-\simulId__diff-vel-arrows-003.pdf}}
   \put(641,481){\includegraphics[viewport= 0 0 641 481, scale=\scale] 
   {flat-frames-\simulId__diff-vel-arrows-004.pdf}}
   \put(  0,000){\includegraphics[viewport= 0 0 641 481, scale=\scale]
   {flat-frames-\simulId__diff-vel-arrows-005.pdf}}
   \put(641,000){\includegraphics[viewport= 0 0 641 481, scale=\scale] 
   {flat-frames-\simulId__diff-vel-arrows-006.pdf}}
\end{picture}
}
\caption{$c$ pattern and $\,\refgrad{\dot{c}}\,$ vector field (lr-tb) (\simulDisplayName).}
\label{simul-s-4-g-2-mixic-disk-15a_diff-vel-frames}
\end{figure}
\clearpage
\renewcommand{\simulId}{2024-01-03_cyl-Cahn-Hill-\simMG-\simulLabelS}
\renewcommand{\scale}{0.19}
\begin{figure}
\setlength{\unitlength}{\scale pt} 
\centering
\boxed{
\begin{picture}(1300,1440)   
   \put(000,980){\includegraphics[viewport= 0 20 641 481, clip, scale=\scale]
   {flat-frames-\simulId__tau-circumf-graph-001.pdf}}
   \put(660,980){\includegraphics[viewport= 0 20 641 481, clip, scale=\scale] 
   {flat-frames-\simulId__tau-circumf-graph-002.pdf}}
   \put(000,490){\includegraphics[viewport= 0 20 641 481, clip, scale=\scale]
   {flat-frames-\simulId__tau-circumf-graph-003.pdf}}
   \put(660,490){\includegraphics[viewport= 0 20 641 481, clip, scale=\scale] 
   {flat-frames-\simulId__tau-circumf-graph-004.pdf}}
   \put(000,000){\includegraphics[viewport= 0 20 641 481, clip, scale=\scale]
   {flat-frames-\simulId__tau-circumf-graph-005.pdf}}
   \put(660,000){\includegraphics[viewport= 0 20 641 481, clip, scale=\scale] 
   {flat-frames-\simulId__tau-circumf-graph-006.pdf}}
\end{picture}}
\caption{Time evolution graphs of $\ctro$ in \eqref{act-bal:090}, 
along the right face boundary (lr-tb) (\simulDisplayName).}
\label{simul-s-4-g-2-mixic-disk-15a_tau-circumf-graph}
\end{figure}
\renewcommand{\simulId}{2024-01-03_cyl-Cahn-Hill-\simMG-\simulLabelS}
\renewcommand{\scale}{0.19}
\begin{figure}
\setlength{\unitlength}{\scale pt} 
\centering
\boxed{
\begin{picture}(1300,1440)   
   \put(000,980){\includegraphics[viewport= 0 20 641 481, clip, scale=\scale]
   {flat-frames-\simulId__isp-spm-circumf-graph-001.pdf}}
   \put(660,980){\includegraphics[viewport= 0 20 641 481, clip, scale=\scale] 
   {flat-frames-\simulId__isp-spm-circumf-graph-002.pdf}}
   \put(000,490){\includegraphics[viewport= 0 20 641 481, clip, scale=\scale]
   {flat-frames-\simulId__isp-spm-circumf-graph-003.pdf}}
   \put(660,490){\includegraphics[viewport= 0 20 641 481, clip, scale=\scale] 
   {flat-frames-\simulId__isp-spm-circumf-graph-004.pdf}}
   \put(000,000){\includegraphics[viewport= 0 20 641 481, clip, scale=\scale]
   {flat-frames-\simulId__isp-spm-circumf-graph-005.pdf}}
   \put(660,000){\includegraphics[viewport= 0 20 641 481, clip, scale=\scale] 
   {flat-frames-\simulId__isp-spm-circumf-graph-006.pdf}}
\end{picture}}
\caption{Time evolution of 
$\spm$ and $\isp$ (dashed lines), $(\spm-\isp)$ (solid line),
along the right face boundary (lr-tb) (\simulDisplayName).}
\label{simul-s-4-g-2-mixic-disk-15a_isp-spm-circumf-graph}
\end{figure}
\clearpage
\renewcommand{\simulId}{2024-01-03_cyl-Cahn-Hill-\simMG-\simulLabelS}
\renewcommand{\scale}{0.25}
\begin{figure}
\setlength{\unitlength}{\scale pt} 
\centering
\boxed{
\begin{picture}(990,725)     
   \put(000,378){\includegraphics[viewport= 00 15 481 361, clip, scale=\scale]
   {flat-\simulId__c-circumf-graph_All.pdf}}
   \put(511,378){\includegraphics[viewport= 00 15 481 361, clip, scale=\scale] 
   {flat-\simulId__c-circumf-graph_Last.pdf}}
   \put(000,000){\includegraphics[viewport= 00 15 481 361, clip, scale=\scale]
   {flat-\simulId__chp-circumf-graph_All.pdf}}
   \put(511,000){\includegraphics[viewport= 00 15 481 361, clip, scale=\scale] 
   {flat-\simulId__chp-circumf-graph_Last.pdf}}
\end{picture}}
\caption{Time evolution graphs of $c$ (top), $\chp$ (bottom), 
along the right face boundary (\simulDisplayName).}
\label{simul-s-4-g-2-mixic-disk-15a_A-circumf-graphs}
\end{figure}
\renewcommand{\simulId}{2024-01-03_cyl-Cahn-Hill-\simMG-\simulLabelS}
\renewcommand{\scale}{0.25}
\begin{figure}
\setlength{\unitlength}{\scale pt} 
\centering
\boxed{
\begin{picture}(990,725)     
   \put(000,378){\includegraphics[viewport= 00 15 481 361, clip, scale=\scale]
   {flat-\simulId__tau-circumf-graph_All.pdf}}
   \put(511,378){\includegraphics[viewport= 00 15 481 361, clip, scale=\scale] 
   {flat-\simulId__tau-circumf-graph_Last.pdf}}
   \put(000,000){\includegraphics[viewport= 00 15 481 361, clip, scale=\scale]
   {flat-\simulId__isp-spm-circumf-graph_All.pdf}}
   \put(511,000){\includegraphics[viewport= 00 15 481 361, clip, scale=\scale] 
   {flat-\simulId__isp-spm-circumf-graph_Last.pdf}}
\end{picture}}
\caption{Time evolution graphs of $\ctro$ (top), 
and $\spm$, $\isp$, $(\spm-\isp)$ (bottom), 
along the right face boundary (\simulDisplayName).}
\label{simul-s-4-g-2-mixic-disk-15a_B-circumf-graphs}
\end{figure}
\clearpage
\renewcommand{\simulLabelS}{mixdd-disk-15d}
\clearpage
\renewcommand{\simulId}{2024-01-03_cyl-Cahn-Hill-\simMG-\simulLabelS}
\renewcommand{\scale}{0.30}
\begin{figure}
\setlength{\unitlength}{\scale pt} 
\centering
\boxed{
\begin{picture}(482,362)
   \put(0,0){\includegraphics[viewport= 0 0 481 361, scale=\scale, clip]
   {flat-\simulId__bnd-traction.pdf}}
\end{picture}
}
\caption{Last frame with $c$ pattern (\simulDisplayName).}
\label{simul-s-4-g-2-mixdd-disk-15d_bnd-traction}
\end{figure}
\renewcommand{\simulId}{2024-01-03_cyl-Cahn-Hill-\simMG-\simulLabelS}
\renewcommand{\scale}{0.30}
\begin{figure}
\setlength{\unitlength}{\scale pt} 
\centering
\boxed{
\begin{picture}(482,362)
   \put(0,0){\includegraphics[viewport= 0 0 481 361, scale=\scale, clip]
   {flat-\simulId__g-arrows.pdf}}
\end{picture}
}
\caption{Last frame with $c$ pattern and $\cgo$ vector field
on the right face (\simulDisplayName).}
\label{simul-s-4-g-2-mixdd-disk-15d_g-arrows}
\end{figure}
\renewcommand{\simulId}{2024-01-03_cyl-Cahn-Hill-\simMG-\simulLabelS}
\renewcommand{\scale}{0.30}
\begin{figure}
\setlength{\unitlength}{\scale pt} 
\centering
\boxed{
\begin{picture}(482,362)
   \put(0,0){\includegraphics[viewport= 0 0 481 361, scale=\scale, clip]
   {flat-\simulId__diff-vel-arrows.pdf}}
\end{picture}
}
\caption{Intermediate frame with $c$ pattern and $\,\refgrad{\dot{c}}\,$ vector field (\simulDisplayName).}
\label{simul-s-4-g-2-mixdd-disk-15d_diff-vel-arrows}
\end{figure}
\renewcommand{\simulId}{2024-01-03_cyl-Cahn-Hill-\simMG-\simulLabelS}
\renewcommand{\scale}{0.30}
\begin{figure}
\setlength{\unitlength}{\scale pt} 
\centering
\begin{picture}(482,362)    
   \put(  0,0){\includegraphics[viewport= 0 12 481 361, clip, scale=\scale]
   {flat-\simulId__strain-energy_Log.pdf}}
\end{picture}
\caption{Log time history of the total strain energy (\simulDisplayName).}
\label{simul-s-4-g-2-mixdd-disk-15d_strain-energy}
\end{figure}
\clearpage
\renewcommand{\simulId}{2024-01-03_cyl-Cahn-Hill-\simMG-\simulLabelS}
\renewcommand{\scale}{0.2}
\begin{figure}
\setlength{\unitlength}{\scale pt} 
\centering
\boxed{
\begin{picture}(1285,1443)
   \put(  0,962){\includegraphics[viewport= 0 0 641 481, scale=\scale]
   {flat-frames-\simulId__cyl-surface-001.pdf}}
   \put(641,962){\includegraphics[viewport= 0 0 641 481, scale=\scale] 
   {flat-frames-\simulId__cyl-surface-002.pdf}}
   \put(  0,481){\includegraphics[viewport= 0 0 641 481, scale=\scale]
   {flat-frames-\simulId__cyl-surface-003.pdf}}
   \put(641,481){\includegraphics[viewport= 0 0 641 481, scale=\scale] 
   {flat-frames-\simulId__cyl-surface-004.pdf}}
   \put(  0,000){\includegraphics[viewport= 0 0 641 481, scale=\scale]
   {flat-frames-\simulId__cyl-surface-005.pdf}}
   \put(641,000){\includegraphics[viewport= 0 0 641 481, scale=\scale] 
   {flat-frames-\simulId__cyl-surface-006.pdf}}
\end{picture}
}
\caption{$c$ pattern time evolution on the right face (lr-tb) (\simulDisplayName).}
\label{simul-s-4-g-2-mixdd-disk-15d_cyl-surface}
\end{figure}
\renewcommand{\simulId}{2024-01-03_cyl-Cahn-Hill-\simMG-\simulLabelS}
\renewcommand{\scale}{0.2}
\begin{figure}
\setlength{\unitlength}{\scale pt} 
\centering
\boxed{
\begin{picture}(1285,1443)
   \put(  0,962){\includegraphics[viewport= 0 0 641 481, scale=\scale]
   {flat-frames-\simulId__diff-vel-arrows-001.pdf}}
   \put(641,962){\includegraphics[viewport= 0 0 641 481, scale=\scale] 
   {flat-frames-\simulId__diff-vel-arrows-002.pdf}}
   \put(  0,481){\includegraphics[viewport= 0 0 641 481, scale=\scale]
   {flat-frames-\simulId__diff-vel-arrows-003.pdf}}
   \put(641,481){\includegraphics[viewport= 0 0 641 481, scale=\scale] 
   {flat-frames-\simulId__diff-vel-arrows-004.pdf}}
   \put(  0,000){\includegraphics[viewport= 0 0 641 481, scale=\scale]
   {flat-frames-\simulId__diff-vel-arrows-005.pdf}}
   \put(641,000){\includegraphics[viewport= 0 0 641 481, scale=\scale] 
   {flat-frames-\simulId__diff-vel-arrows-006.pdf}}
\end{picture}
}
\caption{$c$ pattern and $\,\refgrad{\dot{c}}\,$ vector field (lr-tb) (\simulDisplayName).}
\label{simul-s-4-g-2-mixdd-disk-15d_diff-vel-frames}
\end{figure}
\clearpage
\renewcommand{\simulId}{2024-01-03_cyl-Cahn-Hill-\simMG-\simulLabelS}
\renewcommand{\scale}{0.19}
\begin{figure}
\setlength{\unitlength}{\scale pt} 
\centering
\boxed{
\begin{picture}(1300,1440)   
   \put(000,980){\includegraphics[viewport= 0 20 641 481, clip, scale=\scale]
   {flat-frames-\simulId__tau-circumf-graph-001.pdf}}
   \put(660,980){\includegraphics[viewport= 0 20 641 481, clip, scale=\scale] 
   {flat-frames-\simulId__tau-circumf-graph-002.pdf}}
   \put(000,490){\includegraphics[viewport= 0 20 641 481, clip, scale=\scale]
   {flat-frames-\simulId__tau-circumf-graph-003.pdf}}
   \put(660,490){\includegraphics[viewport= 0 20 641 481, clip, scale=\scale] 
   {flat-frames-\simulId__tau-circumf-graph-004.pdf}}
   \put(000,000){\includegraphics[viewport= 0 20 641 481, clip, scale=\scale]
   {flat-frames-\simulId__tau-circumf-graph-005.pdf}}
   \put(660,000){\includegraphics[viewport= 0 20 641 481, clip, scale=\scale] 
   {flat-frames-\simulId__tau-circumf-graph-006.pdf}}
\end{picture}}
\caption{Time evolution graphs of $\ctro$ in \eqref{act-bal:090}, 
along the right face boundary (lr-tb) (\simulDisplayName).}
\label{simul-s-4-g-2-mixdd-disk-15d_tau-circumf-graph}
\end{figure}
\renewcommand{\simulId}{2024-01-03_cyl-Cahn-Hill-\simMG-\simulLabelS}
\renewcommand{\scale}{0.19}
\begin{figure}
\setlength{\unitlength}{\scale pt} 
\centering
\boxed{
\begin{picture}(1300,1440)   
   \put(000,980){\includegraphics[viewport= 0 20 641 481, clip, scale=\scale]
   {flat-frames-\simulId__isp-spm-circumf-graph-001.pdf}}
   \put(660,980){\includegraphics[viewport= 0 20 641 481, clip, scale=\scale] 
   {flat-frames-\simulId__isp-spm-circumf-graph-002.pdf}}
   \put(000,490){\includegraphics[viewport= 0 20 641 481, clip, scale=\scale]
   {flat-frames-\simulId__isp-spm-circumf-graph-003.pdf}}
   \put(660,490){\includegraphics[viewport= 0 20 641 481, clip, scale=\scale] 
   {flat-frames-\simulId__isp-spm-circumf-graph-004.pdf}}
   \put(000,000){\includegraphics[viewport= 0 20 641 481, clip, scale=\scale]
   {flat-frames-\simulId__isp-spm-circumf-graph-005.pdf}}
   \put(660,000){\includegraphics[viewport= 0 20 641 481, clip, scale=\scale] 
   {flat-frames-\simulId__isp-spm-circumf-graph-006.pdf}}
\end{picture}}
\caption{Time evolution of 
$\spm$ and $\isp$ (dashed lines), $(\spm-\isp)$ (solid line),
along the right face boundary (lr-tb) (\simulDisplayName).}
\label{simul-s-4-g-2-mixdd-disk-15d_isp-spm-circumf-graph}
\end{figure}
\clearpage
\renewcommand{\simulId}{2024-01-03_cyl-Cahn-Hill-\simMG-\simulLabelS}
\renewcommand{\scale}{0.25}
\begin{figure}
\setlength{\unitlength}{\scale pt} 
\centering
\boxed{
\begin{picture}(990,725)     
   \put(000,378){\includegraphics[viewport= 00 15 481 361, clip, scale=\scale]
   {flat-\simulId__c-circumf-graph_All.pdf}}
   \put(511,378){\includegraphics[viewport= 00 15 481 361, clip, scale=\scale] 
   {flat-\simulId__c-circumf-graph_Last.pdf}}
   \put(000,000){\includegraphics[viewport= 00 15 481 361, clip, scale=\scale]
   {flat-\simulId__chp-circumf-graph_All.pdf}}
   \put(511,000){\includegraphics[viewport= 00 15 481 361, clip, scale=\scale] 
   {flat-\simulId__chp-circumf-graph_Last.pdf}}
\end{picture}}
\caption{Time evolution graphs of $c$ (top), $\chp$ (bottom), 
along the right face boundary (\simulDisplayName).}
\label{simul-s-4-g-2-mixdd-disk-15d_A-circumf-graphs}
\end{figure}
\renewcommand{\simulId}{2024-01-03_cyl-Cahn-Hill-\simMG-\simulLabelS}
\renewcommand{\scale}{0.25}
\begin{figure}
\setlength{\unitlength}{\scale pt} 
\centering
\boxed{
\begin{picture}(990,725)     
   \put(000,378){\includegraphics[viewport= 00 15 481 361, clip, scale=\scale]
   {flat-\simulId__tau-circumf-graph_All.pdf}}
   \put(511,378){\includegraphics[viewport= 00 15 481 361, clip, scale=\scale] 
   {flat-\simulId__tau-circumf-graph_Last.pdf}}
   \put(000,000){\includegraphics[viewport= 00 15 481 361, clip, scale=\scale]
   {flat-\simulId__isp-spm-circumf-graph_All.pdf}}
   \put(511,000){\includegraphics[viewport= 00 15 481 361, clip, scale=\scale] 
   {flat-\simulId__isp-spm-circumf-graph_Last.pdf}}
\end{picture}}
\caption{Time evolution graphs of $\ctro$ (top), 
and $\spm$, $\isp$, $(\spm-\isp)$ (bottom), 
along the right face boundary (\simulDisplayName).}
\label{simul-s-4-g-2-mixdd-disk-15d_B-circumf-graphs}
\end{figure}
\clearpage
\renewcommand{\inputParamTable}[1]{}
\clearpage
\clearpage
\section{Appendix -- Homogeneous deformations coupled with diffusion}
\newcommand{\ustrE}{\tilde{\strE}}
\subsection{Uniaxial deformations\label{sect:uniax}} 
Let us consider a cylinder undergoing \emph{homogeneous deformations} as described 
by the following deformation gradient matrix
\begin{equation}\label{uniax:010}
   [\refF] = \beta^\frac{1}{3}\,[\iF] = \beta^\frac{1}{3}%
   \begin{pmatrix}
   \lambda & 0 & 0 \\ 0 & \dfrac{1}{\sqrt{\lambda}} & 0\\ 0 & 0 & \dfrac{1}{\sqrt{\lambda}}
   \end{pmatrix}\,,
\end{equation}
in an orthonormal basis \mbox{$\{\ve_1, \ve_2, \ve_3\}$}, with $\ve_1$ denoting the axis direction.
Let the cylinder, after the inclusion of an amount of cells up to a uniform concentration $\,c\,$,
be stretched by a couple of axial forces on opposite faces defined by 
the traction \mbox{$\,\curt=\trac\,\ve_1$}.
When selecting \emph{affine test velocity fields}, as appropriate to homogeneous deformations, 
the force balance law \eqref{spd-bal:060} simplifies to 
\begin{equation}\label{uniax:020}
   \tT = \trac\,\ve_1\otimes\ve_1\,.
\end{equation}
Assuming that the material is \emph{incompressible} (\mbox{i.e.} $J=1$), the stress is characterized by
\begin{equation}
   \tT = \tTf(\eF) - \ps\,\Id\,,\label{uniax:025}
\end{equation}
while the strain energy \eqref{spd-freeEn:200}, by \eqref{spd-strainEn:200} and \eqref{spd-strainEn:210}, 
takes the expression 
\begin{equation}\label{uniax:030}
  \strE_\elas(\eF) = \strE_\Is(\iF) = {\ustrE}_\Is(\lambda) = k_\Is\,(\bar{I}_1-3) = k_\Is\,\left(\lambda^2 + \frac{2}{\lambda} - 3\right)\,.
\end{equation}
The response function, consistent with \eqref{spd-strainEn:260}, turns out to be
\begin{equation}
   \tTf(\iF) = 
     2\,k_\Is\,\dev(\iF\,\iF^\trp)\,,\label{uniax:050}
\end{equation}
with
\begin{equation}\label{uniax:080}
   [\dev(\iF\,\iF^\trp)] =  \frac{2}{3}\left(\lambda^2-\frac{1}{\lambda}\right)%
   \begin{pmatrix}
   1 & 0 & 0 \\[\jot] 0 & -\dfrac{1}{2} & 0\\[\jot] 0 & 0 & -\dfrac{1}{2}
   \end{pmatrix}\,.
\end{equation}
From the equations above we finally get
\begin{align}
   \trac&= 2\,k_\Is\left(\lambda^2-\frac{1}{\lambda}\right) \,,\label{uniax:100}\\[2\jot]
   \ps  &= -\frac{1}{3}\,\trac                \,. \label{uniax:110}
\end{align}
It is worth noting that in a uniaxial test on a cylinder with a circular cross section of radius $R$ 
the total force $N$ applied to each end face turns out to be
\begin{equation}\label{uniax:130}
   N = \trac\,\pi\,R^2\,\beta^{2/3}\,\frac{1}{\lambda}\,,
\end{equation}
on the right side and the opposite on the left side. 
Therefore, though the traction $\trac$ and the concentration $\,c\,$ are uncoupled, 
the total force $N$ does depend on $\,c\,$ through $\beta$.

It is interesting to look also at the chemical potential expression from \eqref{spd-gfreeEn:060}
\begin{equation}\label{uniax:140}
   \chp =  \chpf(c) -\isp + \frac{\alpha}{\rhoo}\big(\ps+\ustrE_\Is(\lambda)\big)
        =  \chpf(c) -\isp + \frac{\alpha}{\rhoo}\,k_\Is\,\left(\frac{\lambda^2}{3}+\frac{8}{3\,\lambda}-3\right)\,,
\end{equation}
which highlights the contribution of the deformation. 

\subsection{Cylinder in an open container\label{sect:uniax-oc}} 
Let us consider a cylinder undergoing homogeneous deformations as in the previous sections, 
but constrained inside a cylindrical container with open ends, leaving it free to stretch 
in the axis direction while preventing any lateral deformation.
Such a constraint is described, through \eqref{uniax:010}, by the condition
\begin{equation}\label{uniax-oc:180}
   \refF\,\ve_2=\ve_2\quad\&\quad\refF\,\ve_3=\ve_3
   \quad\Rightarrow\quad\beta^\frac{1}{3}\, \dfrac{1}{\sqrt{\lambda}}=1
   \quad\Rightarrow\quad\lambda=\beta^{\frac{2}{3}} \,.
\end{equation}
The force balance law changes to
\begin{equation}\label{uniax-oc:190}
   \tT = \trac\,(\ve_2\otimes\ve_2 + \ve_3\otimes\ve_3)\,,
\end{equation}
leading, by \eqref{uniax:025}, \eqref{uniax:050} and \eqref{uniax:080}, to the expressions for the reactive traction and pressure
\begin{align}
   \trac&= -2\,k_\Is\left(\lambda^2-\frac{1}{\lambda}\right) \,,\label{uniax-oc:200}\\[2\jot]
   \ps  &= -\frac{2}{3}\,\trac                \,. \label{uniax-oc:210}
\end{align}
It is worth noting that because of the constraint expression for $\lambda$ in \eqref{uniax-oc:180},
both $\trac$ and $\ps$ depend on the concentration $c$ through $\beta$.

Finally, 
by the reactive traction and
pressure expressions \eqref{uniax-oc:200}, \eqref{uniax-oc:210},
the strain energy density expression \eqref{uniax:030}, 
and the constraint expression for $\lambda$ in \eqref{uniax-oc:180}, 
we get from \eqref{spd-gfreeEn:060},
still neglecting the dissipative term,
the expression for the chemical potential, 
\begin{equation}\label{uniax-oc:225}
   \chp =  \chpf(c) - \isp + \frac{\alpha}{\rhoo}\big(\ps+\ustrE_\Is(\lambda)\big)
        =  \chpf(c) - \isp + \frac{\alpha}{\rhoo}\,k_\Is\,
        \left(
        \frac{7}{3}\,\lambda^2+\frac{2}{3\,\lambda} -3 
        \right),
\end{equation}
which, although differently from \eqref{uniax:140}, highlights the contribution of the deformation.

We also highlight how the Eshelby stress in \eqref{uniax-oc:225} can be derived from 
the strain energy expression \eqref{uniax:030},
by replacing \eqref{uniax-oc:200}, \eqref{uniax-oc:210}, and \eqref{uniax-oc:180}, 
into the following differentiation chain
\begin{equation}\label{uniax-oc:230}
   \frac{d}{d\beta}\,\big(\beta\,\ustrE_\Is(\lambda)\big) 
   = 
   \beta\left(\frac{d}{d\lambda}\,\ustrE_\Is(\lambda)\right)\frac{d\lambda}{d\beta}
   +
   \ustrE_\Is(\lambda)
   = 
   \ps + \ustrE_\Is(\lambda) \,.
\end{equation}
From the expression above we can also get, by \eqref{spd-def:050}, the time derivative
\begin{equation}\label{uniax-oc:240}
   \frac{d}{dt}\,\big(\beta\,\ustrE_\Is(\lambda)\big) 
   = 
   \frac{d}{d\beta}\,\big(\beta\,\ustrE_\Is(\lambda)\big)\,\dot{\beta} 
   = 
   \alpha\,(\ps + \ustrE_\Is(\lambda))\,\dot{c} \,,
\end{equation}
consistent with \eqref{spd-freeEn:020} and \eqref{spd-freeEn:030}.
The expression above will turn out to be useful, through \eqref{uniax:030}, 
when dealing with the free energy expression \eqref{spd-gfreeEn:010}
entering the dissipation inequality \eqref{spd-gbal:110}.

\subsubsection{Two isolated cylinders in an open container\label{sect:uniax-diff-2cyl-iso}} 

In order to describe diffusion and phase separation with a single interface,
let us consider a couple of cylinders of the same size and made up of the same material,
denoted by ${}^\bbn{+}$ and ${}^\bbn{-}$, sharing the same axis 
and undergoing independent \emph{homogeneous deformations} 
described, as in \secref{sect:uniax},  by the deformation gradient matrices
\begin{equation}\label{uniax-diff:010}
   [\refF^\bn{\pm}] = {\beta^\bn{\pm}}^\frac{1}{3}\,[\iF^\bn{\pm}] = 
   {\beta^\bn{\pm}}^\frac{1}{3}%
   \begin{pmatrix}
   \lambda^\bn{\pm} & 0 & 0 \\ 0 & \dfrac{1}{\sqrt{\lambda^\bn{\pm}}} & 0\\ 0 & 0 & \dfrac{1}{\sqrt{\lambda^\bn{\pm}}}
   \end{pmatrix}\,,
\end{equation}
with $\ve_1$ tangent to the shared axis.
The strain energy, from \eqref{uniax:030}, takes the expression 
\begin{equation}\label{uniax-diff:030}
  \strE_\elas(\eF^\bn{\pm}) = \strE_\Is(\iF^\bn{\pm}) = \ustrE_\Is(\lambda^\bn{\pm}) 
  = k_\Is\,(\bar{I}_1^\bn{\pm} - 3) 
  = k_\Is\,\left({\lambda^\bn{\pm}}^2 + \frac{2}{\lambda^\bn{\pm}} - 3\right)\,.
\end{equation}
The response function, from \eqref{uniax:050}, will be
\begin{equation}
   \tTf(\iF^\bn{\pm}) = 
     2\,k_\Is\,\dev(\iF^\bn{\pm}\,{\iF^\bn{\pm}}^\trp)\,,\label{uniax-diff:050}
\end{equation}
with
\begin{equation}\label{uniax-diff:080}
   [\dev(\iF^\bn{\pm}\,{\iF^\bn{\pm}}^\trp)] =  \frac{2}{3}\left({\lambda^\bn{\pm}}^2-\frac{1}{\lambda^\bn{\pm}}\right)%
   \begin{pmatrix}
   1 & 0 & 0 \\[\jot] 0 & -\dfrac{1}{2} & 0\\[\jot] 0 & 0 & -\dfrac{1}{2}
   \end{pmatrix}\,.
\end{equation}

Let us consider both cylinders constrained inside the same
rigid cylindrical container with open ends, leaving them free to stretch separately
in the axis direction while preventing any lateral deformation, as in \secref{sect:uniax}.
Such a constraint is described, as in \eqref{uniax-oc:180}, by the conditions
\begin{equation}\label{uniax-diff-oc:180}
   \refF^\bn{\pm}\,\ve_2=\ve_2\quad\&\quad\refF^\bn{\pm}\,\ve_3=\ve_3
   \quad\Rightarrow\quad{\beta^\bn{\pm}}^\frac{1}{3}\, \dfrac{1}{\sqrt{\lambda^\bn{\pm}}}=1
   \quad\Rightarrow\quad{\lambda^\bn{\pm}}={\beta^\bn{\pm}}^{\frac{2}{3}} \,.
\end{equation}
The force balance laws for the \emph{non interacting cylinders}
\begin{equation}\label{uniax-diff-oc:190}
   \tT^\bn{\pm} = \trac^\bn{\pm}\,(\ve_2\otimes\ve_2 + \ve_3\otimes\ve_3)\,,
\end{equation}
with the stress characterized by
\begin{equation}
   \tT^\bn{\pm} = \tTf(\eF^\bn{\pm}) - \ps^\bn{\pm}\,\Id\,,\label{uniax-diff-oc:225}
\end{equation}
lead, by \eqref{uniax-diff:050} and \eqref{uniax-diff:080}, to the following expressions for the 
corresponding reactive traction and pressure
\begin{align}
   \trac^\bn{\pm} &= -2\,k_\Is\left({\lambda^\bn{\pm}}^2-\frac{1}{\lambda^\bn{\pm}}\right) \,,\label{uniax-diff-oc:200}
   \\[2\jot]
   \ps^\bn{\pm}  &= -\frac{2}{3}\,\trac^\bn{\pm} \,. \label{uniax-diff-oc:210}
\end{align}

In our kinetic model,
consistent with the assumption of homogeneous deformations, 
the concentrations $\,c^\bn{+}\,$ and $\,c^\bn{-}\,$,
as well as the chemical potentials $\,\chp^\bn{+}\,$ and $\,\chp^\bn{-}\,$, 
will be assumed to be uniform inside each cylinder at any time.

As a consequence, by the uniformity property for $c$, 
the interfacial free energy \eqref{spd-gfreeEn:020} will disappear and,
by extension of that property to the corresponding test fields, 
we will get a reduced form of \eqref{act-bal:150} leading to
\begin{equation}\label{uniax-diff-iso:080}
\isp^\bn{\pm} - \spm^\bn{\pm} = 0\,.
\end{equation}

Finally, we get the chemical potential expression by transforming \eqref{spd-gfreeEn:060},
through \eqref{uniax-diff:030}, \eqref{uniax-diff-oc:180},  
\eqref{uniax-diff-oc:200}, and \eqref{uniax-diff-oc:210}, into
\begin{equation}\label{uniax-diff-oc:230}
   \begin{aligned}
   \chp^\bn{\pm} 
   &=  \chpf(c^\bn{\pm}) - \isp^\bn{\pm} + \frac{\alpha}{\rhoo}\big(\ps^\bn{\pm}+\ustrE_\Is(\lambda^\bn{\pm})\big) \\
   &=  \chpf(c^\bn{\pm}) - \isp^\bn{\pm} + \frac{\alpha}{\rhoo}\,k_\Is\,
       \left(
       \frac{7}{3}\,{\lambda^\bn{\pm}}^2+\frac{2}{3\,\lambda^\bn{\pm}} -3 
       \right).
	\end{aligned}
\end{equation}
Even the strain energy time derivative \eqref{uniax-oc:240} can be extended to both the isolated cylinders
\begin{equation}\label{uniax-diff-oc:240}
   \frac{d}{dt}\,\big(\beta^\bn{\pm}\,\ustrE_\Is(\lambda^\bn{\pm})\big) 
   = 
   \alpha\,(\ps^\bn{\pm} + \ustrE_\Is(\lambda^\bn{\pm}))\,\dot{c}^\bn{\pm} \,.
\end{equation}

%
\newcommand{\SpinMin}{4.03}
\newcommand{\SpinMax}{7.97}
\newcommand{\CoexSolMP}{{1.77, 10.23}}
\newcommand{\CoexSolM}{1.77}
\newcommand{\CoexSolP}{10.23}
\newcommand{\Igio}{{0.0033875}}
\newcommand{\Igo}{{0.00}}
\newcommand{\ItauStar}{{0.0033875}}
\newcommand{\ImuxJmusJ}{{-0.0036805}}
\newcommand{\IeshJ}{{0.00029303}}
%
\newcommand{\CoexSolAltMP}{{10.23, 1.77}}
\newcommand{\CoexSolAltM}{10.23}
\newcommand{\CoexSolAltP}{1.77}
\newcommand{\IgioAlt}{{-0.0033875}}
\newcommand{\IgoAlt}{{0.00}}
\newcommand{\ItauStarAlt}{{-0.0033875}}
\newcommand{\ImuxJmusJAlt}{{0.0036805}}
\newcommand{\IeshJAlt}{{-0.00029303}}
%
\newcommand{\RshiftCoexSolMP}{{0.89, 11.11}}
\newcommand{\RshiftCoexSolM}{0.89}
\newcommand{\RshiftCoexSolP}{11.11}
\newcommand{\RshiftIgio}{{0.0040912}}
\newcommand{\RshiftIgo}{{1.20}}
\newcommand{\RshiftItauStar}{{-1.1959}}
\newcommand{\RshiftImuxJmusJ}{{1.1956}}
\newcommand{\RshiftIeshJ}{{0.00035387}}
%
\newcommand{\RshiftCoexSolAltMP}{{9.26, 2.74}}
\newcommand{\RshiftCoexSolAltM}{9.26}
\newcommand{\RshiftCoexSolAltP}{2.74}
\newcommand{\RshiftIgioAlt}{{-0.0026048}}
\newcommand{\RshiftIgoAlt}{{1.20}}
\newcommand{\RshiftItauStarAlt}{{-1.2026}}
\newcommand{\RshiftImuxJmusJAlt}{{1.2028}}
\newcommand{\RshiftIeshJAlt}{{-0.00022535}}
%
\newcommand{\LshiftCoexSolMP}{{2.74, 9.26}}
\newcommand{\LshiftCoexSolM}{2.74}
\newcommand{\LshiftCoexSolP}{9.26}
\newcommand{\LshiftIgio}{{0.0026048}}
\newcommand{\LshiftIgo}{{-1.20}}
\newcommand{\LshiftItauStar}{{1.2026}}
\newcommand{\LshiftImuxJmusJ}{{-1.2028}}
\newcommand{\LshiftIeshJ}{{0.00022535}}
%
\newcommand{\LshiftCoexSolAltMP}{{11.11, 0.89}}
\newcommand{\LshiftCoexSolAltM}{11.11}
\newcommand{\LshiftCoexSolAltP}{0.89}
\newcommand{\LshiftIgioAlt}{{-0.0040912}}
\newcommand{\LshiftIgoAlt}{{-1.20}}
\newcommand{\LshiftItauStarAlt}{{1.1959}}
\newcommand{\LshiftImuxJmusJAlt}{{-1.1956}}
\newcommand{\LshiftIeshJAlt}{{-0.00035387}}
\subsubsection{Two interconnected cylinders\label{sect:uniaxial-diff-2cyl}} 

Let us consider the two cylinders joined together while constrained inside the open rigid cylinder.
Assuming that the exterior boundary is impermeable, let us infuse an amount of cells up to a uniform concentration.
The cylinders will stretch according to \eqref{uniax-diff-oc:180} and \eqref{spd-def:050}.

After that we rule out any subsequent supply (\mbox{$\refhsrc^\bn{\pm}=0$}) but allow diffusion
which we characterize by a normal flux through the interface between the two cylinders  
\begin{equation}\label{uniax-diff-iso:050}
   \refhfluxs=\refhflux\cdot\refn\,,
\end{equation}
a concentration jump 
\begin{equation}
   \jump{c}=(c^\bn{+}-c^\bn{-})\,,\label{uniax-diff-iso:060} 
\end{equation}
and a chemical potential jump
\begin{equation}
   \jump{\chp}=(\chp^\bn{+}-\chp^\bn{-})\,.\label{uniax-diff-iso:070}
\end{equation}
If the interface is impermeable then \mbox{$\,\refhfluxs=0\,$},
whatever the chemical potentials $\,\chp^\bn{+}\,$ and $\,\chp^\bn{-}\,$.

If we infused a different amount of cells without removing the impermeability 
condition at the interface, we would get different concentrations,
as well as different stretches and tractions.
The impermeability condition at the interface between the cylinders, together with
the given kinematical constraints and no applied traction, 
amounts indeed to their being isolated, \mbox{i.e.} exchanging neither
matter nor forces. 
This allowed us in the previous section to restrict the free energy, 
as well as the balance laws, to each of them.

Let us consider now the two cylinders joined together
but with the impermeability condition at the interface removed.
In fact it is just one cylinder with piecewise uniform concentration and chemical potential,
an impermeable exterior boundary and possibly a flux $\,\refhfluxs\,$ across the internal interface. 

We should lay down the force balance law, the species balance law, the microforce balance law, 
and also a dissipation inequality consistent with an appropriate free energy expression.
This should be done for the two cylinders joined together as well as for each of them separately.

\subsubsection{Reduced balance laws\label{sect:uniax-diff-2cyl-bal}} 

Just for short, let us skip the derivation of the \emph{force balance law} \eqref{uniax-diff-oc:190}, 
which would get by allowing only piecewise \emph{affine test velocity fields} in \eqref{spd-bal:060}.

We can show instead how to specialize the \emph{species power balance law} \eqref{spd-bal:040}
\begin{equation*}
  \int_{\refshapeP}\test{\chp}\,\dot{c}\,\rhoo\,dV = 
  -\int_{\partial\refshapeP}\test{\chp}\,%
  \refhflux\cdot\refn\,dA 
+ \int_{\refshapeP}\refhflux\cdot\refgrad\test{\chp}\,dV
+ \int_{\refshapeP}\test{\chp}\,\refhsrc\,dV
  \qquad\forall\test{\chp}
\end{equation*}
to  any piecewise \emph{uniform chemical potential test field},
while assuming an impermeable exterior boundary and no bulk supply \mbox{($\refhsrc=0$)}.
We get for each cylinder, with just a permeable end face and no gradients,
\begin{align}
  {\refV^\bn{+}}\,(\rhoo\,\dot{c}^\bn{+}\,\test{\chp}^\bn{+}) &= 
  -{\refA^\bn{+}}\,(\refhfluxn^\bn{+}\,\test{\chp}^\bn{+}) 
  \qquad\forall\test{\chp}^\bn{+}, \label{uniaxial-diff-2cyl:041}
  \\[2\jot]
  {\refV^\bn{-}}\,(\rhoo\,\dot{c}^\bn{-}\,\test{\chp}^\bn{-}) &= 
  -{\refA^\bn{-}}\,(\refhfluxn^\bn{-}\,\test{\chp}^\bn{-}) 
  \qquad\forall\test{\chp}^\bn{-}, \label{uniaxial-diff-2cyl:042}
\end{align}
and for the two cylinders joined together, with an internal interface
and an impermeable exterior boundary,
\begin{equation}  
  {\refV^\bn{+}}\,(\rhoo\,\dot{c}^\bn{+}\,\test{\chp}^\bn{+}) + 
  {\refV^\bn{-}}\,(\rhoo\,\dot{c}^\bn{-}\,\test{\chp}^\bn{-}) = 
  {\ith\refV}\,\refhfluxs\,\jump{\test{\chp}}\,\frac{1}{\ith L}
  \qquad\forall\test{\chp}^\bn{+},\, \forall\test{\chp}^\bn{-}, \label{uniaxial-diff-2cyl:043}
\end{equation}
where $\ith L$ and $\ith\refV$ stand for a \emph{fictitious interface thickness} 
and a \emph{fictitious interface volume}, with
\begin{equation}\label{uniaxial-diff-2cyl:044}
   \begin{aligned}
   \refA^\bn{\pm}&=\refA\,,\\
   L^\bn{\pm}&=L\,,\\
   \refV^\bn{\pm}&=\refA^\bn{\pm}\,L^\bn{\pm}=\refV\,,\\
   \ith\refV&=\refA\,\ith L\,.
   \end{aligned}
\end{equation}
By substituting \eqref{uniaxial-diff-2cyl:041} and \eqref{uniaxial-diff-2cyl:042} into \eqref{uniaxial-diff-2cyl:043}
we get, by \eqref{uniaxial-diff-2cyl:044},
\begin{equation}\label{uniaxial-diff-2cyl:045}
   \begin{aligned}
   \refhfluxn^\bn{+}&=-\refhfluxs\,,\\
   \refhfluxn^\bn{-}&=\refhfluxs\,,
   \end{aligned}
\end{equation}
leading to the \emph{reduced species balance laws}
\begin{align}
  \rhoo\,L\,\dot{c}^\bn{+}\ &= \refhfluxs\,,\label{uniaxial-diff-2cyl:046}
  \\[\jot]
  \rhoo\,L\,\dot{c}^\bn{-} &= -\refhfluxs\,.\label{uniaxial-diff-2cyl:047}
\end{align}
As a consequence
\begin{equation}\label{uniaxial-diff-2cyl:048}
   \dot{c}^\bn{+} + \dot{c}^\bn{-}=0\,.
\end{equation}

We can also specialize the extended \emph{microforce balance law} \eqref{act-bal:150}
\begin{equation*}
    \int_{\refshapeP}\rhoo\,\isp\,\test{\dot{c}}\,dV 
   +\int_{\refshapeP}\icgo\cdot\refgrad{\test{\dot{c}}}\,dV  
   =
   \underbrace{
   \int_{\partial\refshapeP}\ctro\,\test{\dot{c}}\,dA
   +\int_{\refshapeP}\rhoo\,\spm\,\test{\dot{c}}\,dV
   +\int_{\refshapeP}\cgo\cdot\refgrad{\test{\dot{c}}}\,dV}_{
   \text{\emph{microforce external power}}}  
   \qquad\forall\test{\dot{c}}
\end{equation*}
to any piecewise \emph{uniform concentration test field},
while assuming no boundary microforce \mbox{($\ctro=0$)} on the exterior boundary.
We get for each cylinder, where the gradient terms are zero,
\begin{align}
  {\refV^\bn{+}}\,(\rhoo\,\isp^\bn{+}\,\test{\dot{c}}^\bn{+}) 
  &= 
  \underbrace{
  {\refA^\bn{+}}\,(\ctro^\bn{+}\,\test{\dot{c}}^\bn{+})
  + {\refV^\bn{+}}\,(\rhoo\,\spm^\bn{+}\,\test{\dot{c}}^\bn{+})}_{
  \text{\emph{microforce external power}}}  
  \qquad\forall\test{\dot{c}}^\bn{+},  \label{uniaxial-diff-2cyl:051}
  \\[2\jot]
  {\refV^\bn{-}}\,(\rhoo\,\isp^\bn{-}\,\test{\dot{c}}^\bn{-}) 
  &= 
  \underbrace{
  {\refA^\bn{-}}\,(\ctro^\bn{-}\,\test{\dot{c}}^\bn{-}) 
  + {\refV^\bn{-}}\,(\rhoo\,\spm^\bn{-}\,\test{\dot{c}}^\bn{-})}_{
  \text{\emph{microforce external power}}}  
  \qquad\forall\test{\dot{c}}^\bn{-}, \label{uniaxial-diff-2cyl:052}
\end{align}
and for the two cylinders joined together, with the internal boundary term removed,
\begin{equation}\label{uniaxial-diff-2cyl:053}
  \begin{aligned}
  {\refV^\bn{+}}\,(\rhoo\,\isp^\bn{+}\,\test{\dot{c}}^\bn{+}) 
  &+ {\refV^\bn{-}}\,(\rhoo\,\isp^\bn{-}\,\test{\dot{c}}^\bn{-})
  + {\ith\refV}\,\icgos\,\jump{\test{\dot{c}}}\,\frac{1}{\ith L}
  = \\[\jot]
  &\quad
  \underbrace{
  {\refV^\bn{+}}\,(\rhoo\,\spm^\bn{+}\,\test{\dot{c}}^\bn{+})
  + {\refV^\bn{-}}\,(\rhoo\,\spm^\bn{-}\,\test{\dot{c}}^\bn{-}) 
  + {\ith\refV}\,\cgos\,\jump{\test{\dot{c}}}\,\frac{1}{\ith L}}_{
  \text{\emph{microforce external power}}}  
  \qquad\forall\test{\dot{c}}^\bn{+},\, \forall\test{\dot{c}}^\bn{-},
  \end{aligned}
\end{equation}
where
\begin{equation}\label{uniaxial-diff-2cyl:060}
   \begin{split}
   \icgos&=\icgo\cdot\refn\,,\\
   \cgos&=\cgo\cdot\refn\,.
   \end{split}
\end{equation}
By substituting \eqref{uniaxial-diff-2cyl:051} and \eqref{uniaxial-diff-2cyl:052} into \eqref{uniaxial-diff-2cyl:053}
we get
\begin{equation}\label{uniaxial-diff-2cyl:070}
  {\refA^\bn{+}}\,(\ctro^\bn{+}\,\test{\dot{c}}^\bn{+})
  + {\refA^\bn{-}}\,(\ctro^\bn{-}\,\test{\dot{c}}^\bn{-}) 
  = 
  - {\ith\refV}\,(\icgos-\cgos)\,\jump{\test{\dot{c}}}\,\frac{1}{\ith L}
  \qquad\forall\test{\dot{c}}^\bn{+},\, \forall\test{\dot{c}}^\bn{-},
\end{equation}
and, in turn,
\begin{align}
  {\refA^\bn{+}}\,(\ctro^\bn{+}\,\test{\dot{c}}^\bn{+})
  &= 
  - \frac{\ith\refV}{\ith L}\,(\icgos-\cgos)\,\test{\dot{c}}^\bn{+}
  \qquad\forall\test{\dot{c}}^\bn{+}, \label{uniaxial-diff-2cyl:080}
  \\[2\jot]
  {\refA^\bn{-}}\,(\ctro^\bn{-}\,\test{\dot{c}}^\bn{-}) 
  &= 
   \frac{\ith\refV}{\ith L}\,(\icgos-\cgos)\,\test{\dot{c}}^\bn{-}
  \qquad\forall\test{\dot{c}}^\bn{-}, \label{uniaxial-diff-2cyl:085}
\end{align}
which simplify, by \eqref{uniaxial-diff-2cyl:044}, to
\begin{align}
  \ctro^\bn{+} &= -(\icgos - \cgos) \,,\label{uniaxial-diff-2cyl:090}
  \\
  \ctro^\bn{-} &=  (\icgos - \cgos) \,.\label{uniaxial-diff-2cyl:095}
\end{align}
It is convenient to define now, somewhat consistent with \eqref{act-bal:090} and by  \eqref{uniaxial-diff-2cyl:060},
\begin{equation}\label{uniaxial-diff-2cyl:098}
   \ctros=(\icgo-\cgo)\cdot\refn = (\icgos-\cgos)\,,
\end{equation}
as a power conjugate quantity to $\jump{\test{\dot{c}}}$ in \eqref{uniaxial-diff-2cyl:070}. 
Therefore the \emph{reduced microforce balance laws} derived from \eqref{uniaxial-diff-2cyl:051} 
and \eqref{uniaxial-diff-2cyl:052} can be written as
\begin{align}
\rhoo\,(\isp^\bn{+} - \spm^\bn{+})&=\frac{1}{L^\bn{+}}\,\ctro^\bn{+}= -\frac{1}{L}\,\ctros\,,\label{uniax-diff-iso:110}\\[\jot]
\rhoo\,(\isp^\bn{-} - \spm^\bn{-})&=\frac{1}{L^\bn{-}}\,\ctro^\bn{-}=  \frac{1}{L}\,\ctros\,,\label{uniax-diff-iso:111}
\end{align}
from which we get the $\isp$ jump
\begin{equation}
\isp^\bn{+}-\isp^\bn{-} = \spm^\bn{+}-\spm^\bn{-}
- \frac{2}{\rhoo\,L}\,\ctros\,.\label{uniax-diff-iso:120}
\end{equation}
Expressions \eqref{uniax-diff-iso:110} and \eqref{uniax-diff-iso:111} should be compared with \eqref{uniax-diff-iso:080}, 
which we derived for isolated cylinders and may recover here again by removing $\ctro^\bn{+}$ and $\ctro^\bn{-}$ from 
\eqref{uniaxial-diff-2cyl:051} and \eqref{uniaxial-diff-2cyl:052}.

\subsubsection{Reduced dissipation inequality\label{sect:uniaxial-diff-2cyl-dis}} 

We should also specialize 
to \emph{piecewise affine deformations} and \emph{piecewise uniform concentration fields}
the dissipation inequality \eqref{spd-gbal:100}
\begin{equation*} 
\begin{split}
    \int_{\refshapeP}\refS\cdot\refFdot\,dV
  + \int_{\refshapeP}\rhoo\,(\chp+\isp)\,\dot{c}\,dV 
  + \int_{\refshapeP}\icgo\cdot\refgrad{\dot{c}}\,dV \\[\jot]
  - \int_{\refshapeP}\refhflux\cdot\refgrad\chp\,dV 
  - \frac{d}{dt}\int_{\refshapeP}\freeE\,dV \ge 0
  \qquad\forall\refshapeP 
  \,,
\end{split}
\end{equation*}
comparing the \emph{internal power expenditure} and the \emph{rate of change of the free energy} $\freeE$.

To this end we derive the appropriate expression for 
the \emph{interfacial energy} \eqref{spd-gfreeEn:020} by assuming
\begin{equation}\label{uniaxial-diff-2cyl-dis:010}
   \int_{\refshapeP}\strE_\gras(\refgrad{c})dV
   = (\ith\refV)\,\frac{1}{2}\,(\ith\kg)\,\jump{c}^2\frac{1}{(\ith L)^2}
   = \refV\,\tilde{\strE}_\gras(\jump{c})
   \,,
\end{equation}
where $\ith L$ and $\ith\refV$ stand for a \emph{fictitious interface thickness} 
and a \emph{fictitious interface volume}, as in \secref{sect:uniax-diff-2cyl-bal},
while $\ith\kg$ makes the expression above independent of them both, resulting in
\begin{equation}\label{uniaxial-diff-2cyl-dis:040}
   \refV\,\tilde{\strE}_\gras(\jump{c}) 
   = \refV\,\frac{1}{2}\,\kg\,\jump{c}^2\frac{1}{L^2}\,.
\end{equation}

While recalling the expression \eqref{spd-gfreeEn:030} for the rate of change of the free energy $\freeE$, 
let us neglect for clarity the dissipative stress $\eS^{+}$ and microstress $\cgodiss$,
as defined in \eqref{spd-gfreeEn:050}, and replace the remaining terms with expressions appropriate
to the reduced model, like \eqref{uniax-diff-oc:240}.
For each cylinder, where both gradients $\refgrad\chp$ and $\refgrad{\dot{c}}$ are zero, we get
respectively the conditions
\begin{align}
  {\refV^\bn{+}}\,\rhoo\left(\chp^\bn{+} + \isp^\bn{+}\right)\dot{c}^\bn{+} - 
  {\refV^\bn{+}}\left(\rhoo\,\chpf(c^\bn{+}) +
  {\alpha}\,\big(\ps^\bn{+} + \ustrE_\Is(\lambda^\bn{+})\big)\right)\,\dot{c}^\bn{+}
  &\ge0 
  \,, 
  \label{uniaxial-diff-2cyl-dis:041}
  \\[2\jot]
  {\refV^\bn{-}}\,\rhoo\left(\chp^\bn{-} + \isp^\bn{-}\right)\dot{c}^\bn{-} - 
  {\refV^\bn{-}}\left(\rhoo\,\chpf(c^\bn{-}) +
  {\alpha}\,\big(\ps^\bn{-} + \ustrE_\Is(\lambda^\bn{-})\big)\right)\,\dot{c}^\bn{-}
  &\ge0 
  \,.
  \label{uniaxial-diff-2cyl-dis:042}
\end{align}
For the two cylinders joined together, we should consider also terms with either $\refgrad\chp$ or $\refgrad{\dot{c}}$
and get
\begin{equation} 
   \begin{aligned}
  &
    {\refV^\bn{+}}\,\rhoo\left(\chp^\bn{+} + \isp^\bn{+}\right)\dot{c}^\bn{+} 
  + {\refV^\bn{-}}\,\rhoo\left(\chp^\bn{-} + \isp^\bn{-}\right)\dot{c}^\bn{-} 
  \\[2\jot]
  &
  + {\ith\refV}\,\icgos\,\jump{\dot{c}}\,\frac{1}{\ith L} 
  - {\ith\refV}\,\refhfluxs\,\jump{\chp}\,\frac{1}{\ith L}
  \\[2\jot]
  &
  - {\refV^\bn{+}}\left(\rhoo\,\chpf(c^\bn{+}) 
     + {\alpha}\,\big(\ps^\bn{+} 
     + \ustrE_\Is(\lambda^\bn{+})\big)\right)\,\dot{c}^\bn{+}
  - {\refV^\bn{-}}\left(\rhoo\,\chpf(c^\bn{-}) 
     + {\alpha}\,\big(\ps^\bn{-} 
     + \ustrE_\Is(\lambda^\bn{-})\big)\right)\,\dot{c}^\bn{-}
  \\[2\jot]
  &
  - {\ith\refV}\,(\ith\kg)\,\jump{c}\,\jump{\dot{c}}\,\frac{1}{(\ith L)^2} 
  \ge0 
  \,,
  \end{aligned}
  \label{uniaxial-diff-2cyl-dis:043}
\end{equation}
where the last term is the time derivative of the interfacial energy \eqref{uniaxial-diff-2cyl-dis:010}.

Let us set, in order to fulfill the inequalities above, the following constitutive expressions
\begin{equation}\label{uniaxial-diff-2cyl-dis:060}
   \refhfluxs=-\frac{M}{L}\,\jump{\chp}\,,\qquad M\ge0\,,
\end{equation}
and
\begin{equation}\label{uniaxial-diff-2cyl-dis:070}
   \icgos=\frac{\kg}{L}\,\jump{c}\,,
\end{equation}
as well as 
\begin{align}
  \chp^\bn{+} &= \chpf(c^\bn{+}) - \isp^\bn{+} +
  \frac{\alpha}{\rhoo}\,\big(\ps^\bn{+} + \ustrE_\Is(\lambda^\bn{+})\big)
  \,, 
  \label{uniaxial-diff-2cyl-dis:051}
  \\[2\jot]
  \chp^\bn{-} &= \chpf(c^\bn{-}) - \isp^\bn{-} +
  \frac{\alpha}{\rhoo}\,\big(\ps^\bn{-} + \ustrE_\Is(\lambda^\bn{-})\big)
  \,, 
  \label{uniaxial-diff-2cyl-dis:052}
\end{align}
identical to the expressions in \eqref{uniax-diff-oc:230} we already derived from \eqref{spd-gfreeEn:060}.

\subsubsection{Balance law summary and phase separation\label{sect:uniaxial-diff-2cyl-bal}} 

Let us collect all the balance equations and the constitutive characterizations derived in the
previous sections.
The \emph{species balance laws} \eqref{uniaxial-diff-2cyl:046} and \eqref{uniaxial-diff-2cyl:047}, 
by \eqref{uniaxial-diff-2cyl-dis:060} become
\begin{align}
  \dot{c}^\bn{+}\ &= -\frac{M}{\rhoo\,L^2}\,\jump{\chp}\,,\label{uniaxial-diff-2cyl:081}
  \\[\jot]
  \dot{c}^\bn{-} &= \frac{M}{\rhoo\,L^2}\,\jump{\chp}\,,\label{uniaxial-diff-2cyl:082}
\end{align}
where, by \eqref{uniaxial-diff-2cyl-dis:051} and \eqref{uniaxial-diff-2cyl-dis:052},
\begin{equation}\label{uniaxial-diff-2cyl-dis:090}
  \jump{\chp}= \big(\chpf(c^\bn{+})-\chpf(c^\bn{-})\big)
  - (\isp^\bn{+}-\isp^\bn{-})
  + \frac{\alpha}{\rhoo}\,\big(
   \big(\ps^\bn{+} + \ustrE_\Is(\lambda^\bn{+})\big) 
  -\big(\ps^\bn{-} + \ustrE_\Is(\lambda^\bn{-})\big)
  \big)\,.
\end{equation}
After substituting the $\isp$ jump expression \eqref{uniax-diff-iso:120}, 
from the \emph{microforce balance laws}
and, in turn, the constitutive expression \eqref{uniaxial-diff-2cyl-dis:070},
we end up with two evolution equations depending just on the external microforce 
$\spm$ jump and the external micro couple $\cgos$ which we designated 
as quantities characterizing \emph{cell activity}. 
Consistently with \eqref{act-diff:110}, \eqref{act-diff:120}, and \eqref{act-diff:122}, we set
\begin{equation}\label{uniaxial-diff-2cyl-dis:071}
   \chpf(c^\bn{\pm}) = \chpf_\cvx(c^\bn{\pm}) = -k_\cvx\,\arctanh(1-{2\,\bar{c}^\bn{\pm}})\,,
\end{equation}
with \mbox{$\,\bar{c}^\bn{\pm}={c^\bn{\pm}}/{c_{max}}$}, and
\begin{equation}\label{uniaxial-diff-2cyl-dis:072}
   \spm^\bn{\pm} =\spmf(c^\bn{\pm},\gamma) 
   = k_\spn\,\frac{\gamma\,(c^\bn{\pm} - c_\spn)}{\exp\big(\kdel\,(c^\bn{\pm} - c_\spn)^2\big)}
   = -\chpf_\spn(c^\bn{\pm},\gamma)
   \,.
\end{equation}
The final expression for the chemical potential jump reads
\begin{equation}\label{uniaxial-diff-2cyl-dis:100}
  \begin{aligned}
  \jump{\chp}&= \big(
   \chpf_\cvx(c^\bn{+})
  +\chpf_\spn(c^\bn{+},\gamma) 
  \big)
  - \big(
   \chpf_\spn(c^\bn{-},\gamma)
  +\chpf_\cvx(c^\bn{-})
  \big)
  + \frac{2}{\rhoo\,L}\,\ctros\\[\jot]
  &\quad
  + \frac{\alpha}{\rhoo}\,
   \big(\ps^\bn{+} + \ustrE_\Is(\lambda^\bn{+})\big) 
  - \frac{\alpha}{\rhoo}\,
   \big(\ps^\bn{-} + \ustrE_\Is(\lambda^\bn{-})\big)
  \,.
  \end{aligned}
\end{equation}
We could describe a time evolution of $c^\bn{+}$ and $c^\bn{-}$, corresponding to given values 
for $\gamma$ and $\cgos$, by integrating \eqref{uniaxial-diff-2cyl:081}
and \eqref{uniaxial-diff-2cyl:082}.

Otherwise, since by \eqref{uniaxial-diff-2cyl:048} the average concentration  $\,c_{m}\,$ is independent of time,
we may substitute 
\begin{equation}\label{uniaxial-diff-2cyl-dis:110}
   c^\bn{-} = 2\,c_{m} - c^\bn{+}
\end{equation}
into \eqref{uniaxial-diff-2cyl-dis:100} and search for its zeros to find stationary solutions.

\renewcommand{\scale}{0.8}
\begin{figure}[!t]
\setlength{\unitlength}{\scale pt} 
\centering
\boxed{
\begin{picture}(261,162)
{\includegraphics[viewport= 0 0 260 161, scale=\scale, clip]{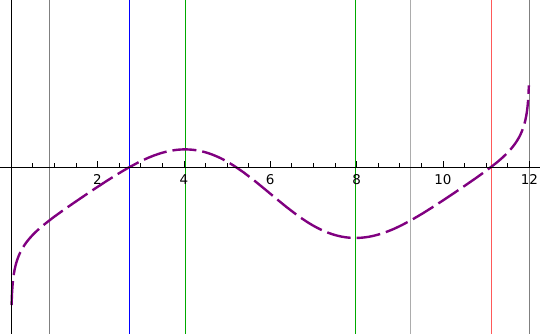}}
\end{picture}
}
\caption{Chemical potential jump $\jump{\chp}$ vs $c^\bn{+}$\,, with
 \mbox{$\cgos = \RshiftIgo$},
 \mbox{$\gamma = 0.25$},
 \mbox{$c_{max} = 12$}, 
 \mbox{$c_\spn = 0.5\,c_{max}$}, 
 \mbox{$\kdel = 0.1$}, 
 \mbox{$k_\cvx/k_\spn = 0.2$}. 
 Vertical lines mark $c^\bn{+}$ values,
 $c_{12}=\RshiftCoexSolP$ (red line) and 
 $c_{22}=\RshiftCoexSolAltP$ (blue line),
 where $\jump{\chp}=0$,
 as well as the complementary $c^\bn{-}$ values, 
 $c_{21}=\RshiftCoexSolM$ and
 $c_{11}=\RshiftCoexSolAltM$, 
 \mbox{i.e.} the corresponding coexistent concentrations.
 Vertical green lines mark the spinodal interval $[\SpinMin, \SpinMax]$.}
\label{fig-2cyl:1-muPlot0}
\end{figure}

\begin{figure}[t!]
\setlength{\unitlength}{\scale pt} 
\centering
\boxed{
\begin{picture}(261,162)
   \put(0,0){\includegraphics[viewport= 0 0 260 161, scale=\scale, clip]{flat-fig-2cyl-P_1-muPlot0.pdf}}
   \put(0,0){\includegraphics[viewport= 0 0 260 161, scale=\scale, clip]{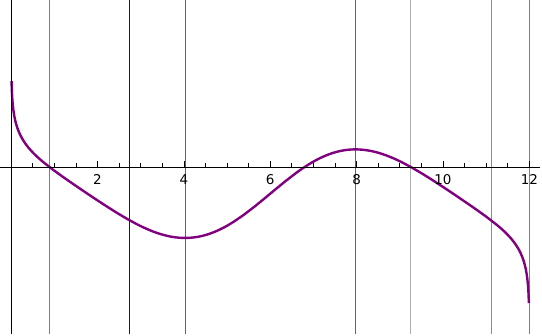}}
\end{picture}
}
\caption{Graph of the chemical potential jump $\jump{\chp}$ vs $c^\bn{-}$ (solid line)\,, 
with
 \mbox{$\cgos = \RshiftIgo$},
 \mbox{$\gamma = 0.25$},
 \mbox{$c_{max} = 12$}, 
 \mbox{$c_\spn = 0.5\,c_{max}$}, 
 \mbox{$\kdel = 0.1$}, 
 \mbox{$k_\cvx/k_\spn = 0.2$}, 
 overlapped to the graph of the chemical potential jump $\jump{\chp}$ vs $c^\bn{+}$ (dashed line)
 in \figref{fig-2cyl:1-muPlot0}.
 Vertical lines mark $c^\bn{-}$ values,
 $c_{21}=\RshiftCoexSolM$ and
 $c_{11}=\RshiftCoexSolAltM$, 
 where $\jump{\chp}=0$ in the first graph (solid line),
 as well as the complementary $c^\bn{+}$ values, 
 $c_{12}=\RshiftCoexSolP$ (red line) and 
 $c_{22}=\RshiftCoexSolAltP$ (blue line),
 \mbox{i.e.} the corresponding coexistent concentrations,
 where $\jump{\chp}=0$ in the second graph (dashed line).
 The two graphs are meant to  describe a complete set of solutions,
 as well as to highlight their symmetry, as summarized by the first row in Table~\ref{table-2cyl:summary-1-2}.}
\label{flat-fig-2cyl-P-M:1-muPlot0}
\end{figure}

\begin{figure}[t!]
\setlength{\unitlength}{\scale pt} 
\centering
\boxed{
\begin{picture}(261,162)
{\includegraphics[viewport= 0 0 260 161, scale=\scale, clip]{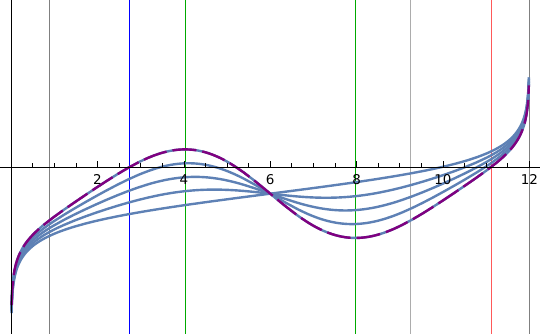}}
\end{picture}
}
\caption{Chemical potential jump $\jump{\chp}$ vs $c^\bn{+}$\,, with
 \mbox{$\cgos = \RshiftIgo$},
 \mbox{$\gamma \in [0, 0.25]$},
 \mbox{$c_{max} = 12$}, 
 \mbox{$c_\spn = 0.5\,c_{max}$}, 
 \mbox{$\kdel = 0.1$}, 
 \mbox{$k_\cvx/k_\spn = 0.2$}.
 Vertical lines mark $c^\bn{+}$ values,
 $c_{12}=\RshiftCoexSolP$ (red line) and
 $c_{22}=\RshiftCoexSolAltP$ (blue line),
 where $\jump{\chp}=0$,
 as well as the complementary $c^\bn{-}$ values, 
 $c_{11}=\RshiftCoexSolAltM$ and  
 $c_{21}=\RshiftCoexSolM$,
 \mbox{i.e.} the corresponding coexistent concentrations.}
\label{fig-2cyl:1-muPlot1}
\end{figure}

\begin{figure}[t]
\setlength{\unitlength}{\scale pt} 
\centering
\boxed{
\begin{picture}(261,162)
{\includegraphics[viewport= 0 0 260 161, scale=\scale, clip]{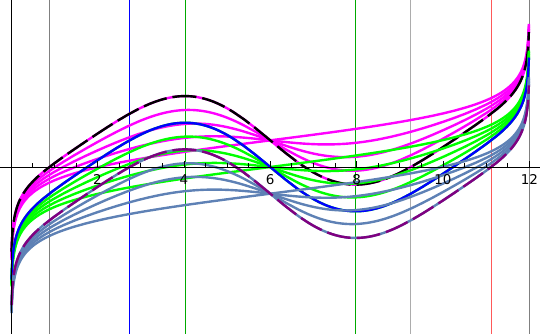}}
\end{picture}
}
\caption{Chemical potential jump $\jump{\chp}$ vs $c^\bn{+}$\,, with
 \mbox{$\cgos = \RshiftIgo$} (below), \mbox{$\cgos = \Igo$} (middle, dashed), \mbox{$\cgos = \LshiftIgo$} (above),
 \mbox{$\gamma \in [0, 0.25]$},
 \mbox{$c_{max} = 12$}, 
 \mbox{$c_\spn = 0.5\,c_{max}$}, 
 \mbox{$\kdel = 0.1$}, 
 \mbox{$k_\cvx/k_\spn = 0.2$}.}
\label{fig-2cyl:1-muPlot2}
\end{figure}

\subsubsection{Chemical potential primitive function\label{sect:uniaxial-diff-2cyl-ene}} 

A different way of looking at the stationary solutions relies on the graph of the \emph{total free energy}.
Let us start again from \eqref{uniaxial-diff-2cyl-dis:051} and \eqref{uniaxial-diff-2cyl-dis:052} and
expand them by replacing \eqref{uniax-diff-iso:110} and \eqref{uniax-diff-iso:111} 
while referring to definitions
\eqref{uniaxial-diff-2cyl-dis:071} and  \eqref{uniaxial-diff-2cyl-dis:072}
\begin{align}
  \chp^\bn{+} &= 
  \chpf_\cvx(c^\bn{+}) 
  - \spmf(c^\bn{+},\gamma) 
  + \frac{1}{\rhoo\,L}\,(\icgos-\cgos)
  + \frac{\alpha}{\rhoo}\,\big(\ps^\bn{+} + \ustrE_\Is(\lambda^\bn{+})\big)
  \,,\label{uniaxial-diff-2cyl-ene:051}
  \\
  \chp^\bn{-} &= 
  \chpf_\cvx(c^\bn{-}) - \spmf(c^\bn{-},\gamma) 
  - \frac{1}{\rhoo\,L}\,(\icgos-\cgos)
  + \frac{\alpha}{\rhoo}\,\big(\ps^\bn{-} 
  + \ustrE_\Is(\lambda^\bn{-})\big)
  \,.\label{uniaxial-diff-2cyl-ene:052}
\end{align}
That is just what we did already to get to the jump expression \eqref{uniaxial-diff-2cyl-dis:100}.
It is worth recalling that the constitutive expressions above, as well as the more general \eqref{act-gfreeEn:170},
originate from a dissipation inequality based on the free energy expression \eqref{spd-gfreeEn:010},
and on a constitutive characterization of the external microforce $\,\spm\,$ 
and the external micro couple $\,\cgos\,$,
which we designated as quantities accounting for the migrating cell activity in \secref{sect:act-chp} and \secref{sect:go}.

Let us now go backward, hopefully up to the free energy expression \eqref{spd-gfreeEn:010}, 
and dig out how it is related to the zeros of the chemical potential jump. 
To be clear, let us look for a primitive function of the chemical potential $\chp$ taking values
$\chp^\bn{+}$ at $c^\bn{+}$ and $\chp^\bn{-}$ at $c^\bn{-}$.

Recall first that we assigned the expression \eqref{act-diff:100} to the energy density $\,\strE_\cvx\,$ 
and derived the chemical potential $\,\chpf_\cvx\,$, such that by \eqref{act-diff:110} 
\begin{equation}\label{uniaxial-diff-2cyl-ene:060}
   \rhoo\,\chpf_\cvx(c^\bn{\pm})\,\dot{c}^\bn{\pm} = \frac{d}{dt}\,\strE_\cvx(c^\bn{\pm})  \,. 
\end{equation}
Afterward we assigned the expression \eqref{act-diff:120} to the microforce $\,\spm\,$ 
and defined the energy density $\,\strE_\spn\,$, such that by \eqref{act-diff:122} and \eqref{act-diff:150}  
\begin{equation}\label{uniaxial-diff-2cyl-ene:070}
   \rhoo\,\chpf_\spn(c^\bn{\pm},\gamma) = -\rhoo\,\spmf(c^\bn{\pm},\gamma)\,\dot{c}^\bn{\pm} = 
   \frac{d}{dt}\,\strE_\spn(c^\bn{\pm},\gamma) \,. 
\end{equation}

Let us recall also the strain energy derivative relations \eqref{uniax-diff-oc:240}
\begin{equation*}   
   {\alpha}\,\big(\ps^\bn{\pm} + \ustrE_\Is(\lambda^\bn{\pm})\big)\,\dot{c}^\bn{\pm} 
   = 
   \frac{d}{dt}\,\big(\beta^\bn{\pm}\,\ustrE_\Is(\lambda^\bn{\pm})\big) \,.
\end{equation*}

Further, by the gradient energy expression \eqref{uniaxial-diff-2cyl-dis:040} 
and the constitutive expression \eqref{uniaxial-diff-2cyl-dis:070}, it turns out
\begin{equation}\label{uniaxial-diff-2cyl-ene:090}
   \icgos\,\jump{\dot{c}}\,\frac{1}{L} 
   = 
   \frac{d}{dt}\,\tilde{\strE}_\gras(\jump{c}) 
   \,. 
\end{equation}

From \eqref{uniaxial-diff-2cyl-ene:051}, \eqref{uniaxial-diff-2cyl-ene:052} 
we can write down an expression for the \emph{total chemical potential power} 
\begin{equation}\label{uniaxial-diff-2cyl-ene:105}
   \begin{aligned}
   &{\refV^\bn{+}}\,\rhoo\,\chp^\bn{+}\,\dot{c}^\bn{+} + {\refV^\bn{-}}\,\rhoo\,\chp^\bn{-}\,\dot{c}^\bn{-} 
   \\
   &\qquad\qquad
   =
   {\refV^\bn{+}}\,\rhoo\,\big(
    \chpf_\cvx(c^\bn{+}) + \chpf_\spn(c^\bn{+},\gamma) 
   +\frac{\alpha}{\rhoo}\,\big(\ps^\bn{+} + \ustrE_\Is(\lambda^\bn{+})\big)
   \big)\,\dot{c}^\bn{+}
   \\
   &\qquad\qquad +
   {\refV^\bn{-}}\,\rhoo\,\big(
    \chpf_\cvx(c^\bn{-}) + \chpf_\spn(c^\bn{-},\gamma) 
   +\frac{\alpha}{\rhoo}\,\big(\ps^\bn{-} + \ustrE_\Is(\lambda^\bn{-})\big)
   \big)\,\dot{c}^\bn{-}
   \\
   &\qquad\qquad\quad +
   {\refV^\bn{+}}\,\frac{1}{L}\,(\icgos-\cgos)\,\dot{c}^\bn{+}
   \\ 
   &\qquad\qquad\quad -
   {\refV^\bn{-}}\,\frac{1}{L}\,(\icgos-\cgos)\,\dot{c}^\bn{-}
   \,.
   \end{aligned}
\end{equation}
By \eqref{uniaxial-diff-2cyl:044} we can get rid of the volume and simplify to
\begin{equation}\label{uniaxial-diff-2cyl-ene:106}
   \begin{aligned}
   \rhoo\,\chp^\bn{+}\,\dot{c}^\bn{+} + \rhoo\,\chp^\bn{-}\,\dot{c}^\bn{-} 
   &=
   \rhoo\,\big(\chpf_\cvx(c^\bn{+}) + \chpf_\spn(c^\bn{+},\gamma) 
   +\frac{\alpha}{\rhoo}\,\big(\ps^\bn{+} + \ustrE_\Is(\lambda^\bn{+})\big)
   \big)\,\dot{c}^\bn{+}
   \\
   &+
   \rhoo\,\big(\chpf_\cvx(c^\bn{-}) + \chpf_\spn(c^\bn{-},\gamma) 
   +\frac{\alpha}{\rhoo}\,\big(\ps^\bn{-} + \ustrE_\Is(\lambda^\bn{-})\big)
   \big)\,\dot{c}^\bn{-}
   \\
   &\quad +
   \frac{1}{L}\,\icgos\,\jump{\dot{c}}
   -
   \frac{1}{L}\,\cgos\,\jump{\dot{c}}
   \,.
   \end{aligned}
\end{equation}
By replacing the energy relations
\eqref{uniaxial-diff-2cyl-ene:060},
\eqref{uniaxial-diff-2cyl-ene:070},
\eqref{uniaxial-diff-2cyl-ene:090},
\eqref{uniax-diff-oc:240},
we get the previous expression transformed into

\begin{equation}\label{uniaxial-diff-2cyl-ene:110}
   \begin{aligned}
   \rhoo\,\chp^\bn{+}\,\dot{c}^\bn{+} + \rhoo\,\chp^\bn{-}\,\dot{c}^\bn{-} 
   &=\frac{d}{dt}\,\Big(\strE_\cvx(c^\bn{+}) + \strE_\cvx(c^\bn{-})\Big)\\
   &+\frac{d}{dt}\,\Big(\strE_\spn(c^\bn{+},\gamma) + \strE_\spn(c^\bn{-},\gamma)\Big)\\
   &+\frac{d}{dt}\,\Big(\beta^\bn{+}\,\ustrE_\Is(\lambda^\bn{+}) + \beta^\bn{-}\,\ustrE_\Is(\lambda^\bn{-})\Big)\\
   &+\frac{d}{dt}\,\Big(\tilde{\strE}_\gras(\jump{c})\Big)\\
   &-\frac{1}{L}\,\cgos\,\jump{\dot{c}}   
   \,.
   \end{aligned}
\end{equation}
\subsubsection{Common tangent construction\label{sect:uniaxial-diff-2cyl-}} 
By \eqref{uniaxial-diff-2cyl:048}, we can further transform \eqref{uniaxial-diff-2cyl-ene:110} into
\begin{equation}\label{uniaxial-diff-2cyl-ene:120}
   \begin{aligned}
   \rhoo\,\jump{\chp}\,\dot{c}^\bn{+} 
   &=\frac{d}{dt}\,\Big(\strE_\cvx(c^\bn{+}) + \strE_\cvx(c^\bn{-})\Big)\\
   &+\frac{d}{dt}\,\Big(\strE_\spn(c^\bn{+},\gamma) + \strE_\spn(c^\bn{-},\gamma)\Big)\\
   &+\frac{d}{dt}\,\Big(\beta^\bn{+}\,\ustrE_\Is(\lambda^\bn{+}) + \,\beta^\bn{-}\,\ustrE_\Is(\lambda^\bn{-})\Big)\\
   &+\frac{d}{dt}\,\Big(\tilde{\strE}_\gras(\jump{c})\Big)\\
   &-\frac{2}{L}\,\cgos\,\dot{c}^\bn{+}   
   \,,
   \end{aligned}
\end{equation}
showing its relation to \eqref{uniaxial-diff-2cyl-dis:100}.
This allows us to define the \emph{total free energy density} 
\begin{equation}\label{uniaxial-diff-2cyl-ene:130}
   \begin{aligned}
     \hat\freeE_\effs(c^\bn{+},\gamma) &=\big(\strE_\cvx(c^\bn{+}) + \strE_\cvx(c^\bn{-})\big)
    +\big(\strE_\spn(c^\bn{+},\gamma) + \strE_\spn(c^\bn{-},\gamma)\big)\\[2\jot]
   &+\big(\beta^\bn{+}\,\ustrE_\Is(\lambda^\bn{+}) + \,\beta^\bn{-}\,\ustrE_\Is(\lambda^\bn{-})\big)
    +\tilde{\strE}_\gras(\jump{c}) \,,
   \end{aligned}
\end{equation}
with \mbox{$\,c^\bn{-} = 2\,c_{m} - c^\bn{+}$}, as in \eqref{uniaxial-diff-2cyl-dis:110}, and such that
\begin{equation}\label{uniaxial-diff-2cyl-ene:140}
   \rhoo\,\jump{\chp}\,\dot{c}^\bn{+} =
   \frac{d}{dt}\,\hat\freeE_\effs(c^\bn{+},\gamma)
   -\frac{2}{L}\,\cgos\,\dot{c}^\bn{+}\,.
\end{equation}

Therefore the values of $c^\bn{+}$ where $\,\jump{\chp}=0$ are characterized by the property
\begin{equation}\label{uniaxial-diff-2cyl-ene:150}
   \left.\frac{d}{dt}\,\hat\freeE_\effs(c^\bn{+},\gamma)\right|_{\jump{\chp}=0}
   =\frac{2}{L}\,\cgos\,\dot{c}^\bn{+}\,.
\end{equation}
No properties have been assumed for $\cgos$ until now. If $\,\cgos\,\dot{c}^\bn{+}\,$ is integrable then we
can define an \emph{augmented free energy} $\hat\freeE_\augs$ such that
\begin{equation}\label{uniaxial-diff-2cyl-ene:160}
   \frac{d}{dt}\,\hat\freeE_\augs(c^\bn{+}, \gamma, \cgos)=
   \frac{d}{dt}\,\hat\freeE_\effs(c^\bn{+}, \gamma)
   -\frac{2}{L}\,\cgos\,\dot{c}^\bn{+}\,.
\end{equation}



\renewcommand{\scale}{0.8}
\begin{figure}[t!]
\setlength{\unitlength}{\scale pt} 
\centering
\boxed{
\begin{picture}(261,162)
{\includegraphics[viewport= 0 0 260 161, scale=\scale, clip]{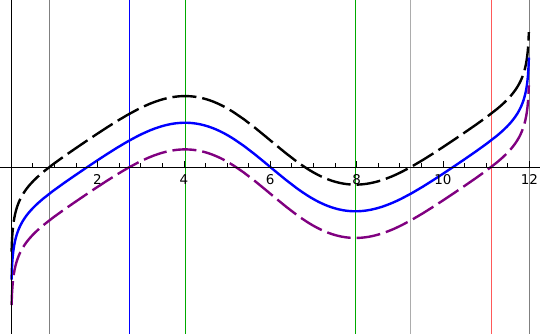}}
\end{picture}
}
\caption{Chemical potential jump $\jump{\chp}$ vs $c^\bn{+}$\,, with
 \mbox{$\cgos = \RshiftIgo$} (below, dashed line), 
 \mbox{$\cgos = \Igo$} (middle, solid line), 
 \mbox{$\cgos = \LshiftIgo$} (above. dashed line),
 \mbox{$\gamma = 0.25$}.
 Vertical lines are inherited from the previous graphs.
 }
\label{fig-2cyl:1-muPlot20}
\end{figure}

\begin{figure}[h!]
\setlength{\unitlength}{\scale pt} 
\centering
\boxed{
\begin{picture}(261,162)
{\includegraphics[viewport= 0 0 260 161, scale=\scale, clip]{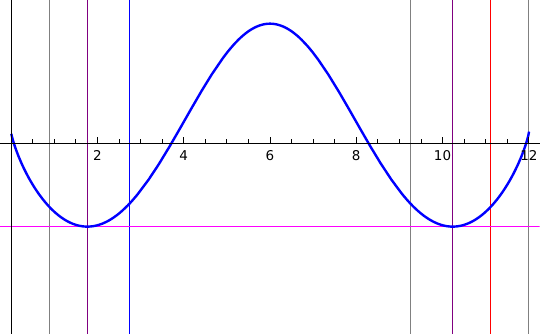}}
\end{picture}
}
\caption{Double well total free energy $\hat\freeE_\effs$ vs $c^\bn{+}$\,, with
 \mbox{$\gamma = 0.25$},
 \mbox{$c_{max} = 12$}, 
 \mbox{$c_\spn = 0.5\,c_{max}$}, 
 \mbox{$\kdel = 0.1$}, 
 \mbox{$k_\cvx/k_\spn = 0.2$}.
 Two additional vertical (purple) lines mark the local minima at 
 $c_{01}=\CoexSolM$ and 
 $c_{02}=\CoexSolAltM$.}
\label{fig-2cyl:1-EnPlot0}
\end{figure}

It is worth noting that stationary solutions to \eqref{uniaxial-diff-2cyl:081} with $c^\bn{+}$ lying inside 
the spinodal interval are linearly unstable since the slope of the $\jump{\chp}$ graph is negative.

\begin{figure}[!t]
\setlength{\unitlength}{\scale pt} 
\centering
\boxed{
\begin{picture}(261,162)
{\includegraphics[viewport= 0 0 260 161, scale=\scale, clip]{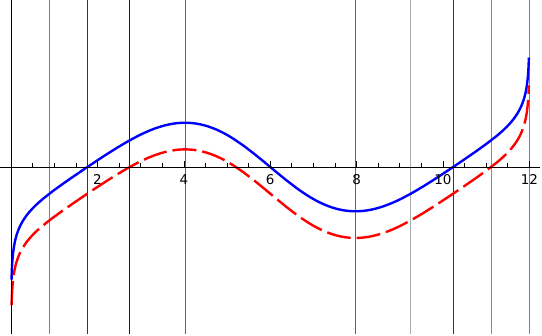}}
\end{picture}
}
\caption{Chemical potential jump $\jump{\chp}$ vs $c^\bn{+}$ (below, dashed line),
with \mbox{$\cgos = \RshiftIgo$}, 
compared to the derivative of the total free energy $\hat\freeE_\effs$ (above, solid line)\,, 
with
 \mbox{$\gamma = 0.25$},
 \mbox{$c_{max} = 12$}, 
 \mbox{$c_\spn = 0.5\,c_{max}$}, 
 \mbox{$\kdel = 0.1$}, 
 \mbox{$k_\cvx/k_\spn = 0.2$}.
 The gap is \mbox{$-2\,\cgos/L$}.
 All vertical lines are inherited from the previous graphs.}
\label{fig-2cyl:1-muPlot3}
\end{figure}

\begin{figure}[h]
\setlength{\unitlength}{\scale pt} 
\centering
\boxed{
\begin{picture}(261,162)
{\includegraphics[viewport= 0 0 260 161, scale=\scale, clip]{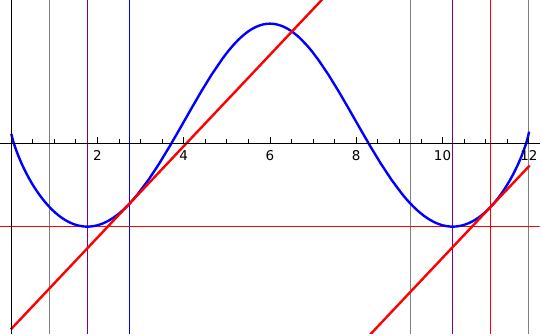}}
\end{picture}
}
\caption{Double well total free energy graph $\hat\freeE_\effs$ vs $c^\bn{+}$\,, with
 \mbox{$\cgos = \RshiftIgo$},
 \mbox{$\gamma = 0.25$},
 \mbox{$c_{max} = 12$}, 
 \mbox{$c_\spn = 0.5\,c_{max}$}, 
 \mbox{$\kdel = 0.1$}, 
 \mbox{$k_\cvx/k_\spn = 0.2$}.
 The two tangent lines, with the same slope $2\,\cgos/L$, touch the graph at  
 $c_{12}=\LshiftCoexSolAltM$ (red vertical line), and 
 $c_{22}=\LshiftCoexSolM$ (blue vertical line).}
\label{fig-2cyl:1-EnPlot2}
\end{figure}

\begin{figure}
\setlength{\unitlength}{\scale pt} 
\centering
\boxed{
\begin{picture}(261,162)
{\includegraphics[viewport= 0 0 260 161, scale=\scale, clip]{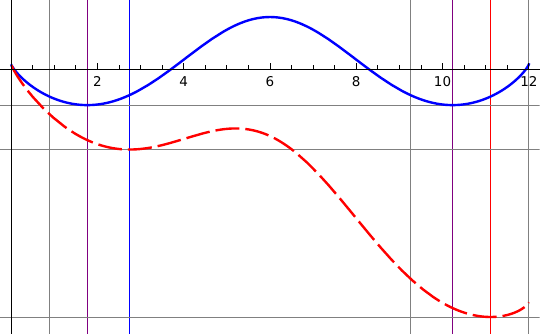}}
\end{picture}
}
\caption{Double well augmented free energy graph $\hat\freeE_\augs$  vs $c^\bn{+}$ (below)
compared to the total free energy $\hat\freeE_\effs$ (above, rescaled)\,, with
 \mbox{$\cgos = \RshiftIgo$},
 \mbox{$\gamma = 0.25$},
 \mbox{$c_{max} = 12$}, 
 \mbox{$c_\spn = 0.5\,c_{max}$}, 
 \mbox{$\kdel = 0.1$}, 
 \mbox{$k_\cvx/k_\spn = 0.2$}.
 The local minima are at  
 $c_{12}=\LshiftCoexSolAltM$ (red vertical line) and 
 $c_{22}=\LshiftCoexSolM$ (blue vertical line).}
\label{fig-2cyl:1-EnPlot3}
\end{figure}

\clearpage
 
\renewcommand{\scale}{0.8}
\begin{figure}
\setlength{\unitlength}{\scale pt} 
\centering
\boxed{
\begin{picture}(261,162)
{\includegraphics[viewport= 0 0 260 161, scale=\scale, clip]{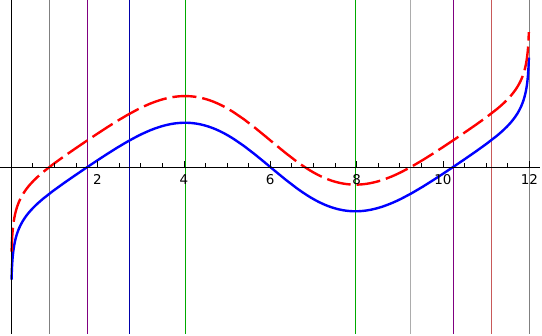}}
\end{picture}
}
\caption{Chemical potential jump $\jump{\chp}$ \,vs\, $c^\bn{+}$ (above, dashed line),
with \mbox{$\cgos = \LshiftIgo$},
compared to the derivative of the total free energy $\hat\freeE_\effs$ (below, solid line)\,, 
with
 \mbox{$\gamma = 0.25$},
 \mbox{$c_{max} = 12$}, 
 \mbox{$c_\spn = 0.5\,c_{max}$}, 
 \mbox{$\kdel = 0.1$}, 
 \mbox{$k_\cvx/k_\spn = 0.2$}.
 The gap is \mbox{$-2\,\cgos/L$}.
 All vertical lines are inherited from the previous graphs.}
\label{fig-2cyl:2-muPlot3}
\end{figure}

\begin{figure}[t!]
\setlength{\unitlength}{\scale pt} 
\centering
\boxed{
\begin{picture}(261,162)
{\includegraphics[viewport= 0 0 260 161, scale=\scale, clip]{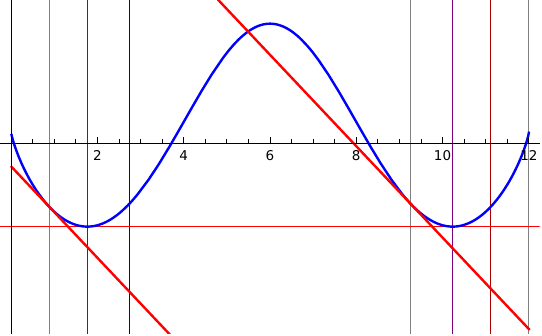}}
\end{picture}
}
\caption{Double well total free energy graph $\hat\freeE_\effs$ \,vs\, $c^\bn{+}$\,, with
 \mbox{$\cgos = \LshiftIgo$},
 \mbox{$\gamma = 0.25$},
 \mbox{$c_{max} = 12$}, 
 \mbox{$c_\spn = 0.5\,c_{max}$}, 
 \mbox{$\kdel = 0.1$}, 
 \mbox{$k_\cvx/k_\spn = 0.2$}.
 The two tangent lines, with the same slope $2\,\cgos/L$, touch the graph at  
 $c_{11}=\RshiftCoexSolAltM$ (gray vertical line), and 
 $c_{21}=\RshiftCoexSolM$ (gray vertical line).}
\label{fig-2cyl:2-EnPlot2}
\end{figure}

\begin{figure}
\setlength{\unitlength}{\scale pt} 
\centering
\boxed{
\begin{picture}(261,162)
{\includegraphics[viewport= 0 0 260 161, scale=\scale, clip]{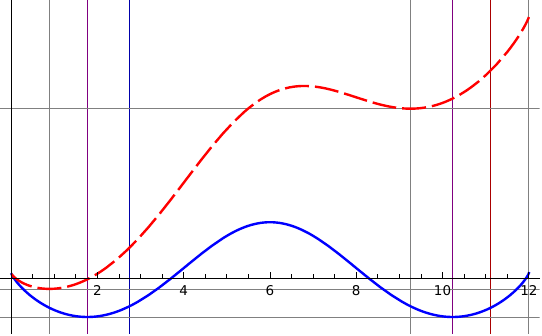}}
\end{picture}
}
\caption{Double well augmented free energy graph $\hat\freeE_\augs$ \,vs\, $c^\bn{+}$ (above)
compared to the total free energy $\hat\freeE_\effs$ (below, rescaled)\,, with
 \mbox{$\cgos = \LshiftIgo$},
 \mbox{$\gamma = 0.25$},
 \mbox{$c_{max} = 12$}, 
 \mbox{$c_\spn = 0.5\,c_{max}$}, 
 \mbox{$\kdel = 0.1$}, 
 \mbox{$k_\cvx/k_\spn = 0.2$}.
 The local minima are at 
 $c_{11}=\RshiftCoexSolAltM$ (gray vertical line) and 
 $c_{21}=\RshiftCoexSolM$ (gray vertical line).}
\label{fig-2cyl:2-EnPlot3}
\end{figure}

 
\clearpage
\begin{table}
\centering
\boxed{
\begin{tabular}{cc}
		\multicolumn{2}{c}{\rule{0pt}{4ex}$\cgos = \RshiftIgo$}\\
	\begin{tabular}{ccc}
		\rule{0pt}{3ex}
		  \makebox[2ex][l]{$\bbn{-}$} 
		& \makebox[2cm][c]{$\longrightarrow$} 
		& \makebox[2ex][l]{$\bbn{+}$}
	\end{tabular}
&	
	\begin{tabular}{ccc}
		\rule{0pt}{3ex}
		  \makebox[2ex][l]{$\bbn{-}$} 
		& \makebox[2cm][c]{$\longrightarrow$} 
		& \makebox[2ex][l]{$\bbn{+}$}
	\end{tabular}
	\\
	\begin{tabular}{|c|c|}
		\hline
		\rule{0pt}{3ex}
		  \makebox[2cm][c]{$c_{21}$} 
		& \makebox[2cm][c]{$c_{12}$}\\
		\rule{0pt}{3ex}
		  \makebox[2.6cm][c]{$\RshiftCoexSolM$} 
		& \makebox[2.6cm][c]{$\RshiftCoexSolP$}\\
		\hline
	\end{tabular}
&
	\begin{tabular}{|c|c|}
		\hline
		\rule{0pt}{3ex}
		  \makebox[2cm][c]{$c_{11}$} 
		& \makebox[2cm][c]{$c_{22}$}\\
		\rule{0pt}{3ex}
		  \makebox[2.6cm][c]{$\RshiftCoexSolAltM$} 
		& \makebox[2.6cm][c]{$\RshiftCoexSolAltP$}\\
		\hline
	\end{tabular}
	\\ 
	\begin{tabular}{c}
		\rule{0pt}{4.0ex}
		  \makebox[6.5cm][c]{$\ctros=\RshiftIgio-\RshiftIgo=\RshiftItauStar$}\\
		\rule{0pt}{3.0ex}
		  \makebox[6.5cm][c]{$\jump{\chp}=\RshiftImuxJmusJ + \RshiftIeshJ + \ctros$}
	\end{tabular}
&	
	\begin{tabular}{c}
		\rule{0pt}{4.0ex}
		  \makebox[6.5cm][c]{$\ctros=\RshiftIgioAlt-\RshiftIgoAlt=\RshiftItauStarAlt$}\\
		\rule{0pt}{3.0ex}
		  \makebox[6.5cm][c]{$\jump{\chp}=\RshiftImuxJmusJAlt + (\RshiftIeshJAlt) + \ctros$}
	\end{tabular}
	\vspace{2ex}\\
	\hline
	\multicolumn{2}{c}{\rule{0pt}{4ex}$\cgos = \Igo$}\\
	\begin{tabular}{ccc}
		\rule{0pt}{3.4ex}
		  \makebox[2ex][l]{$\bbn{-}$} 
	    & \makebox[2cm][c]{}
	    & \makebox[2ex][l]{$\bbn{+}$}
	\end{tabular}
&	
	\begin{tabular}{ccc}
		\rule{0pt}{3.4ex}
		  \makebox[2ex][l]{$\bbn{-}$} 
		& \makebox[2cm][c]{}
		& \makebox[2ex][l]{$\bbn{+}$}
	\end{tabular}
	\\
	\begin{tabular}{|c|c|}
		\hline
		\rule{0pt}{3ex}
		  \makebox[2cm][c]{$c_{01}$} 
		& \makebox[2cm][c]{$c_{02}$}\\
		\rule{0pt}{3ex}
		  \makebox[2.6cm][c]{$\CoexSolM$} 
		& \makebox[2.6cm][c]{$\CoexSolP$}\\
		\hline
	\end{tabular}
&	
	\begin{tabular}{|c|c|}
		\hline
		\rule{0pt}{3ex}
		  \makebox[2cm][c]{$c_{02}$} 
		& \makebox[2cm][c]{$c_{01}$}\\
		\rule{0pt}{3ex}
		  \makebox[2.6cm][c]{$\CoexSolAltM$} 
		& \makebox[2.6cm][c]{$\CoexSolAltP$}\\
		\hline
	\end{tabular}
	\\ 
	\begin{tabular}{c}
		\rule{0pt}{4.0ex}
		  \makebox[6.5cm][c]{$\ctros=\Igio-\Igo=\ItauStar$}\\
		\rule{0pt}{3.0ex}
		  \makebox[6.5cm][c]{$\jump{\chp}=\ImuxJmusJ + \IeshJ + \ctros$}
	\end{tabular}
&	
	\begin{tabular}{c}
		\rule{0pt}{4.0ex}
		  \makebox[6.5cm][c]{$\ctros=\IgioAlt-\IgoAlt=\ItauStarAlt$}\\
		\rule{0pt}{3.0ex}
		  \makebox[6.5cm][c]{$\jump{\chp}=\ImuxJmusJAlt + (\IeshJAlt) + \ctros$}
	\end{tabular}
	\vspace{2ex}\\   
	\hline
	\multicolumn{2}{c}{\rule{0pt}{4ex}$\cgos = \LshiftIgo$}\\
	\begin{tabular}{ccc}
		\rule{0pt}{3.4ex}
		  \makebox[2ex][l]{$\bbn{-}$} 
		& \makebox[2cm][c]{$\longleftarrow$} 
		& \makebox[2ex][l]{$\bbn{+}$}
	\end{tabular}
&	\begin{tabular}{ccc}
		\rule{0pt}{3.4ex}
		  \makebox[2ex][c]{$\bbn{-}$} 
		& \makebox[2cm][c]{$\longleftarrow$} 
		& \makebox[2ex][c]{$\bbn{+}$}
	\end{tabular}
	\\
	\begin{tabular}{|c|c|}
		\hline
		\rule{0pt}{3ex}
		  \makebox[2cm][c]{$c_{22}$} 
		& \makebox[2cm][c]{$c_{11}$}\\
		\rule{0pt}{3ex}
		  \makebox[2.6cm][c]{$\LshiftCoexSolM$} 
		& \makebox[2.6cm][c]{$\LshiftCoexSolP$}\\
		\hline
	\end{tabular}
&	\begin{tabular}{|c|c|}
		\hline
		\rule{0pt}{3ex}
		  \makebox[2cm][c]{$c_{12}$} 
		& \makebox[2cm][c]{$c_{21}$}\\
		\rule{0pt}{3ex}
		  \makebox[2.6cm][c]{$\LshiftCoexSolAltM$} 
		& \makebox[2.6cm][c]{$\LshiftCoexSolAltP$}\\
		\hline
	\end{tabular}
	\\ 
	\begin{tabular}{c}
		\rule{0pt}{4.0ex}
		  \makebox[6.5cm][c]{$\ctros=\LshiftIgio-(\LshiftIgo)=\LshiftItauStar$}\\
		\rule{0pt}{3.0ex}
		  \makebox[6.5cm][c]{$\jump{\chp}=\LshiftImuxJmusJ + \LshiftIeshJ + \ctros$}
	\end{tabular}
&	
	\begin{tabular}{c}
		\rule{0pt}{4.0ex}
		  \makebox[6.5cm][c]{$\ctros=\LshiftIgioAlt-(\LshiftIgoAlt)=\LshiftItauStarAlt$}\\
		\rule{0pt}{3.0ex}
		  \makebox[6.5cm][c]{$\jump{\chp}=\LshiftImuxJmusJAlt + (\LshiftIeshJAlt) + \ctros$}
	\end{tabular}
	\vspace{2ex}\\  
\end{tabular}	
}
\caption{Summary of stationary solutions ($\jump{\chp}=0$), with $\gamma=0.25$ and
$\cgos=\LshiftIgo,\, \Igo,\, \RshiftIgo\,$, according to the graphs in \figref{fig-2cyl:1-muPlot20}.
Each panel shows two solutions arranged in two columns,
with boxes depicting $\bbn{+}$ and $\bbn{-}$ cylinders;
$\bbn{+}$ boxes show the values of $c^\bn{+}$ which are zeros for $\jump{\chp}$ outside the spinodal interval, 
while $\bbn{-}$ boxes show the complementary values $(c^\bn{-}=c_{max}-c^\bn{+})$,
\mbox{i.e.} the corresponding coexistent concentrations.
The arrow running above the boxes stands for the $\cgo$ direction
(notice how $\cgo$ breaks the symmetry of the solutions displayed in the middle panel).
The value of $\ctros=(\icgos-\cgos)$, as defined in \eqref{uniaxial-diff-2cyl:098}, is displayed on the first line below the boxes, 
together with a scaled version of \eqref{uniaxial-diff-2cyl-dis:100} on the second line, 
where the second term (almost negligible) is the contribution of the mechanical stress (the Eshelby stress)
to the chemical potential jump. 
The panels are arranged in such a way to highlight the resulting symmetries of the solutions, which have been computed separately.}
\label{table-2cyl:summary-1-2}
\end{table}


\clearpage
\newpage

%
 
\end{document}